\documentclass[a4paper,11pt]{article}
\pdfoutput=1

\usepackage{jheppub}

\usepackage[T1]{fontenc}
\usepackage{graphicx}
\usepackage{epstopdf}
\usepackage{bm}
\usepackage{epsfig}
\usepackage{graphics}
\usepackage{xspace}
\usepackage{hyperref}
\usepackage{slashed}
\usepackage{xcolor}
\usepackage{longtable}
\usepackage[normalem]{ulem}
\usepackage{subcaption}
\usepackage{multirow}
\usepackage{orcidlink}
\usepackage{float}
\usepackage{booktabs}

\newcommand{\nl}{n_{\ell}}

\newcommand{\pole}{\mathrm{pole}}
\newcommand{\MSb}{\overline{\mathrm{MS}}}
\newcommand{\df}{{\rm d}}
\newcommand{\Ord}{\mathcal{O}}
\newcommand{\LQCD}{\Lambda_\mathrm{QCD}}

\title{\boldmath A Precise $\alpha_s$ Determination from the R-improved QCD Static Energy}

\author{Jose M. Mena-Valle\orcidlink{0009-0006-8966-1711},}
\author{Vicent Mateu\orcidlink{0000-0003-0902-5012}}
\author{and Pablo~G.~Ortega\orcidlink{0000-0002-1580-2706}}

\affiliation{Departamento de F\'isica Fundamental and IUFFyM, Universidad de Salamanca\\
Plaza de la Merced S/N, E-37008 Salamanca, Spain}

\emailAdd{jmmena@usal.es}
\emailAdd{vmateu@usal.es}
\emailAdd{pgortega@usal.es}

\abstract{The strong coupling $\alpha_s$ is determined with high precision from fits to lattice QCD simulations on the static energy. Our theoretical setup relies on \mbox{R-improving} the \mbox{three-loop} fixed-order prediction for the static energy
by removing its $u=1/2$ renormalon and summing up the associated large (infrared) logarithms which, in combination with radius-dependent renormalization scales (called profile functions) extends the validity of perturbation theory to distances up to $\sim 0.5\,$fm.
Furthermore, we resum large ultrasoft logarithms to N$^3$LL accuracy using
renormalization group evolution. We have checked that the standard four-loop R-evolution treats N$^4$LL and higher remnants in a non-symmetric way, hence we also account for this potential bias.
Our estimate of the perturbative uncertainty is based on a random scan over the parameters specifying the profile functions and the treatment of R-evolution.
We also devise a method to statistically combine into a single dataset results from independent simulations which use different lattice spacing and cover various ranges, which can be used to carry out
fits in a much faster way.
We explore the dependence of the extracted $\alpha_s$ value on the smallest and largest distances included in the dataset, on how R-evolution is treated,
on how the fit is performed, and on the accuracy of ultrasoft resummation. From our final analysis, after evolving to the $Z$-pole we obtain $\alpha^{(n_f=5)}_s(m_Z)=0.1166\pm 0.0009$, compatible with the world average with similar incertitude.}

\begin{document}
\maketitle
\flushbottom

\section{Introduction}\label{sec:intro}
The strong coupling $\alpha_s$ governs Quantum Chromodynamics (QCD for short) and, among other things, determines below which energy the strong interactions become confining. More importantly, precise $\alpha_s$ determinations are paramount for accurate predictions of cross sections at lepton and hadron colliders, which in turn are necessary to disentangle new physics signals from the Standard Model background. It also plays an important role in other contexts such as Higgs decays, B-physics, and quarkonia.

While the determination of $\alpha_s$ is not a new subject, in recent years tremendous progress has been achieved in reducing its uncertainty, either through improving standard methods or figuring out entirely new strategies. However, these efforts have partially obscured the landscape since a wide range of (sometimes) incompatible values are quoted (see e.g.\ Refs.~\cite{Salam:2017qdl,Pich:2018lmu,dEnterria:2022hzv} for reviews). In particular, some determinations from event shapes using analytic power corrections yield small but accurate results \cite{Gehrmann:2010uax,Abbate:2010xh,Abbate:2012jh,Gehrmann:2012sc,Hoang:2015hka,Bell:2023dqs,Benitez:2024nav,Benitez:2025vsp}, which have been excluded form the PDG 2024 world average. While traditionally $\alpha_s$ has been determined by figuring out observables highly sensitive to this parameter and comparing experimental measurements to the corresponding (perturbative) computations, lattice QCD has, since the moment unquenched simulations became possible, entered the game becoming one of the main characters. Lattice determinations are becoming so accurate that overly dominate the PDG wold average~\cite{ParticleDataGroup:2024cfk}. Some of these precise results come from Wilson loops~\cite{Davies:2008sw} or current-current charm pseudo-scalar correlators~\cite{McNeile:2010ji,HPQCD:2014aca,Nakayama:2016atf,Maezawa:2016vgv,Petreczky:2019ozv,Petreczky:2020tky},\footnote{See Refs.~\cite{Boito:2019pqp,Boito:2020lyp} for a similar strategy using experimental data on the vector-current charm- and bottom-quark correlators.} but there exist other strategies as well~\cite{Blossier:2013ioa,Bruno:2017gxd,Zafeiropoulos:2019flq,DallaBrida:2022eua}. For an extensive review on strong-coupling determinations from lattice QCD, see Ref.~\cite{FlavourLatticeAveragingGroupFLAG:2024oxs}.

An interesting
strategy to determine the strong coupling from lattice QCD simulations uses the QCD static energy, an (up to a constant) observable dominated by UV dynamics at small distances, which is known to
$\mathcal{O}(\alpha_s^4)$ in perturbation theory and is relatively simple to compute in the lattice. For these analyses, only three dynamical light quarks are required in the simulation. This observable can be thought of as the potential energy between a heavy quark-antiquark pair at a distance $r$, both at rest.\footnote{We will also refer to this distance as ``radius'' in the remainder of the article.} It plays a central role in describing quarkonium bound states and the production of a top-antitop pair at threshold, and provides a qualitative explanation for infrared slavery. Since the pioneering study of Ref.~\cite{Bazavov:2012ka}, based on the data of \cite{Bazavov:2011nk} and employing renormalon subtractions from the exact asymptotic form of the series (without resumming the associated logarithms), substantial progress has been achieved on the theoretical ---\,with more refined renormalon subtractions\,--- and lattice ---\,reaching higher precision at shorter distances\,--- sides. This has led to even more accurate determinations, such as those obtained in Refs.~\cite{Bazavov:2014soa,Bazavov:2019qoo} which, however, carry out fits restricted to
$r \leq 0.073\,$fm. To obtain the final result of Ref.~\cite{Bazavov:2019qoo} $\alpha_s^{(n_\ell=5)}(m_Z) = 0.1166(8)$ a dataset of only $19$ points defined by $a\leq r \leq 0.073\,$fm is used, employing simulations with different lattice spacing $a=\{0.0249, 0.0295, 0.035, 0.041, 0.049, 0.0606\}\,$fm corresponding to $\beta=\{8.4, 8.2, 8.0, 7.825, 7.596, 7.373\}$. Here $\beta=10/g^2$ where $g$ denotes the lattice QCD gauge coupling, so that larger $\beta$ corresponds to smaller $a$. Data with $\beta<8$ was released in Ref.~\cite{Bazavov:2014pvz}, whereas sets with smaller lattice spacing can be found in \cite{Bazavov:2017dsy}. The earlier analysis~\cite{Bazavov:2014soa} uses data from Ref.~\cite{Bazavov:2014pvz} only.

To the best of our knowledge, the most recent analysis using this method is Ref.~\cite{Ayala:2020odx}, which removes the two leading renormalons with $u=1/2$ and $u=3/2$ through a hyperasymptotic approximation.\footnote{See also Ref.~\cite{Ananthanarayan:2020umo} for an analysis where higher-order estimates from Pad\'e approximants are used.} Unlike the $u=1/2$ ambiguity that can be removed expressing the heavy quark mass in a short-distance scheme, the $u=3/2$ renormalon is related to a non-local condensate and is formally removed by its suitable redefinition. While the associated high-order behavior is removed in Ref.~\cite{Ayala:2020odx}, the condensate's effects are not accounted for. Data taken from Ref.~\cite{Bazavov:2019qoo} (corresponding to the results of Ref.~\cite{Bazavov:2017dsy} with updated errors, see also \cite{Bazavov:2018wmo} for more details)
restricted to \mbox{$r>0.0697\,$fm} is used, and four upper limits for the distance are considered \mbox{$r_{\rm max}=\{0.0985,0.1208,0.1579,0.1973\}\,$fm}, containing 8, 17, 31 and 50 points, respectively.\footnote{In fact, \cite{Ayala:2020odx} only uses data with results for the smallest lattice spacing $a\simeq 0.0249\,$fm.}
Their final value \mbox{$\alpha_s^{(n_f=5)}(m_Z)=0.1181(9)$} is obtained with the smallest dataset. Finally, in their fits correlations between lattice predictions for different distances were ignored, arguing they have a tiny impact on the strong coupling.

Whereas these analyses use HotQCD lattice results, based on MILC gluon ensembles and highly improved staggered quarks (HISQ), another study~\cite{Takaura:2018lpw,Takaura:2018vcy} uses JLQCD results based on the Symanzik gauge and M\"obius domain-wall quark action, fitting in a significantly larger window reaching distances $r\leq 0.4\,$fm encompassing a dataset with $22$ points. On the theory side, while the HotQCD analyses of Refs.~\cite{Bazavov:2014soa,Bazavov:2019qoo} remove the leading renormalon through a numerical integration of the force, using canonical renormalization scales which restrict the applicability range, the JLQCD analysis removes the two most important renormalons using the OPE framework, see Ref.~\cite{Sumino:2020mxk}, summing up large logarithms through a numerical Fourier transform.
A combined fit is carried out, which determines the strong coupling and performs the continuum extrapolation simultaneously, finding $\alpha_s^{(n_\ell=5)}(m_Z)=0.1180(15)$, larger than the HotQCD result and closer to the world average, although with somewhat larger uncertainties.

In Ref.~\cite{Mateu:2018zym} it was shown that the validity range of the static potential could be extended to larger distances if a)~the renormalon is removed with appropriate short-distance subtractions, b)~large logarithms associated to the renormalon are summed up using e.g.\ \mbox{R-evolution}, and c)~the dependence of renormalization scales on the radius is chosen such that perturbation theory is well behaved. It is demonstrated that, after implementing these requirements, the static potential agrees with HotQCD results up to distances of the order of $1$\,fm, reproducing the linear-rising behavior of the Cornell potential (obtained through fits to quarkonium data) at large distances. A rudimentary parametrization of the renormalization scales ---\,denoted as profile functions in what follows\,--- was employed, constructed with piecewise functions patched together in a smooth way, which implements a canonical scaling for small distances, smoothly freezing out at large $r$ values. Even though this simple functional form was enough to proof the concept, in this article we introduce smoother and more sophisticated parametrizations, depending on some
parameters that allow to estimate perturbative uncertainties through random scans. Since Ref.~\cite{Bazavov:2019qoo} uses canonical profiles and estimates the truncation uncertainty by varying the scale by factors of $2$ and $1/2$
their dataset is limited to $r< 0.14\,$fm as larger distances ruin perturbative convergence.\footnote{In the earlier analysis of Ref.~\cite{Bazavov:2014soa} the scale is varied between $\sqrt{2}$ and $1/\sqrt{2}$.} Our profile functions are more flexible and
our random scan for renormalization scales roughly covers the standard variation in the small $r$ region.

In this article we extract the strong coupling from fits to lattice results on the static energy using our R-improved theoretical prediction, carefully examining how the outcome depends on restrictions imposed on data,
e.g.\ on the minimum and maximum allowed distances,
denoted $r_{\rm min}$ and $r_{\rm max}$, respectively.
We explore a number of standard subtraction schemes, such as the MSR~\cite{Hoang:2008yj,Hoang:2017suc}
and PS~\cite{Beneke:1998rk} masses, for which \mbox{R-evolution} is implemented. We have observed that the latter is sensitive to infrared physics and hence not adequate for high-precision analyses. In Ref.~\cite{Hoang:2017suc}, a similar observation was made on the PS mass
when determining the normalization of the \mbox{$u=1/2$} renormalon, dubbed $N_{1/2}$.
In addition, we show that integrating the force, approach followed in Ref.~\cite{Bazavov:2019qoo}, is analytically equivalent to R-evolution if the position-space potential at a given radius is used to subtract the renormalon and canonical scales are implemented. Hence, unlike R-evolution, this method is also sensitive to the infrared physics it inherits from the static potential, and not ideal for high-precision. We design an optimized fitting algorithm based on a $\chi^2$ function marginalized with respect to the constants up to which the static energy predictions for various lattice runs are defined (from now on referred to as ``offsets''). We also consider fits with a common offset for all lattice ensembles. This marginalization can be performed analytically even when multiple offsets appear in the $\chi^2$, which makes our fitting routines very fast and flexible. Building on the idea that predictions for different lattice simulations should coincide up to additive constants,
we devise a recombination algorithm to merge various lattice results into a single prediction for the static potential. Our method is based on the algorithm used in Refs.~\cite{Hagiwara:2003da,Dehnadi:2011gc} to reconstruct the total hadronic cross section, but implements a smoother interpolation. Fitting on the recombined dataset is statistically equivalent to using all individual datasets, but is much faster and decouples two distinct physical effects. We perform fits for both strategies finding very similar results, but consider the determination from the recombined dataset as a validation of our final outcome.

This paper is organized as follows: In Sec.~\ref{sec:theory} we give an overview of the theoretical ingredients our analysis builds upon, both the static energy (Sec.~\ref{sec:statpot})
and renormalon subtractions (Sec.~\ref{sec:renormalon}). Renormalization scale setting is outlined in Sec.~\ref{sec:profiles}, where profile functions are introduced. A thorough analysis of the perturbative properties of our \mbox{R-improved} static energy is carried out in Sec.~\ref{sec:pert}. Lattice data is reviewed
in Sec.~\ref{sec:data}, where we present our method to statistically combine all available data into a single dataset with a reduced amount of points. In Sec.~\ref{sec:procedure} we lay out our strategy to implement $\chi^2$ fits in an optimal way by analytically marginalizing with respect to one or several
offsets. A critical analysis of the various renormalization schemes is conducted in Sec.~\ref{sec:RSlam}, where we justify why MSRn is our preferred choice. In Sec.~\ref{sec:fit} we perform fits for the strong coupling and examine the impact of dataset selection and methodological choices on the outcome. Our final results are presented in Sec.~\ref{sec:final}, which are compared to previous analyses in Sec.~\ref{sec:comparison}. Conclusions are to be found in Sec.~\ref{sec:conclusions}. A number of technical aspects have been relegated to the appendices.

\section{Theoretical Ingredients}\label{sec:theory}

In this section we review the theoretical input that goes into our analysis: the perturbative expression for the static potential, the ultrasoft contributions to the static energy and the resummation of the associated large logarithms, renormalon subtractions, and R-evolution. All the theory ingredients described in this section have been implemented into independent \texttt{Python}~\cite{python3} and \texttt{C++}~\cite{gccmakeNTU} codes, which agree within $15$ decimal places.\footnote{In preliminary stages of this projects we also created a \texttt{Fortran 2008}~\cite{gfortran} version of the code, which agrees with the other two, but has not been used in our fits.} We use the former to generate interpolation tables that are used to perform the fits, whereas with the latter the $\chi^2$ function is minimized directly.
While the \texttt{Python} code uses \texttt{REvolver}~\cite{Hoang:2021fhn} to compute $\alpha_s$ and the MSR mass, its \texttt{C++} counterpart uses a modified version of this public package that
can R-evolve the PS mass and other subtractions based on the position-space static potential. Furthermore, we have slightly optimized $\mu$- and R-evolution, expressing both in terms of the perturbative result for $\Lambda_{\rm QCD}$, and took advantage of the object-oriented paradigm of \texttt{C++} to reduce the number of evaluations necessary for
fitting the strong coupling.
We evolve $\alpha_s(\mu)$ using the five-loop beta function~\cite{Larin:1993tp, vanRitbergen:1997va, Czakon:2004bu,Luthe:2017ttg} and match at the various quark thresholds with four-loop accuracy~\cite{Chetyrkin:1997un,Chetyrkin:2005ia,Schroder:2005hy}.
All renormalon subtractions have been coded up to $\mathcal{O}(\alpha_s^4)$ and large logarithms appearing in the subtractions, along with ultrasoft logs, are summed up to N$^3$LL accuracy.
The \texttt{C++} code is executed using a \texttt{Python} wrapper (generated with \texttt{SWIG}~\cite{swig}), which is convenient to make phenomenological analyses, generate plots and perform fits, making advantage of \texttt{Python}'s vast repertoire of libraries. In particular, we make extensive use of \texttt{numpy}~\cite{harris2020array}, employ root-finding, interpolation and minimization functions included in \texttt{scipy}~\cite{2020SciPy-NMeth}, and produce all plots with \texttt{matplotlib}~\cite{Hunter:2007}.
In our codes we measure energies in GeV and distances in fm, using the numerical value \mbox{$\hbar c=0.1973269804\,{\rm GeV}\times{\rm fm}$}
to convert distances into energies and vice-versa.

\subsection{Static Potential and Static Energy}\label{sec:statpot}
The (singlet) static energy $E_{\rm s}(r,\mu)$ is a scheme- and scale-independent non-perturbative quantity, defined as the ground state energy of a (static) quark-antiquark pair at a distance~$r$.\footnote{Even if the static energy does not depend on the renormalization scale $\mu$, there is some residual dependence left since only a finite number of terms in perturbation theory are known. Accordingly, we keep $\mu$ as a second argument. We shall do the same for other RG-invariant quantities such as the force, soft static potential, and ultrasoft static energy.} It is particularly convenient to compute this quantity in the frame of potential non-relativistic QCD (pNRQCD for short)~\cite{Pineda:1997bj,Brambilla:1999xf}, an effective field theory (EFT) that builds on NRQCD~\cite{Lepage:1987gg}.\footnote{See e.g.\ Ref.~\cite{Luke:1999kz} for an alternative formulation known as vNRQCD.} The static potential $V_{s}(r,\mu)$, on the other hand, has infrared (IR in what follows) divergences and therefore depends on the scheme and scale at which its ultrasoft contributions are defined. These arise from retardation effects that invalidate the frame-independent static limit~\cite{Appelquist:1977tw,Appelquist:1977es}. In an effective field theory language, the static potential is the Wilson coefficient of the pNRQCD dimension-$6$ singlet operator, and added to the corresponding contribution from ultrasoft gluons, referred to as by $\delta_{\rm us}(r,\mu)$, conforms the static energy.

The static potential $V_{s}(r,\mu)$ [\,denoted sometimes by $V_{\rm QCD}(r,\mu)$\,] can be defined in a non-relativistic language as the color-singlet potential between two infinitely heavy quarks in the fundamental representation. It can be computed in perturbation theory and so far is known up to $\mathcal{O}(\alpha_s^4)$, precision at which ultrasoft terms show up for the first time, making additional regularization necessary.
Adding $V_{s}(r,\mu)$ and $\delta_{\rm us}(r,\mu)$,
the divergence cancels out along with the scheme and scale dependence, but leaves behind large ultrasoft logarithms. In order to sum them up, and following the usual lore in EFTs, it is convenient to renormalize the Wilson coefficient and matrix element separately, deriving the corresponding renormalization group evolution (RGE for short) equation for the former and choosing renormalization scales $\mu$ and $\mu_0$ that minimize the (otherwise large) logarithms of both matrix element and RGE boundary condition. The RGE will this way sum up logs of $\mu/\mu_0$.

\subsubsection{Fixed-order results}
Even though the static potential is computed in momentum space, in the spirit of the analysis carried out in this article we quote its form and numerical coefficients in position space. Converting between the momentum and position representations is trivial, and we provide closed expressions in App.~\ref{sec:AppPS}. Considering $n_\ell$ light quarks which are taken as massless [\,quark mass corrections start at $\mathcal{O}(\alpha_s^2)$ as secondary gluon splitting\,]
the static potential and energy can be written as the following perturbative expansions\footnote{Similarly, one defines an effective coupling for the octet potential $V^{\rm octet}_{\rm QCD}(r,\mu)\equiv \alpha_{V_{\rm o}}(r,\mu)/(2r N_c)$ with $\alpha_{V_o}$ and $\alpha_{V_s}$ equal at lowest order.}
\begin{align}\label{eq:static}
V_{s}(r,\mu) =\,& V_{s}^{\rm soft}(r,\mu)+V_{s}^{\rm us}(r,\mu)\equiv -C_{\!F}\frac{\alpha_{V_{s}}(r,\mu)}{r}\,,\\
E_{\rm s}(r,\mu) =\,& V_{s}(r,\mu)+\delta_{\rm us}(r,\mu)=
V_{s}^{\rm soft}(r,\mu)+ E^{\rm us}_{\rm s}(r,\mu)
\equiv -C_{\!F}\frac{\alpha_{E_{\rm s}}(r)}{r}\,,\nonumber\\
V_{s}^{\rm soft}(r,\mu) =&-\! C_{\!F}\frac{\alpha_s(\mu)}{r}\sum_{i=0} \biggl[\frac{\alpha_s(\mu)}{4\pi}\biggr]^{\!i} \sum_{j=0}^{i} a_{ij} L_r^j
\equiv -C_{\!F}\frac{\alpha_{V^{\rm soft}_{\rm s}}(r)}{r}\,,
\nonumber
\end{align}
where for convenience we have defined $L_r\equiv \log(r\mu e^{\gamma_E})$, being $\gamma_E$ the Euler–Mascheroni constant. In the rest of this article we consider only $n_\ell=3$, integrating out charm, bottom and top quarks. If the outer sum in $V_{s}^{\rm soft}$ (dubbed the soft static potential) is truncated at $i=n$ the $\mathcal{O}(\alpha_s^{n+1})$ potential reaches N$^n$LO (or $n$-loop) precision. The tree-level (Coulomb-like) result is recovered with $a_{00}=1$. The perturbative coefficients have been computed to $\mathcal{O}(\alpha_s^4)$~\cite{Fischler:1977yf,Billoire:1979ih,Schroder:1998vy,Pineda:1997hz, Brambilla:1999qa,Kniehl:2002br,Penin:2002zv,Smirnov:2008pn,Smirnov:2009fh,Anzai:2009tm,Lee:2016cgz} and, therefore, the potential is known to N$^3$LO (or \mbox{$3$-loop}) accuracy. In practice, one only needs to know the non-logarithmic terms $a_{i0}$, since the rest can be expressed in terms of those and the QCD beta function coefficients imposing $\mu$-independence on $V_{s}^{\rm soft}(r,\mu)$, resulting in the following recursion relation first obtained in Ref.~\cite{Mateu:2017hlz}:
\begin{equation}\label{eq:getLogs}
a_{ij} = \frac{2}{j}\sum_{k=j}^{i}k\,a_{k-1,j-1}\beta_{i-k}\,.
\end{equation}
The values for $a_{i0}$ up to $\mathcal{O}(\alpha_s^4)$ and the QCD beta function coefficients up to five loops are given in numerical form in Table~\ref{tab:coef}. The ultrasoft potential $V_{s}^{\rm us}(r,\mu)$ has an IR divergence in fixed-order (FO for short) perturbation theory. This was pointed out first in Refs.~\cite{Appelquist:1977tw,Appelquist:1977es}, and originally computed in~\cite{Brambilla:1999qa}. The result in momentum space, as given in \cite{Kniehl:2002br,Anzai:2009tm} reads
\begin{equation}\label{eq:usPotMom}
\bigl[\tilde V_{s}^{\rm us}(q,\mu)\bigr]_{\rm FO}=-C_{\!A}^3C_{\!F}\frac{[\alpha_s(\mu)]^4}{q^2}
\biggl[\frac{1}{6\varepsilon}+\log\biggl(\frac{\mu}{q}\biggr)\!\biggr],
\end{equation}
where $q=|\vec q\,|$, being $\vec q$ the 3-momentum transferred between the heavy quark and antiquark (there is no energy transmission in the static limit).
This result is customarily employed to define the PS mass at four loops and, accordingly, will be used to specify the potential-based subtractions in Sec.~\ref{sec:renormalon}. Ref.~\cite{Kniehl:2002br} claims that their coefficient for the logarithm in Eq.~\eqref{eq:usPotMom}
is a factor of three larger than the results quoted in Refs.~\cite{Appelquist:1977es,Kniehl:1999ud,Brambilla:1999qa}. It is argued that the difference is caused by using dimensional regularization in all integrals, not just the IR-divergent ones. We regard this as a scheme difference, with no consequence in physical predictions. As pointed out in Ref.~\cite{Brambilla:2006wp}, since Eq.~\eqref{eq:usPotMom} has an IR divergence (regulated in dimensional regularization) left after QCD renormalization, to be consistent one needs to perform the Fourier transform in $d=3-2\varepsilon$ dimensions to obtain the position-space ultrasoft potential. The integral one needs is (see App.~\ref{sec:AppIR} for a detailed derivation)
\begin{align}
\int\! \frac{\df^{3 - 2\varepsilon} \vec{q}}{(2 \pi)^{3 - 2 \varepsilon}} \frac{\tilde{\mu}^{2 \varepsilon}}{q^2}\,e^{i \vec{r} \cdot\vec{q}}
={}&\frac{1}{4 \pi r} \bigl[1 + 2 \varepsilon \log (\mu re^{\gamma_E})\bigr] + \mathcal{O}(\varepsilon^2)\,,\\
\int\! \frac{\df^{3 - 2\varepsilon} \vec{q}}{(2 \pi)^{3 - 2 \varepsilon}} \frac{\tilde{\mu}^{2 \varepsilon}}{q^2}\,e^{i \vec{r} \cdot\vec{q}}\log\biggl(\frac{\mu}{q}\biggr)
={}&\frac{1}{4 \pi r} \log (\mu re^{\gamma_E})+ \mathcal{O}(\varepsilon)\,,\nonumber
\end{align}
with $\mu^2 \equiv 4 \pi \tilde{\mu}^2 e^{- \gamma_E}$. Using this result in Eq.~\eqref{eq:usPotMom} we find
\begin{equation}\label{eq:usPotPT}
\bigl[V_{s}^{\rm us}(r,\mu)\bigr]_{\rm FO}=-\frac{C_{\!A}^3C_{\!F}}{12\pi}\,\frac{[\alpha_s(\mu)]^4}{r}
\biggl[\frac{1}{2\varepsilon}+4\log(\mu r e^{\gamma_E})\biggr].
\end{equation}
Within pNRQCD $\bigl[V_{s}^{\rm us}(r,\mu)\bigr]_{\rm FO}$ is identified with the bare ultrasoft potential $\bigl[V_{s}^{\rm us}(r,\mu)\bigr]_{\rm bare}$ i.e., before it is renormalized (in the $d$-dimensional scheme, different from the one employed in Sec.~\ref{sec:usoft} that shall be dubbed three-dimensional). According to Ref.~\cite{Anzai:2009tm}, the divergence in Eq.~\eqref{eq:usPotPT} is an artifact of using FO,
and is absent in resummed perturbation theory, being regularized by the energy difference of color singlet and octet intermediate states.\footnote{Such resummation is of diagrammatic type and not related to the renormalization group.} We identify this resummed prediction with the static energy $E_s(r,\mu)$, and its ultrasoft contribution with $E^{\rm us}_s(r,\mu)$. The difference of the resummed and FO ultrasoft contributions, which (as pointed out in \cite{Anzai:2009tm}) coincides with the bare matrix element (again in the \mbox{$d$-dimensional} scheme) $[\delta_{\rm us}]_{\rm bare}=E^{\rm us}_s - [V_s^{\rm us}]_{\rm bare}$ as defined in pNRQCD, is also divergent and scheme dependent. It can be computed from the results in Refs.~\cite{Kniehl:1999ud,Brambilla:1999qa} and in the pure $d$-dimensional scheme can be found in Eq.~(8) of Ref.~\cite{Anzai:2009tm}
\begin{equation}\label{eq:usDiff}
[\delta_{\rm us}(r,\mu)]_{\rm bare}=
\frac{C_{\!A}^3C_{\!F}}{12\pi}\,\frac{[\alpha_s(\mu)]^4}{r}\biggl\{\frac{1}{2\varepsilon}+4\log(\mu r e^{\gamma_E})
-\log\bigl[C_{\!A}\alpha_s(\mu)e^{\gamma_E}\bigr]+\frac{5}{6}\biggr\}.
\end{equation}
Therefore,
the ultrasoft contribution to the static energy is finite and renormalization-scale independent:
\begin{equation}\label{eq:usoft-RS}
E_{\rm s}^{\rm us}(r,\mu) = - \frac{C_{\!A}^3C_{\!F}}{12\pi}\,\frac{[\alpha_s(\mu)]^4}{r}
\biggr\{\!\log\Bigl[C_{\!A}\alpha_s(\mu)e^{\gamma_E}\Bigr]-\frac{5}{6}\biggl\},
\end{equation}
with a factor $\alpha_s$ in the logarithm's argument. The same result is obtained in the pNRQCD framework, as shown below in
Sec.~\ref{sec:usoft}. At $\mathcal{O}(\alpha_s^4)$ all logarithmic terms are known, either powers of $L_r$ (through
$\mu$-independence) or $\log(C_{\!A}\,\alpha_s/2)$ (using pNRQCD RGE equations).

From the static energy one can derive the (singlet) static force $F_{\rm s}(r) \equiv \df E_{\rm s}(r)/\df r$,\,\footnote{Note that in this context the static
force is defined with an opposite sign as compared to Newton's second law. The static force may be viewed as the magnitude of the vector force.}
which is free from the $u=1/2$ renormalon as long as the derivative with respect to $r$ is taken with $r$-independent renormalization scales (that is,
scale setting should be performed after the derivative is taken). The static energy is, due to confinement, monotonically increasing for all distances, hence the static force is always positive. In terms of the $a_{ij}$ coefficients, it takes the following simple form $F_{\rm s}(r,\mu) = F_{\rm s}^{\rm soft}(r,\mu) + F_{\rm s}^{\rm us}(r,\mu)$ with
\begin{align}\label{eq:stat-en-FO}
F^{\rm soft}_{\rm s}(r,\mu) =\, & C_{\!F}\frac{\alpha_s (\mu)}{r^2} \sum_{i=0}
\biggl[ \frac{\alpha_s (\mu)}{4\pi} \biggr]^{\!i} \biggl\{a_{ii}L_r^i+\sum_{j = 0}^{i - 1} \,\bigl[a_{ij} - (j + 1) a_{i, j + 1}\bigr] L_r^j\biggr\} ,\\
F_{\rm s}^{\rm us}(r,\mu) =\,& \frac{C_{\!A}^3C_{\!F}}{12\pi}\,\frac{[\alpha_s(\mu)]^4}{r^2}\biggl\{\log\!\Bigl[C_{\!A}\alpha_s(\mu)e^{\gamma_E}\Bigr]-\frac{5}{6}\biggr\}.\nonumber
\end{align}
Of course, defining $f_{ij}\equiv a_{ij} -
(j + 1) a_{i, j + 1}$,
one can again compute the logarithm-independent coefficients $f_{i0}=
a_{i0}-2
\sum_{k=0}^{i-1} (k+1)\, a_{k0}\,\beta_{i-k-1}$ (where the renormalon on each term cancels in the difference) and obtain $f_{ij}$ with $j\geq 1$ from Eq.~\eqref{eq:getLogs} replacing $a\to f$. For convenience, we show the numerical values for $f_{i0}$ in Table~\ref{tab:coef}. In Ref.~\cite{Bazavov:2019qoo}, the force is used with $\mu=\xi/r$ varying $\xi$ between $2$ and $1/2$ ($\sqrt{2}$ and $1/\sqrt{2}$ in \cite{Bazavov:2014soa}) being $\xi=1$ the default value. After the scales are set, the force is numerically integrated to obtain a renormalon-free prediction for the static energy. We shall show in the next section that this procedure is analytically equivalent to using R-evolution for a particular choice of the renormalon-subtraction scale.

\subsubsection{Ultrasoft Resummation}\label{sec:usoft}
Although the contribution from the ultrasoft potential is very small, to achieve the precision we aim at, one must
sum up its large logarithms to all orders using RGE
equations.
If one counts the large logarithm
$L_{\rm us}\equiv\log(\alpha_s)\sim\mathcal{O}(1/\alpha_s)$, then $V_{s}^{\rm us}$ is of the same order as the $\mathcal{O}(\alpha_s^3)$ terms and resummation is a priori already necessary at N$^2$LO. Terms of the form $c_{nm}\alpha_s^{m}(\alpha_sL_{\rm us})^n$ are summed up for all $n\geq 0$, with $m=3$ and $4$ (denoted in the usual counting by N$^2$LL and N$^3$LL) with leading and subleading ultrasoft resummation, respectively. Therefore, within this counting, a full-fledged N$^3$LO theoretical expression needs subleading ultrasoft resummation. In Ref.~\cite{Bazavov:2014soa} it was argued that $L_{\rm us}$ is not as large as in the na\"\i ve counting and, therefore, at N$^3$LO only the leading ultrasoft resummation is necessary. We will make a dedicated numerical study when performing our $\alpha_s$ fits to show the impact of N$^2$LL and N$^3$LL resummation.

Theoretical expressions for the leading and subleading ultrasoft resummation have been worked out in Refs.~\cite{Pineda:2000gza,Brambilla:2009bi}, and we review the derivation here. It is then convenient to use the same renormalization scheme as in these references, which uses $d=3-2\varepsilon$ dimensions in the computations, keeping the potential in the ultrasoft loops three-dimensional. Finally, in this section we use $\mu_s$ to denote the soft scale, previously referred to as $\mu$. With this notation, the renormalized fixed-order static potential and the corresponding matrix element of the singlet operator read\footnote{After Eq.~\eqref{eq:usPotPT2} we will drop the label pNRQCD from $V_{s}$ and $V^{\rm us}_{\rm s}$ since the presence of a third argument
should be enough to make the distinction.}
\begin{align}\label{eq:usPotPT2}
[V_{s}(r,\mu_s,\mu_{\rm us})]_{\rm pNRQCD}={} &V_{s}^{\rm soft}(r,\mu_s)+[V_{s}^{\rm us}(r,\mu_s,\mu_{\rm us})]_{\rm pNRQCD} \,,\\
[V_{s}^{\rm us}(r,\mu_s,\mu_{\rm us})]_{\rm pNRQCD}={}&\!-\!\frac{C_{\!A}^3C_{\!F}}{12\pi}\,\frac{[\alpha_s(\mu_s)]^3\alpha_s(\mu_{\rm us})}{r}\log(\mu_{\rm us} r e^{\gamma_E})\,,\nonumber\\
\delta_{\rm us}(\mu_s,\mu_{\rm us})={}&\!-\!\frac{C_{\!A}^3C_{\!F}}{12\pi}\,\frac{[\alpha_s(\mu_s)]^3\alpha_s(\mu_{\rm us})}{r}
\biggl\{ \log\!\biggl[\frac{C_{\!A}\alpha_s(\mu_s)}{\mu_{\rm us}r}\biggr]-\frac{5}{6}\biggr\}.\nonumber
\end{align}
When these two equations are added with $\mu_{\rm us}=\mu_s=\mu$,
Eq.~\eqref{eq:usoft-RS} is recovered, but since both scales are set equal (that is, one does not disentangle soft and ultrasoft physics), no resummation of large logarithms is performed. Even though the static potential and matrix element are separately $\mu_{\rm us}$ dependent, its sum does not depend on this scale provided the same value is used in both. The necessity of resummation is easily illustrated looking at Eq.~\eqref{eq:usPotPT2}: to keep the
logarithm small in $V_{s}^{\rm us}$
one needs to choose $\mu_{\rm us}\sim 1/r$, but this choice makes the logarithm in $\delta_{\rm us}$ large; on the other hand, if one chooses $\mu_{\rm us}\sim\alpha_s/r$, the matrix element has no large logarithms, but the static potential does. The great
benefit of using EFTs is that Wilson coefficients and matrix elements can be separately renormalized. In this way, Wilson coefficients define an anomalous dimension that, upon solving the corresponding RGE equation, sums up large logarithms. One first matches QCD to NRQCD at the soft scale $\mu_s\sim 1/r$, such that large (soft) logarithms of the type $\log(r\mu_s)$ are summed by ordinary $\alpha_s$ evolution. Then, NRQCD is matched onto pNRQCD at the ultrasoft scale $\mu_{\rm us}\sim \alpha_s/r$, and large ultrasoft logarithms of the type $\log[C_{\!A}\alpha_s(\mu)]$
are summed up through the pNRQCD RGE equations. The differential
equation one needs to solve is rather simple and, up to
subleading ultrasoft summation, reads (here $\mu$ refers to a generic ultrasoft scale)
\begin{align}
\mu\frac{\df V_s(r, \mu_s,\mu)}{\df \mu} =\,& -\!\frac{2C_{\!F}}{3r}\frac{\alpha_s(\mu)}{\pi}\biggl[1+B\frac{\alpha_s(\mu)}{\pi}\,\biggr]\,(r V_o-r V_s)^3\,,\\
B=\,& \frac{C_{\!A}(6 \pi^{2}+47) - 15}{18}\,.\nonumber
\end{align}
At the order we are working, $r(V_o-V_s)$ is independent of $\mu_{\rm us}$ and, therefore, behaves as a constant as far as the RGE is concerned. At this order, the singlet and octet static potentials are identical up to a global factor, hence taking the difference is simple
\begin{equation}
\bigl[r V_o(r, \mu_s)-r V_s(r, \mu_s)\bigr]^3 = \frac{C^3_{\!A}}{8}[\alpha_s(\mu_s)]^3
\biggl\{1+3\frac{\alpha_s(\mu_s)}{4\pi} \Bigl[a_{10}+2\beta_0\log(r\mu_se^{\gamma_E})\Bigr]\!\biggr\}
+\, \ldots,
\end{equation}
where we have strictly truncated to $\mathcal{O}(\alpha_s^4)$. Using the chain rule
to convert the $\mu$ derivative into an $\alpha_s$ derivative, and evolving from $\mu=\mu_s$ to $\mu=\mu_{\rm us}$ we find at N$^3$LL
\begin{align}\label{eq:RGE-sol}
V_s(r, \mu_s,\mu_{\rm us}) =\,& V_s(r, \mu_s,\mu_s) + U_{\rm us}(r, \mu_s,\mu_{\rm us})\,,\\
U_{\rm us}(r, \mu_s,\mu_{\rm us}) =\,& \frac{C_{\!A}^3C_{\!F}}{6\beta_0 r}[\alpha_s(\mu_s)]^3
\biggl\{\!\biggl(\!1+3\frac{\alpha_s(\mu_s)}{4\pi} \Bigl[a_{10}+2\beta_0\log(r\mu_se^{\gamma_E})\Bigr]\!\biggr)\!
\log\!\biggl[\frac{\alpha_s(\mu_{\rm us})}{\alpha_s(\mu_s)}\biggr]\nonumber\\
& +\!\biggl(\!B-\frac{\beta_1}{4\beta_0}\biggr)\!
\biggl[\frac{\alpha_s(\mu_{\rm us})}{\pi}-\frac{\alpha_s(\mu_s)}{\pi}\biggr] \!\biggr\},\nonumber
\end{align}
where we have dropped terms beyond N$^3$LL accuracy. In this equation, $V_s(r, \mu_s,\mu_s)$ can be understood as the matching (or matrix element) contribution, with small logs only, and $ U_{\rm us}(\mu_s,\mu_{\rm us})$ as the running contribution, which sums up large logarithms of the type $\log(\mu_s/\mu_{\rm us})$ and vanishes for $\mu_s=\mu_{\rm us}$. Using the profile functions described in Sec.~\ref{sec:profiles}, as can be seen in Fig.~\ref{fig:MuRatio} the largest value of $\mu_s/\mu_{\rm us}$ attained within the fit range is $\sim 2.5$ and happens at the smallest value of $r$ considered, thus ensuring the logarithms remain small.
The $\mathcal{O}(\alpha_s^4)$ matching term starts contributing at N$^3$LL only. The leading ultrasoft resummation is obtained setting to zero in Eq.~\eqref{eq:RGE-sol} single powers of $\alpha_s$ inside the curly brackets (remaining only the logarithm).
The resummed static energy at N$^3$LL is obtained by adding the ultrasoft contribution
\begin{align}\label{eq:Esum}
E_{\rm s}(r,\mu_s,\mu_{\rm us}) =\,& V^{\rm soft}_{\rm s}(r,\mu_s) +\frac{C_{\!A}^3C_{\!F}}{12}\,\frac{[\alpha_s(\mu_s)]^3}{r}
\biggl\{\frac{2}{\beta_0}\biggl(\!B-\frac{\beta_1}{4\beta_0}\biggr)\! \biggl[\frac{\alpha_s(\mu_{\rm us})}{\pi}-\frac{\alpha_s(\mu_s)}{\pi}\biggr]\\
&+\!\frac{2}{\beta_0}\biggl(\!1+3\frac{\alpha_s(\mu_s)}{4\pi} \Bigl[a_{10}+2\beta_0\log(r\mu_se^{\gamma_E})\Bigr]\!\biggr)\!
\log\!\biggl[\frac{\alpha_s(\mu_{\rm us})}{\alpha_s(\mu_s)}\biggr] \nonumber \\
&-\!\frac{\alpha_s(\mu_{\rm us})}{\pi}\biggl[\log\biggl(\frac{C_{\!A}\alpha_s(\mu_s)}{r\mu_{\rm us}}\biggl)-\frac{5}{6}\biggr]\!
-\frac{\alpha_s(\mu_s)}{\pi}L_r\!\biggr\}.\nonumber
\end{align}
To the best of our knowledge, the static energy with summation at N$^3$LL for arbitrary $\mu_s$ and $\mu_{\rm us}$ scales has been given explicitly before only in Ref.~\cite{Ayala:2020odx}, and we find agreement with the expression therein if subleading terms are dropped. In Ref.~\cite{Brambilla:2009bi} and related articles the choice $\mu_s=1/r$ is adopted. This article does not explicitly work out the difference of the singlet and octet potentials, which we expand and truncate to keep only terms which are strictly N$^3$LL accurate. It also includes a contribution from the residual mass term, inherited from the HQET Lagrangian when switching from the pole mass to a short-distance scheme. This term, even after resummation, and as long as renormalization scales are not set to \mbox{$r$-dependent} values, is constant and therefore vanishes when computing the force, as pointed out in Ref.~\cite{Bazavov:2014soa}. In our approach, the term is absent already in the static energy since we work out resummation in the pole scheme and only switch to a low-scale short-distance mass afterwards, as explained in our previous work~\cite{Mateu:2018zym}. Ref.~\cite{Tormo:2013tha} also sets $\mu_s=1/r$ and includes the potentials to NLO accuracy, but when taking the third power of the difference, no expansion is performed. This is perfectly consistent, but numerically different from our choice. As already pointed out in Ref.~\cite{Ayala:2020odx}, this review has a small inconsistency in the ultrasoft matrix element contribution (which was corrected in 2020, see erratum of \cite{Bazavov:2014soa}), since the ultrasoft $\alpha_s$ is evaluated at the soft scale. This is not justified at N$^3$LL, since expanding $\alpha_s(\mu_{\rm us})$ in terms of $\alpha_s(\mu_s)$ would generate at the next order a single power of $\log(\mu_s/\mu_{\rm us})$, which is large and invalidates the logarithmic counting. By expanding out our N$^3$LL result we recover the $\mathcal{O}(\alpha_s^4)$ fixed-order logarithms computed in Ref.~\cite{Brambilla:2006wp}. With the result in Eq.~\eqref{eq:Esum} one can trivially compute the N$^3$LL expression for the static force\,:
\begin{align}
F_{\rm s}(r,\mu,\mu_{\rm us}) =\,& F^{\rm soft}_{\rm s}(r,\mu) -\frac{C_{\!A}^3C_{\!F}}{12}\,\frac{[\alpha_s(\mu_s)]^3}{r^2}
\biggl\{\frac{2}{\beta_0}\biggl(\!B-\frac{\beta_1}{4\beta_0}\biggr)\! \biggl[\frac{\alpha_s(\mu_{\rm us})}{\pi}-\frac{\alpha_s(\mu_s)}{\pi}\biggr]\\
&+\!\frac{2}{\beta_0}\biggl(\!1+3\frac{\alpha_s(\mu_s)}{4\pi} \Bigl[f_{10}+2\beta_0\log\bigl(r\mu_se^{\gamma_E}\bigr)\Bigr]\!\biggr)\!
\log\!\biggl[\frac{\alpha_s(\mu_{\rm us})}{\alpha_s(\mu_s)}\biggr]\nonumber\\
&-\!\frac{\alpha_s(\mu_{\rm us})}{\pi}\biggl[\log\biggl(\frac{C_{\!A}\alpha_s(\mu_s)}{r\mu_{\rm us}}\biggl)+\frac{1}{6}\biggr]\!
-\frac{\alpha_s(\mu_s)}{\pi}(L_r-1)\biggr\}\,.\nonumber
\end{align}
This result agrees with Eq.~(11) of Ref.~\cite{Bazavov:2014soa} when taking $\mu_s=1/r$, but we find the same inconsistency (only important at N$^3$LL and already fixed) since the ultrasoft $\alpha_s$ is evaluated at the soft scale.\footnote{While this mistake has been corrected in arXiv and pointed out in the corresponding erratum, it is claimed that the code used in their numerical analyses was always correct.} Our result is however different from Eq.~(3) of Ref.~\cite{Bazavov:2019qoo} which only performs leading ultrasoft resummation.

\begin{table}[t!]
\center
\begin{tabular}{|c|ccccc|}
\hline
$i$ & $0$ & $1$ & $2$ & $3$ & $4$ \tabularnewline\hline
$a_{i0}$ & $1$ & $7$ & $535.27675$ & $30374.35776$ & -- \\
$f_{i0}$ & $1$ & $-11$ & $155.27675$ & $-1610.25354$ & -- \\
$\beta_i$ & $9$ & $64$ & $\frac{3863}{6}$ & $12090.37813$ & $130377.90682$ \\
\hline
$\delta^{\rm MSRp}_i$ & -- & $\frac{16}{3}$ & $165.10873$ & $7443.20469$ & $434847.27794$\\
$\delta^{\rm MSRn}_i$ & -- & $\frac{16}{3}$ & $163.45165$ & $7328.88394$ & $434797.97242$\\
$\delta^{\rm PS}_i$ & -- & $\frac{16}{3}$ & $\frac{400}{3}$ & $6916.252979$ & $469325.26623$\\
$\delta^{\rm F}_i$ & -- & $8.37758$ & $145.68513$ & $7226.23573$ & $446723.08099$ \\
$\delta^{\rm V}_i$ & -- & $14.92108$ & $104.44754$ & $7986.90585$ & $440977.36762$ \\
\hline
$\gamma^{\rm MSRp}_i$ & $\frac{16}{3}$ & $69.10873$ & $816.62380$ & $-16228.78588$ & --\\
$\gamma^{\rm MSRn}_i$ & $\frac{16}{3}$ & $67.45165$ & $761.95773$ & $-9680.21386$ & --\\
$\gamma^{\rm PS}_i$ & $\frac{16}{3}$ & $\frac{112}{3}$ & $1433.58631$ & $50471.42010$ & --\\
$\gamma^{\rm F}_i$ & $-8.37758$ & $5.11131$ & $-909.24060$ & $-8423.42605$ & -- \\
$\gamma^{\rm V}_i$ & $-14.92108$ & $164.13185$ & $-2316.89644$ & $36267.49306$ & --
\tabularnewline
\hline
\end{tabular}
\caption{Perturbative coefficients for the position-space static potential
(second row), force (third row), QCD beta function (fourth row), subtraction series (5th to 9th rows) and \mbox{R-anomalous} dimensions (rest). All coefficients are given for $\nl=3$ light active flavors. \label{tab:coef}}
\end{table}

\subsection{Renormalon Subtractions and R-evolution}\label{sec:renormalon}
The static potential in the $\MSb$ scheme suffers from an $\mathcal{O}(\LQCD)$ renormalon ambiguity. If left unsubtracted, predictions made in perturbation theory
are not eligible for precision studies such as top quark pair production near threshold or quarkonium masses. Similarly, one cannot determine $\alpha_s$ from the static energy in this situation. A remarkable feature of this ambiguity is its $r$-independence and, therefore, can be subtracted by an appropriate $r$-independent perturbative series, as long as it has the same asymptotic behavior and is expressed in terms of the strong coupling evaluated at the same renormalization scale~$\mu$.
A second crucial feature is that the ambiguity is identical to that in the series relating the pole and $\MSb$ masses for heavy quarks up to a factor of two and a sign~\cite{Pineda:1998id,Hoang:1998nz,Beneke:1998rk}.

Since the static energy covers a wide range of distances (hence, also of energies), it is important that the renormalon subtraction series depends on an adjustable scale $R$. To relate the subtractions for different $R$ values in a way which does not involve large logarithms one needs to solve the corresponding RGE equations, procedure known as R-evolution. We will explore various $R$-dependent subtraction schemes and will use R-evolution in all of them. Let us work out R-improvement~\cite{Hoang:2009yr} for the general case of a subtraction series $\delta^R(R)$ with the same asymptotic behavior as the static potential:\footnote{See Ref.~\cite{Gracia:2021nut} for a discussion of the MSR mass and R-evolution in the large-$\beta_0$ limit.}
\begin{equation}\label{eq:subtraction}
\delta^R(R) = R\sum_{n=1}^\infty \delta^R_n \biggl[\frac{\alpha_s(R)}{4\pi}\biggr]^{\!n} =
R\sum_{n=1}^\infty \biggl[\frac{\alpha_s(\mu)}{4\pi}\biggr]^{\!n}\sum_{j = 0}^{n-1}\delta^R_{nj}\log^j\!\biggl(\frac{\mu}{R}\biggr)\,,
\end{equation}
with $\delta^R_n\equiv \delta^R_{n0}$ and where $\delta^R_{nj}$ for $j\geq1$ can be obtained in terms of $\delta^R_{n0}$ using Eq.~\eqref{eq:getLogs} with the replacement $a\to \delta^R$. To use R-evolution, it is important that the ambiguity of $\delta^R(R)$ is independent of $R$. Therefore, we can subtract the series from the static energy at any $R$ and the resulting expression will be renormalon free. The series is $\mu$ independent, but in order to cancel the renormalon it has to be expressed in terms of $\alpha_s(\mu)$, as shown in the second equality of Eq.~\eqref{eq:subtraction}. The logarithms present in the rightmost expression may become large if $\mu$ and $R$ are far apart, making resummation necessary. As already anticipated, R-evolution can sum up those large logarithms using an RGE equation. One key feature is that the associated anomalous dimension is free from the leading renormalon, as can be seen from its analytic form
\begin{equation}\label{eq:Rgamma}
\gamma^R(R) = \frac{\df}{\df R}\delta^R(R) = \sum_{n=0}^\infty \gamma^R_n \biggl[\frac{\alpha_s(R)}{4\pi}\biggr]^{\!n+1},
\qquad\gamma_n^R = \delta^R_{n+1}-2\sum_{j=0}^{n-1} (n-j)\beta_j \delta^R_{n-j}\,.
\end{equation}
The ambiguity cancels between the first term and the sum in the second expression, which enter with opposite signs. The renormalon-free static potential is defined by subtracting $\delta^R(R_0)$ with a fixed (and arbitrary) $R_0$. To avoid large powers of $\log(\mu/R_0)$ we perform R-evolution as indicated (for the sake of conciseness, we drop the dependence on the soft and ultrasoft renormalization scales)\footnote{There is a typo in Eq. (25) of our earlier work Ref.~\cite{Mateu:2018zym}: the middle equality should be multiplied by an overall minus sing.}
\begin{align}\label{eq:R-potential}
E^R_{\rm s} (r, R_0) \equiv\, & E_{\rm s} (r) + 2\delta^R(R_0) = \!\bigl[E_{\rm s} (r) + 2\delta^R(R)\bigr] - 2\bigl[\delta^R(R) - \delta^R(R_0)\bigr]\\
=\, & \bigl[E_{\rm s} (r) + 2\delta^R(R)\bigr] - 2\!\int^R_{R_0}\! \df R^\prime \gamma^R(R^\prime)
\equiv \bigl[E_{\rm s} (r) + 2\delta^R(R)\bigr] + 2 \Delta_R(R,R_0)\,,\nonumber
\end{align}
where we have added and subtracted $2\delta^R(R)$ to the second expression such that the terms within square brackets in the first and second lines are separately renormalon-free. To get to the second line we have written the difference of $\delta^R$ terms as an integral over the R-anomalous dimension $\gamma^R$, c.f.\ Eq.~\eqref{eq:Rgamma}, which defines the R-evolution kernel $\Delta_R$. In this way, large logarithms of the type $\log(R/R_0)$ associated to renormalon subtractions are summed up by the R-RGE shown in its integral form in the second line of Eq.~\eqref{eq:R-potential}.\footnote{In our numerical codes we exactly solve (within machine precision) the R-evolution equations using the same semi-analytically algorithm as in \texttt{REvolver}~\cite{Hoang:2021fhn}, which is based on expansions and recursive relations, which we have slightly improved to compute only those pieces which are strictly necessary.} The term within square brackets in the two expressions shown in the second line is renormalon free and does not contain large logarithms if $R$ is chosen similar to the scale $\mu$ appearing in $E_s(r,\mu,\mu_s)$. Of course, the R-improved static energy is formally independent of $\mu$ and $R$.

In what follows, we will consider different choices for the subtraction, the first of them directly linked to using the Force.

\subsubsection*{Force-type subtractions}
In the original articles \cite{Brambilla:2010pp,Bazavov:2012ka}, the exact asymptotic form of the relation between the pole and $\MSb$ masses~\cite{Beneke:1994rs} was used, leaving unresummed logarithms of the type $\log(rR_0)$.
The technology for carrying out such summation within the so-called RS scheme~\cite{Pineda:2001zq} existed and was developed in Ref.~\cite{Bali:2003jq}, but the dataset used in those first analyses covered a sufficiently narrow range of distances to justify using a fixed value of $R_0$.
The exact asymptotic form depends on an overall normalization constant $N_{1/2}$ which is only known numerically within a certain accuracy. In the subsequent articles~\cite{Bazavov:2014soa,Bazavov:2019qoo} lattice results at much smaller distances were used, which invalidated their initial procedure. The problem is general and can summarized as follows: The subtraction has to be performed at a fixed reference scale $R_0$ which cannot be varied once it is chosen. For the renormalon to cancel properly, the subtraction series has to be expressed in terms of $\alpha_s(\mu)$, and therefore powers of $\log(R_0/\mu)$ appear in the series, see Eq.~\eqref{eq:subtraction}. The renormalization scale has to depend on $r$ roughly as $\mu\sim1/r$, turning the subtraction logarithm into $\sim \log(R_0\,r)$ which becomes large
for any $R_0$, making perturbation theory badly behaved. Therefore, one needs to sum up powers of $\log(R_0/\mu)$
to restore convergence. The strategy followed by the more recent HotQCD analyses exploits the \mbox{$r$-independence} of the renormalon in a different way: it
drops when acting with an $r$-derivative
as long as scales do not depend on $r$. The resulting expression is integrated numerically after setting the scales such that large logarithms are absent in the force. (that is, with $\mu\sim 1/r$). As shall be shown next, this procedure effectively sums up large logarithms of $\log(R_0/\mu)$ in a way rather similar to R-evolution.

Since the ambiguity is $r$ independent, another easy way of eliminating it is by simply subtracting the potential at a given, fixed distance $r_0$, see e.g.\ Ref~\cite{Ayala:2020odx}:
\begin{align}\label{eq:force-ren}
E_{\rm s}^F(r, r_0) \equiv \,& E_{\rm s} (r) - E_{\rm s} (r_0) = E_{\rm s}(r) - E_{\rm s}(r_1) +\!
\bigl[E_{\rm s}(r_1) - E_{\rm s} (r_0)\bigr]\\
= \,& E_{\rm s} (r) - E_{\rm s} (r_1) + \!\int_{r_0}^{r_1} \!\df r^\prime F_{\rm s} (r^\prime)
\equiv E_{\rm s} (r) - E_{\rm s} (r_1) + 2\Delta_F(r_0, r_1)\nonumber\,,
\end{align}
where we again add and subtract $E_{\rm s}(r_1)$. The (renormalon-free) expression within square brackets in the first line has been written as an integral over the force in the second. The specific choice $r_1=r$ trivially reproduces the approach of Ref.~\cite{Bazavov:2019qoo}. In our analysis we vary $r_1$ to estimate uncertainties coming from missing higher orders in the renormalon subtraction. In order to connect with the notion of R-evolution we define $R=1/r_1$ and $R_0=1/r_0$ and write the subtraction in Eq.~\eqref{eq:force-ren} as\footnote{Since the ultrasoft potential has nothing to do with the renormalon, in practice one can define the subtraction using only the soft potential $V_{s}^{\rm soft}$. We will introduce ultrasoft corrections later.}
\begin{equation}\label{eq:force-subtractions}
\delta_{\rm soft}^F(R) \equiv -\frac{1}{2}V^{\rm soft}_{\rm s}\!\biggl(\frac{1}{R}\biggr)=2\pi C_F R \sum_{i = 1} \biggl[ \frac{\alpha_s (R)}{4\pi} \biggr]^{\!i} \sum_{j = 0}^{i - 1} a_{i-1, j}\, \gamma_E^j
\equiv R \sum_{i = 1} \biggl[ \frac{\alpha_s (R)}{4\pi} \biggr]^{\!i}\delta^F_i ,
\end{equation}
in which we have set $\mu=R$ without any loss of generality given that the potential is $\mu$ independent.\footnote{Another choice which does not imply large logarithms is $\mu=\lambda R$ with $\lambda$ of $\mathcal{O}(1)$. Varying $\lambda$ between $1/2$ and $2$ provides an estimate for the uncertainty in R-evolution due to missing higher-order terms as shown in Ref.~\cite{Hoang:2017suc}.
In our final analyses, which utilize MSR subtractions, we choose a default value $\lambda\neq1$ to avoid a biased central value, and vary this parameter to have a more conservative theoretical uncertainty estimate, see discussion below.} The R-anomalous dimension can be directly related to the force using the chain rule, and this allows to express $\Delta_F$ as an R-evolution integral
\begin{align}
\gamma_{\rm soft}^F(R) = &-\!\frac{1}{2}\bigl[r^2 F^{\rm soft}_{\rm s}(r)\bigr]_{r = 1 / R}\, ,\\
\Delta^{\rm soft}_F(r_0, r_1) = & \,\frac{1}{2}\!\int_{r_0}^{r_1}\! \frac{\df r^\prime}{(r^\prime)^2} [(r^\prime)^2 F^{\rm soft}_{\rm s}(r^\prime)] =
\int^{R_1}_{R_0} \!\df R^\prime \gamma_{\rm soft}^F(R^\prime)\,.\nonumber
\end{align}
One can include ultrasoft effects in force-type subtractions and, for simplicity, we consider the FO situation only. In this case, one simply has to add to the subtraction in Eq.~\eqref{eq:force-subtractions} and the corresponding R-anomalous dimension the following terms:
\begin{align}
\delta^F \!(R) & = \delta_{\rm soft}^F (R) + R \frac{C_{\!A}^3C_{\!F}}{24\pi}\,[\alpha_s(R)]^4\biggl\{\log\!\Bigl[C_{\!A}\alpha_s(R)e^{\gamma_E}\Bigr]-\frac{5}{6}\biggr\},\\
\gamma^F\! (R) & = \gamma_{\rm soft}^F(R) - \frac{C_{\!A}^3C_{\!F}}{24\pi}\,[\alpha_s(R)]^4
\biggl\{\log\!\Bigl[C_{\!A}\alpha_s(R)e^{\gamma_E}\Bigr]-\frac{5}{6}\biggr\},\nonumber
\end{align}
obtained from Eq.~\eqref{eq:usoft-RS}. The second term in the last line can be easily integrated numerically to carry out \mbox{R-evolution}. In App.~\ref{sec:Revo} we generalize the algorithm developed in Ref.~\cite{Hoang:2021fhn} to account for log-dependent R-anomalous dimensions, such that the evolution can be carried out exactly within machine precision using analytic formulas based on recursive algorithms. We have implement this novel strategy into our numerical codes.

When the subtraction is added to the static energy with $\mu \neq R$, we set the argument of $\alpha_s$ (both as prefactor and inside the logarithm) to $\mu$, discarding higher-order corrections generated when $\alpha_s(\mu)$ is expanded in powers of $\alpha_s(R)$. The same procedure will be followed with other ultrasoft-sensitive subtractions introduced in the remainder of this section. This concludes our proof of the equivalence between \mbox{R-evolution} and integrating the force. For completeness, we provide numerical values for $\delta_i^F$ and $\gamma_i^F$ in Table~\ref{tab:coef}. Since this scheme has a somewhat strong sensitivity to soft physics, it will not be used for our final analysis.

\subsubsection*{Potential-type subtractions}
These are very similar to those in Eq.~\eqref{eq:force-subtractions}, but follow the spirit of the gap subtractions defined in the context of event shapes in Refs.~\cite{Hoang:2008fs,Jain:2008gb,Clavero:2024yav}. Therefore, the series is defined now such that logarithms in position space exactly vanish, setting $r = e^{- \gamma_E} / R$ and $\mu = R$:
\begin{equation}\label{eq:V-type}
\delta_{\rm soft}^V(R) \equiv -\frac{1}{2}V^{\rm soft}_{\rm s}\!\biggl(\frac{e^{- \gamma_E}}{R}\biggr)= 2\pi e^{\gamma_E} C_F R \sum_{i = 1} \biggl[
\frac{\alpha_s (R)}{4\pi} \biggr]^i a_{i-1, 0} \equiv R \sum_{i = 1} \biggl[ \frac{\alpha_s (R)}{4\pi} \biggr]^i \delta^V_i \,,
\end{equation}
with $\delta_i^V=2\pi e^{\gamma_E} C_Fa_{i-1, 0}$. From Eq.~\eqref{eq:V-type} one can derive the associated \mbox{R-anomalous} dimension which is shown, along with the corresponding subtraction, in Table~\ref{tab:coef}. Again, one can add ultrasoft effects from the static energy, implying the following modifications
\begin{align}
\delta^V\!(R) & = \delta_{\rm soft}^V (R) + R\, \frac{C_{\!A}^3C_{\!F}}{24\pi}\,e^{\gamma_E}[\alpha_s(R)]^4\biggl\{\log\!\Bigl[C_{\!A}\alpha_s(R)e^{\gamma_E}\Bigr]-\frac{5}{6}\biggr\},\\
\gamma^V\!(R) & = \gamma_{\rm soft}^F(R) - \frac{C_{\!A}^3C_{\!F}}{24\pi}\,e^{\gamma_E}[\alpha_s(R)]^4\biggl\{\log\!\Bigl[C_{\!A}\alpha_s(R)e^{\gamma_E}\Bigr]-\frac{5}{6}\biggr\}.\nonumber
\end{align}
This scheme is afflicted by ultrasoft effects and, therefore, will be discarded for the final analysis along with other equally sensitive choices such as the force and PS subtractions.

\subsubsection*{MSR-type subtractions}
The MSR mass was succinctly introduced in Ref.~\cite{Hoang:2008yj} and worked out in greater detail in~\cite{Hoang:2017suc}. It has been used in different threshold problems such as quarkonium masses~\cite{Mateu:2017hlz}, calibration of the top quark mass in $e^+e^-$ parton shower Monte Carlo's~\cite{Butenschoen:2016lpz,Hoang:2018zrp,Dehnadi:2023msm}, boosted tops at lepton~\cite{Bachu:2020nqn} and hadron colliders~\cite{Hoang:2019ceu}, and threshold top pair production at a linear collider~\cite{Boronat:2019cgt}. Details on the MSR mass implementation on the static potential were discussed at length in Ref.~\cite{Mateu:2018zym} and will not be repeated here. In our numerical codes, both MSRn and MSRp subtractions have been implemented. For the reader's convenience, the numerical values for the subtraction series and R-anomalous dimensions are displayed in Table~\ref{tab:coef}. As shall be justified later, we take MSRn as our default scheme for renormalon subtractions, which shall be referred to as simply MSR, and use the rest as a validation.

\subsubsection*{PS-type subtractions}
The potential-subtracted scheme is a low-scale short-distance mass specifically designed for threshold problems. It depends on a subtraction scale $\mu_f$ but, in order to maintain a uniform notation, we will rename it as $\mu_f=R$. It was introduced in Ref.~\cite{Beneke:1994rs} in the context of the static potential, and has been used in non-relativistic sum-rules~\cite{Beneke:2014pta}, quarkonium masses~\cite{Beneke:2005hg} and threshold top pair production at a future $e^+e^-$ collider (see e.g.~\cite{Beneke:2015kwa}). To the best of our knowledge, \mbox{R-evolution} (or any other technique which sums up renormalon-type logarithms) has never been used in conjunction with PS-mass subtractions. Therefore, we will do so for the first time in this article. The PS mass is defined from its relation to the pole mass $m_p$ as $m_p-m^{\rm PS}(R)\equiv\delta_{\rm PS}(R)$, with
\begin{equation}\label{eq:PS}
\delta_{\rm PS}
(R)= - \frac{1}{2} \int_{| \vec{q}\, | < R} \frac{\df^3 \vec{q}}{(2 \pi)^3}\, \tilde{V}_{\rm s}^{\rm soft}(q,R)
\equiv C_F R\, \frac{\alpha_s (R)}{\pi} \sum_{i = 0} \biggl[ \frac{\alpha_s (R)}{4\pi} \biggr]^i c_i\,,
\end{equation}
where $\tilde{V}_{\rm s}^{\rm soft}(q,\mu)$ is the momentum-space soft static potential and
\mbox{$\delta_i^{\rm PS}=4C_Fc_{i-1}$}. The definition of $\tilde{V}_{\rm s}^{\rm soft}$ and its perturbative expansion have the following form:\footnote{Here we use a three-dimensional Fourier transform since we use the position-space static potential with the infrared singularity subtracted.}
\begin{align}\label{eq:momV}
\tilde{V}_{\rm s}^{\rm soft}(q,\mu)\equiv& \int \!\df^3 \vec{r}\, e^{- i \vec{r} \cdot \vec{q}} \,V_{s}^{\rm soft}(r)
=- \frac{4 \pi}{q^2} C_F \alpha_s (\mu) \sum_{i = 0} \biggl[ \frac{\alpha_s(\mu)}{4 \pi} \biggr]^i \sum_{j = 0}^i b_{ij} \log^j\! \biggl( \frac{\mu}{q}\biggr)\\
\equiv & - \!\frac{4 \pi}{q^2} C_F \alpha^{\rm soft}_V (q)\,,\nonumber
\end{align}
where for now we ignore the ultrasoft term. In the second line we define the effective coupling $\alpha_V(q)$, whose perturbative expansion in terms of the $b_{ij}$ coefficients follows directly from
the second expression in the first line. Again, $b_{ij}$ with $j\geq 1$ can be generated from $b_{i0}$ using Eq.~\eqref{eq:getLogs} replacing $a\to b$. Since the static potential depends only on the distance $r$, one can directly relate the PS subtraction series and the position-space static potential by inserting Eq.~\eqref{eq:momV} into \eqref{eq:PS} and performing all integrals except the radial variable:
\begin{equation}\label{eq:PS2}
\delta^{\rm PS}_{\rm soft}(R) = \frac{1}{\pi}\! \int_0^{\infty} \!\frac{\df r}{r} [r R \cos (r R) - \sin (r R)] V^{\rm soft}_{\rm s}(r,R)
= \frac{1}{\pi}\! \int_0^{\infty} \!\!{\df r} \frac{\rm d}{{\rm d}r}\!\biggl[\frac{\sin (r R)}{r}\biggr] rV^{\rm soft}_{\rm s}(r,R)\,.
\end{equation}
The integral is convergent at $r\to0$ since $x \cos(x)-\sin(x)=-x^3/3+\mathcal{O}(x^5)$ and at $r\to\infty$ since $rV^{\rm soft}_{\rm s}(r,R)$ only contains powers of $\log(r\mu)$. To see that $\delta^{\rm PS}_{\rm soft}(R)$ will correctly subtract the renormalon one needs to take into account that the ambiguity is $\mu$ and $r$ independent, hence the above integral has to be computed with the replacement $V^{\rm soft}_{\rm s}(r,R)\to 1$. Changing variables to $r=xR$ the dependence on $R$ drops out and the convergence at $x\to\infty$ appears questionable. Cutting the upper limit at $x=x_{\rm cut}$ and integrating by parts one obtains $[\sin(x_{\rm cut})-\text{Si}(x_{\rm cut})]/\pi$ where the sine integral function is defined as
\begin{equation}
\text{Si}(x) = \int_0^x {\rm d}z\frac{\sin(z)}{z}\,.
\end{equation}
One has $\text{Si}(x_{\rm cut}\to\infty) = \pi/2$ and, since $\sin(x_{\rm cut})$ oscillates around zero, the prescription $\sin(x_{\rm cut}\to\infty)\equiv0$ can be adopted. A more direct way to compute the regularized integral is replacing $r^{-1}\to r^{-1-\eta}$ with $\eta$ a positive infinitesimal, and taking $\eta\to0$ after the analytic integration is carried out. Either way, the ambiguity in $\delta^{\rm PS}_{\rm soft}(R)$ equals minus one half of that in $V^{\rm soft}_{\rm s}(r,R)$, warranting the renormalon cancellation.

Inserting Eq.~\eqref{eq:static} in \eqref{eq:PS2}, taking $\mu=R$ and making the change of variables $r=xR$ we obtain (ignoring for now the ultrasoft contributions) the following relation
\begin{equation}
c_i=\sum_{j=0}^i a_{ij} h_j\,,\qquad {\rm with}\qquad
h_j = j!\sum_{j = 0}^{\left\lfloor \frac{k}{2} \right\rfloor} \frac{(- 1)^j}{(2 j) !}\Bigl( \frac{\pi}{2} \Bigr)^{\!2 j} \sum_{\ell = 0}^{k - 2 j} \kappa_\ell\,,
\end{equation}
where $\lfloor x \rfloor$ is the floor function of $x$, that is, the greatest integer number smaller than $x$, with $h_j=(1,1,2-\pi^2/12,6 - \pi^2/4 - 2\,\zeta_3)$ for $i=(0,1,2,3)$. Here, $\kappa_i$ are defined from the Taylor expansion of the Euler gamma function, first defined in Eq.~(D4) of Ref.~\cite{Clavero:2024yav}, which can be computed with the following recursion relation:
\begin{equation}\label{eq:kappa}
e^{a \gamma_E} \Gamma (1 + a)
\equiv \sum_{i = 0} \kappa_i a^i
,
\qquad \kappa_{i} =
\frac{1}{i} \Biggl[ (- 1)^i \zeta_{i} - \sum_{j = 1}^{i - 3} (- 1)^j \zeta_{j + 1} \kappa_{i -1 - j} \Biggr] ,
\end{equation}
where $\zeta_s=\sum_{j=1}^\infty j^{-s}$ is the Riemann zeta function and $\kappa_i=(1,0,\pi^2/12,-\zeta_3/3)$ for \mbox{$i=(0,1,2,3)$} (see Table~4 of Ref.~\cite{Clavero:2024yav} for more values).

Let us now deal with the ultrasoft contributions. The definition of the PS mass at $\mathcal{O}(\alpha_s^4)$ given in Ref.~\cite{Beneke:2005hg} includes ultrasoft effects based on the scheme given in Eq.~\eqref{eq:usPotMom}. Applying Eq.~\eqref{eq:PS2} to the ultrasoft position-space term shown in Eq.~\eqref{eq:usPotPT} we find
\begin{equation}\label{eq:usPS}
\delta_{\rm PS}^{\rm us}(R,\mu)=R\frac{C_{\!A}^3C_{\!F}}{12\pi^2}[\alpha_s(\mu)]^4
\biggl\{\frac{1}{2\varepsilon}+4\Bigl[1 + \log\Bigl(\frac{\mu}{R}\Bigr)\Bigr]\!\biggr\},
\end{equation}
which has an IR divergence and a non-residual $\mu$ dependence. In Ref.~\cite{Beneke:2005hg} the divergence is dropped along with the logarithmic contribution. We follow Ref.~\cite{Hoang:2017suc} and include the IR-sensitive logarithm, therefore modifying $\delta_4^{\rm PS}$ [\,c.f.\ \eqref{eq:subtraction}\,] as follows:
\begin{equation}
\delta_4^{\rm PS}(R,\mu_{\rm us}) = 4C_F \Bigl\{c_3 + 4 C_A^3\pi^2\Bigl[1+\log\Bigl(\frac{\mu_{\rm us}}{R}\Bigr)\Bigr]\Bigr\},
\end{equation}
where we have renamed $\mu\to \mu_{\rm us}$ to distinguish it from the ``soft'' renormalization scale. The PS subtraction coefficients with $\mu_{\rm us}=R$ (that is, for the original four-loop definition given in Ref.~\cite{Beneke:2005hg}) are collected in Table~\ref{tab:coef}. As pointed out in Ref.~\cite{Hoang:2017suc}, the PS mass thus defined has a rather large IR sensitivity at four-loop order. As we shall see, this has a sizable impact in the determination of $\alpha_s$ making it less precise than those which employ other short-distance schemes. In order to somewhat alleviate this IR sensitivity we propose a slightly modified PS-mass definition that makes use of the static energy rather than the static potential.\footnote{In this regard, the name could be changed to ES-mass (Energy Subtracted).} This implies the following modification of the four-loop coefficient:
\begin{equation}
\delta_4^{\rm ES}(R) = 4C_{\!F} \biggl\{c_3 + \frac{16\pi^2 C_{\!A}^3}{3}
\biggl[\log\Bigl(C_{\!A}\alpha_s(R)e^{\gamma_E}\Bigl)-\frac{5}{6}\biggr]\biggr\},
\end{equation}
which makes the newly defined ES mass in line with e.g.\ the 1S mass, since it contains logarithms of $\alpha_s$ which might be summed up with RGE techniques. At four loops, the R-anomalous dimension gets an additional ultrasoft term:
\begin{equation}
\gamma^{\rm ES}(R) = \gamma^{\rm PS}_{\rm soft}(R) - \frac{C_{\!F}C_{\!A}^3}{12\pi^2}[\alpha_s(R)]^4\biggr[
\log\Bigl(C_{\!A}\alpha_s(R)e^{\gamma_E}\Bigr)-\frac{5}{6}\biggl].
\end{equation}
Unless otherwise stated, we shall use the energy-subtracted mass in our numerical explorations.

\subsubsection*{RS-type subtractions}
\begin{figure}[t!]
\centering
\begin{subfigure}{0.445\textwidth}
\centering
\includegraphics[width=\linewidth]{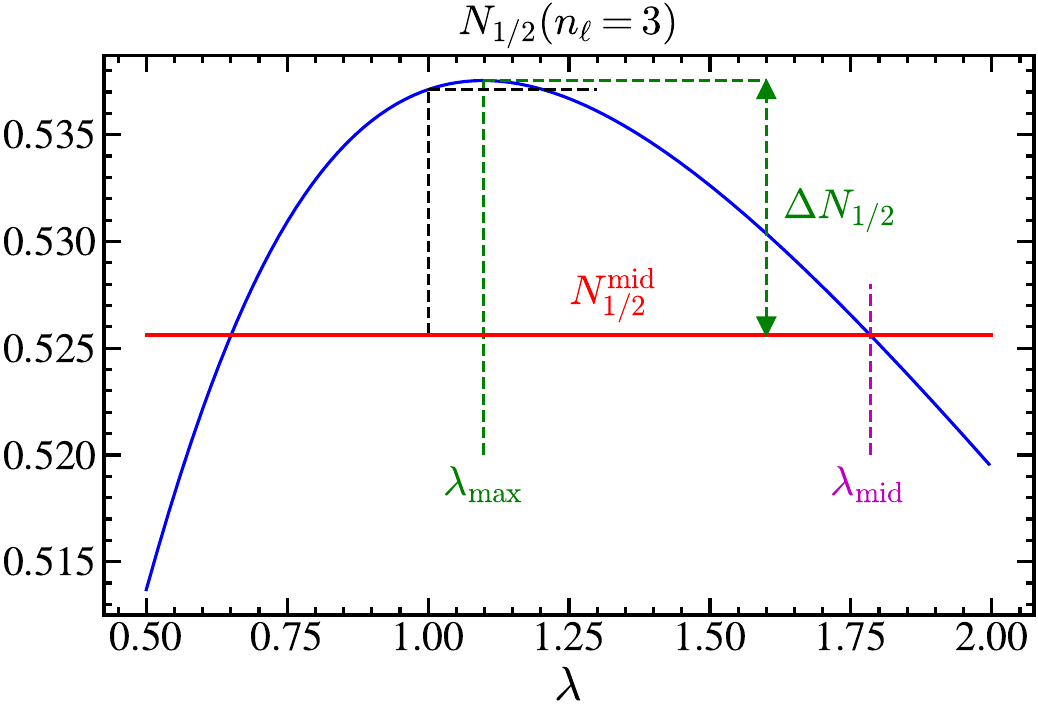}
\caption{}
\label{fig:N12}
\end{subfigure}
\hfill
\begin{subfigure}{0.43\textwidth}
\centering
\includegraphics[width=\linewidth]{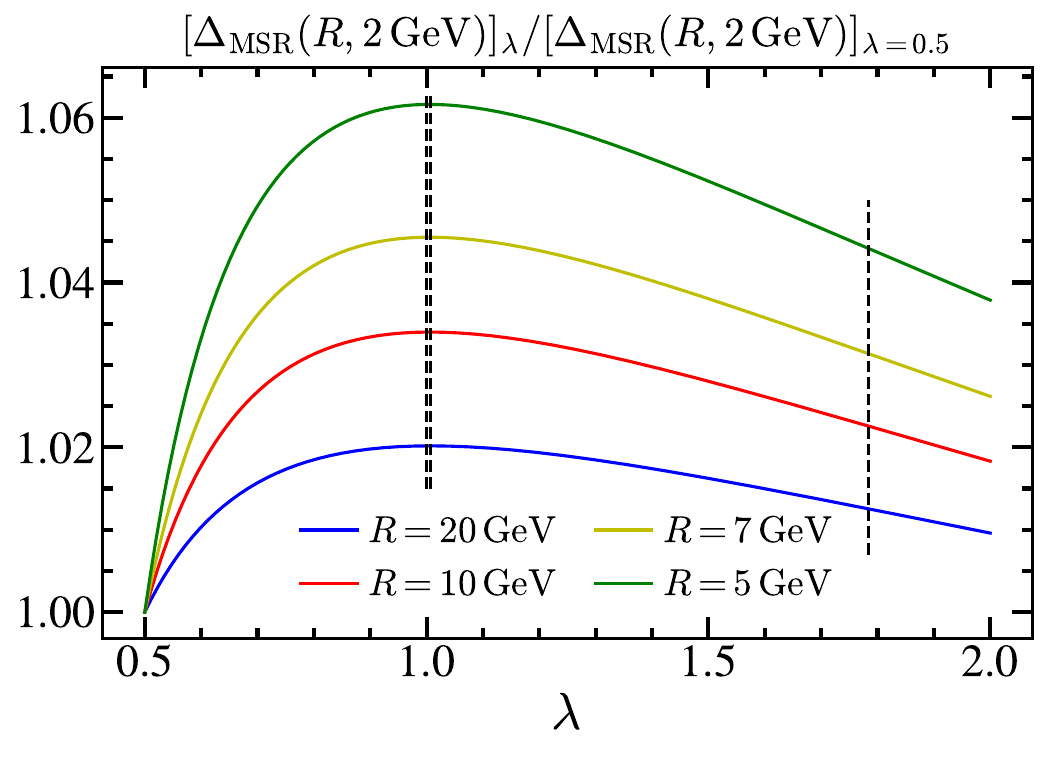}
\caption{}
\label{fig:DeltaMSR}
\end{subfigure}
\vspace*{-0.1cm}
\caption{\label{fig:lambda} Left panel: Dependence of the four-loop estimate for the renormalon normalization $N_{1/2}(n_\ell=3)$ with the dimensionless parameter $\lambda$. The green dashed lines show the position of the maximum while the red, solid, horizontal line indicates the central value. The green, dashed, double-pointed arrow shows the uncertainty on $N_{1/2}$ while the magenta vertical line corresponds to the value of $\lambda>1$ for which the central value is reproduced. Finally, the black dashed lines show the value of $N_{1/2}$ for the default value $\lambda = 1$. Right panel: Dependence of the four-loop MSR-mass R-evolution kernel with $\lambda$ for $n_\ell=3$ active flavors between $R$ and $R_0$ for various values of $R$ (20, 10, 7 and 4\,GeV are shown in blue, red, yellow and green, respectively) and a fixed boundary condition $R_0=2\,$GeV. The vertical dashed black lines correspond, from left to right, to $\lambda=1$ and the position of the maxima (which does not depend on $R$) and $\lambda_{\rm mid}$. The lines are normalized to their respective value at $\lambda=0.5$ and generated with $\alpha_s^{(3)}(m_\tau)=0.305$.}
\end{figure}

The RS scheme for heavy quark masses, first introduced in Ref.~\cite{Pineda:2001zq}, is defined from the pole mass by subtracting its leading asymptotic behavior. Following the notation of Ref.~\cite{Hoang:2017suc} it can be written as follows:
\begin{equation}\label{eq:RSmass}
m_Q^\pole - m_Q^{\rm RS}(R) = \frac{2\pi}{\beta_0}R\,N_{1/2}\sum_{n=1}^\infty\biggl[\frac{\beta_0\alpha_s(R)}{2\pi}\biggr]^{\!n}
\sum_{\ell=0}^\infty g_\ell\, (1+\hat b_1)_{n - 1 -\ell}\,,
\end{equation}
where $\hat b_1=\beta_1/(2\beta^2_0)$, \mbox{$(a)_n=\Gamma(a+n)/\Gamma(a)$} stands for the Pochhammer symbol and, in practice, the inner sum is cut at the highest known value for $g_\ell$, that is $\ell_{\rm max}=3$. Recursion relations to obtain $g_\ell$ in terms of $\beta_i$ can be found in Ref.~\cite{Hoang:2017suc}. All in all, one has that \mbox{$\delta_i^{\rm RS}=2^{n+1}\pi\beta_0^{n-1}N_{1/2}\sum_{\ell=0}^\infty g_\ell\, (1+\hat b_1)_{n - 1 -\ell}$}. The constant $N_{1/2}$ is the (scheme- and scale-independent) normalization of the leading renormalon, whose value can be estimated in a number of ways. In Ref.~\cite{Hoang:2017suc}, a formula called ``The $\Ord(\LQCD)$ Renormalon Sum Rule'' was derived which, at $(n+1)$-loop order, reads:
\begin{equation}\label{eq:N-sumrule}
N^{(n)}_{1/2} = \frac{\beta_0}{2\pi}\sum_{k=0}^n \frac{S_k}{(1+\hat b_1)_k}\,.
\end{equation}
The definition and expressions for $S_k$ in terms of the series coefficients defining the relation between the pole and MSR masses can be found in Ref.~\cite{Hoang:2017suc}. The sum is formally independent of rearrangements of the $m_p-m^{\rm MSR}(R)$ perturbative series (parametrized by a dimensionless parameter $\lambda$, which is varied continuously between $0.5$ and $2$ to estimate the truncation error on $N_{1/2}$) if the sum is carried to infinity, see \cite{Hoang:2017suc} for more details. In that reference
it was shown numerically that when cutting the sum at four loops the $\lambda$ dependence becomes already very small. Estimates of $N_{1/2}$ based on the ratio of the $n$-th order to the asymptotic expression yield numerical results almost identical to Eq.~\eqref{eq:N-sumrule}, both central value and uncertainty estimate. For $n_\ell=3$, using Eq.~\eqref{eq:N-sumrule} with coefficients computed from the natural MSR mass one gets
\begin{equation}\label{eq:N12}
N_{1/2}(n_\ell=3) = 0.526\pm 0.012\,.
\end{equation}
The maximum is reached for $\lambda_{\rm max} = 1.098$, which is very close to the default $\lambda = 1$, but the minimum corresponds to the extreme value $\lambda_{\rm min} = 0.5$, see Fig.~\ref{fig:N12}. The central value $N_{1/2}^{\rm mid}$ corresponds to $\lambda_{\rm mid}=1.784$, far away from the default value. Moreover \mbox{$N^{\rm max}_{1/2}(n_\ell=3)-N^{\lambda=1}_{1/2}(n_\ell=3)=0.0004$}, which is much smaller than the uncertainty shown in Eq.~\eqref{eq:N12}.
Numerical analyses carried out in Sec.~\ref{sec:RSlam} reveal that the incertitude of this quantity significantly affects the accuracy $\alpha_s$ can be determined with.
We have checked that this behavior is completely absent at one and two loops, orders at which $N_{1/2}$ is monotonically decreasing (being the latter almost linear). The three-loop result exhibits a maximum at $\lambda\simeq0.7$, significantly far away from the `default' $\lambda=1$, hence we regard this order is also free from the anomalous behavior.

The nice feature of Eq.~\eqref{eq:RSmass} is that its Borel transform can be computed exactly, as long as the inner sum is carried out up to
infinity~\cite{Beneke:1994rs}:
\begin{equation}\label{eq:borelRS}
B_{\alpha_s(R)}\!\!\left[m_Q^\pole - m_Q^{\rm RS}(R)\right]\!(u)= \frac{4\pi R}{\beta_0}N_{1/2}
\sum_{\ell=0}^{\infty}g_\ell\,(1-2u)^{-1-\hat b_1+\ell}\,(1+\hat b_1)_{-\ell} \,.
\end{equation}
The inverse Borel transform (which we call from now on the Borel sum) of Eq.~\eqref{eq:borelRS}, as first noted in Ref.~\cite{Bali:2003jq}, can be computed analytically using some prescription to regulate the singularities at $u=1/2$ (e.g.\ taking the real part of the principal value prescription), such that the difference of Borel sums for two distinct values of $R$ (for which the pole mass drops out) does not depend on the specific prescription taken to perform the integration, and sums up large logarithms associated to the leading renormalon. In fact, the imaginary part of the Borel sum of Eq.~\eqref{eq:borelRS} defined with the principal value prescription is proportional to $\LQCD$ and independent of $R$, therefore canceling out in any difference of Borel sums~\cite{Hoang:2017suc}.
It can be shown that when carrying out the sum to infinity, the principal value prescription and R-evolution are fully identical. Therefore, for simplicity, in our analysis based on the RS mass ---\,which is conducted as a cross check only\,--- we will use R-evolution.

\begin{figure}[t!]
\centering
\begin{subfigure}{0.43\textwidth}
\centering
\includegraphics[width=\textwidth]{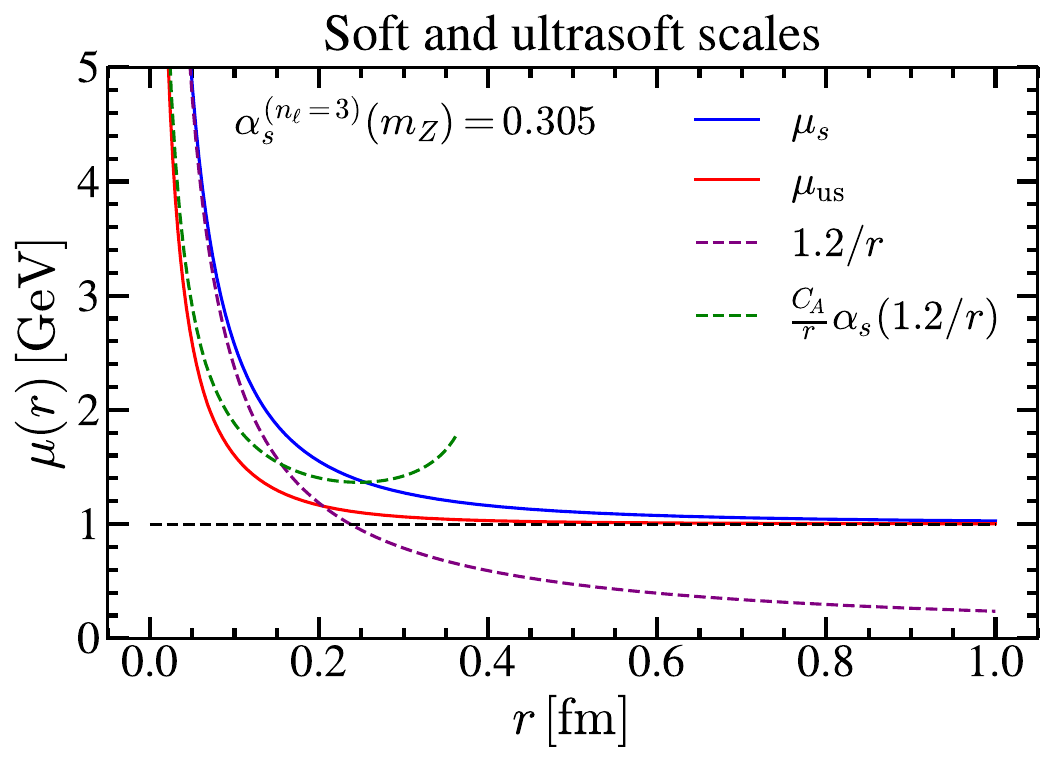}
\caption{}
\label{fig:profile1}
\end{subfigure}
\hfill
\begin{subfigure}{0.455\textwidth}
\centering
\includegraphics[width=\linewidth]{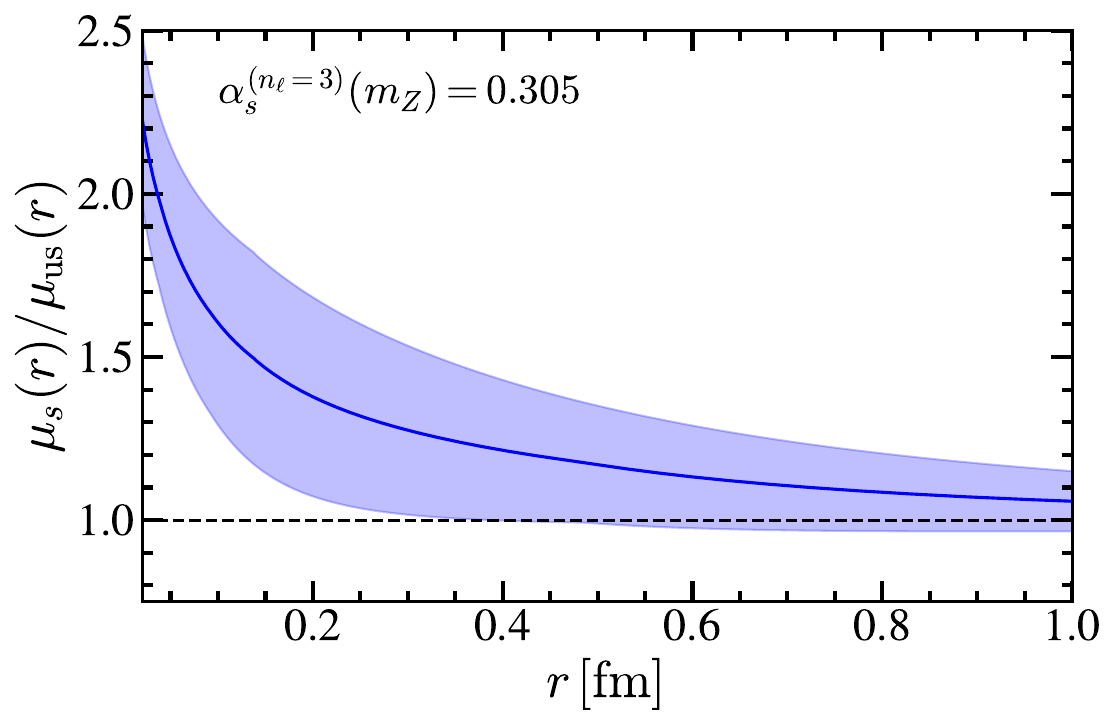}
\caption{}
\label{fig:MuRatio}
\end{subfigure}
\caption{\label{fig:profiles} Left panel: Dependence of the soft and ultrasoft renormalization scales with the distance $r$ for default parameters of the profile functions. The dashed purple line represents a pure canonical choice for the soft scale $\mu_s=1.2/r$ which becomes smaller than $1\,$GeV (marked with a thin solid black line) for $r>0.2\,$fm. The solid blue line shows Eq.~\eqref{eq:assimp} for $\xi=1.2$, $\mu_0=1\,$GeV and $b=0$, and in solid red we show the result of Eq.~\eqref{eq:usoftProf} for the same values plus the choice $\kappa=1$. The dashed green line (which is plotted up to $0.4\,$fm only since for larger $r$ the strong coupling is ill-defined) corresponds to the purely canonical ultrasoft scale. Right panel: ratio of the soft over the ultrasoft scale within the random scan. Both panels use the boundary condition $\alpha_s^{(3)}(m_\tau)=0.305$.}
\end{figure}

The same parameter $\lambda$ can be used to estimate the uncertainty in
$\Delta(R,R_0)$ caused by truncating the R-anomalous dimension series at a finite order, as can be seen in Fig.~\ref{fig:DeltaMSR} for various $R$ and a fixed $R_0$. Even though this functionality is implemented in \texttt{REvolver}, most phenomenological analyses do not add to their error budget the uncertainty coming from RG evolution, as it is assumed to be captured already by regular scale variation. Figure~\ref{fig:lambda} shows that a very similar pattern to that described for $N_{1/2}$ is also observed for $\Delta(R,R_0)$ between any two scales at N$^3$LL, where the minimum happens at $\lambda=0.5$ and the maximum
at the $R$-independent value $\lambda=1.007$, extremely close to the `default' $\lambda=1$. The value of $\lambda$ for which the central value of $\Delta(R,R_0)$ (that is, the average of the largest and smallest values attained in the range $\lambda\in[0.5,2]$) is reproduced exhibits some dependence with $R$ and is in the vicinity of $\lambda=2$, hence not far from $\lambda_{\rm min}$. The comparison with $N_{1/2}$ follows an analogous pattern also at lower orders. Given the similarities between the RS and MSR masses, this is only to be expected, and suggests that using the default value $\lambda=1$
might result in a biased central value. Therefore, for our final analysis at N$^3$LL we will use in the MSR scheme
$\lambda=\lambda_{\rm mid}$ (that we round to $1.8$) and, to be conservative, vary this parameter in the range $\lambda\in[1.5,2.1]$ to account in the error budget
for the arbitrariness of this choice.\footnote{One also reproduces $N_{1/2}^{\rm mid}$ for $\lambda = 0.65$, as can be seen in the figure. However, we do not use this value as our unbiased default since, when it comes to R-evolution, it implies evaluating the strong coupling at unphysically small scales that jeopardize perturbation theory.} The effect of this variation is rather mild, increasing the uncertainty mainly for $r<0.04\,$fm (where there is more R-evolution), decreasing it for intermediate distances (since the strong coupling is evaluated at larger scales) and leaving it essentially unaffected for $r>0.2\,$fm (where there is very little R-evolution). As for central values, it raises the static energy about 0.1\,GeV at $r=0$ but the difference decreases and roughly disappears at $r\simeq 0.1\,$fm.

\section{Renormalization Scale Setting}\label{sec:Renormalization}
In this section we discuss how to implement the $r$ dependence of the various renormalization scales to extend the validity range of the perturbative prediction for the static energy. We also carry out an exploration of its order-by-order behavior to ensure all expected features for a well-behaved series are satisfied. All analyses in this section use the MSR subtraction scheme.

\subsection{Profiles Functions for the Static Energy}\label{sec:profiles}
After removing the leading renormalon, a good perturbative behavior
must be achieved for all the orders considered within the range of distances used in our numerical analysis. This makes sure the perturbative contribution to the static energy dominates over non-local condensates which are not accounted for in our theoretical expressions. Such perturbative behavior is ensured if the logarithms and $\alpha_s(\mu)$ factors appearing in the perturbative expansion of the static energy are kept small (or at least not large) \emph{simultaneously}.

\begin{figure}[t!]
\centering
\begin{subfigure}{0.45\textwidth}
\centering
\includegraphics[width=\textwidth]{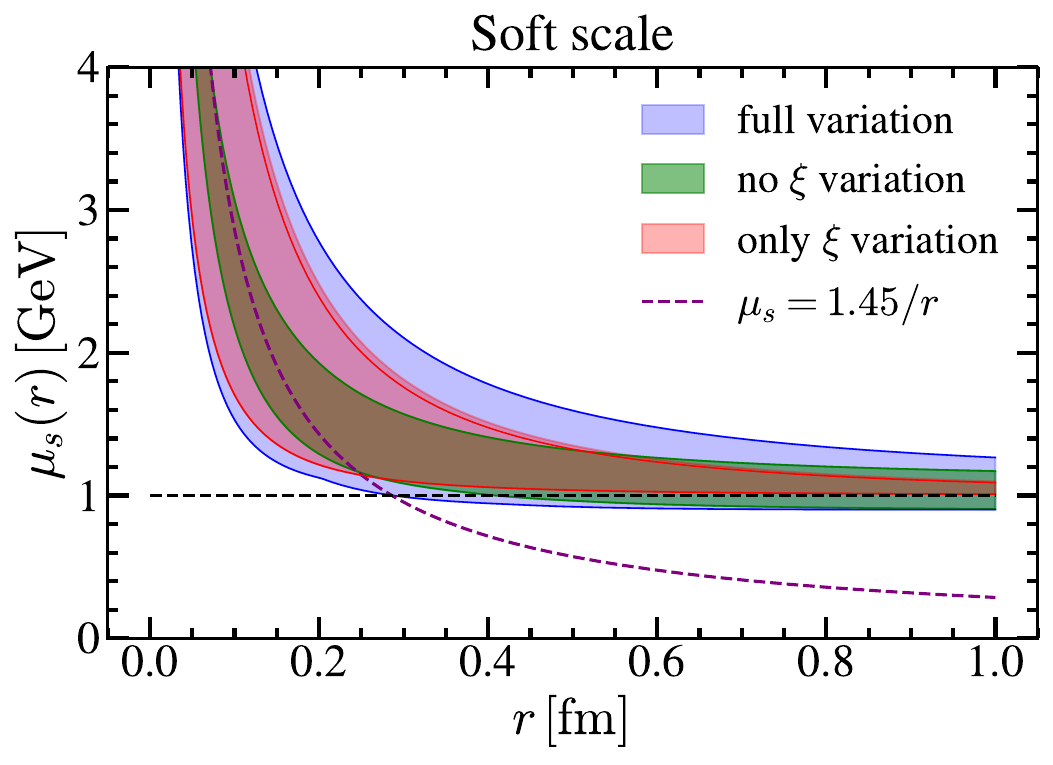}
\caption{}\label{fig:profile2}
\end{subfigure}
\hfill
\begin{subfigure}{0.45\textwidth}
\centering
\includegraphics[width=\textwidth]{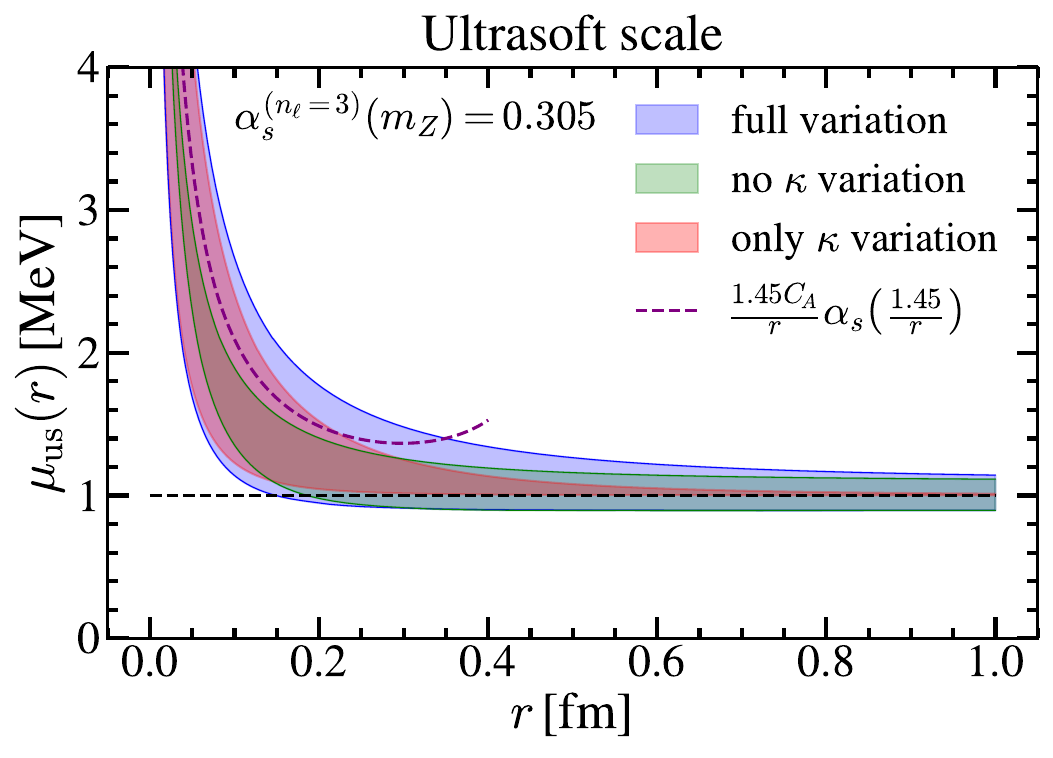}
\caption{}\label{fig:profile3}
\end{subfigure}
\caption{\label{fig:profile}
Soft (left panel) and ultrasoft (right panel) renormalization scale variation obtained by randomly scanning over the parameters that define their analytic forms. The blue bands are obtained by varying all parameters within the ranges specified in Table~\ref{tab:profiles}. The green band for $\mu_s$ ($\mu_{\rm us}$) is obtained by fixing $\xi$ ($\kappa$) to its default value and varying the rest of parameters, while in the red band only $\xi$ ($\kappa$) is varied, with the remaining profile parameters set to their default values. The red band in panel (b) is not centered for small $r$ because $\xi$ is fixed to the default rather than to its middle value. For reference, the canonical scales $\mu_s=1.45/r$ and $\mu_{\rm us}=1.45C_A\alpha_s(1.45/r)/r$ (dashed purple lines) as well as the freeze-out scale $1\,$GeV (dashed black line) are shown. Both panels use $\alpha_s^{(3)}(m_\tau)=0.305$.}
\end{figure}

As can be seen in Eq.~\eqref{eq:static}, for fixed values of $\mu_s$ the $r$ dependence of $V_{\rm QCD}$ ---\,besides the overall $1/r$ factor\,--- is dominated by powers of $\log(r\mu)$. On the one hand, a constant renormalization scale $\mu_s\sim 1/r_0$ with $r_0\ll 1\,$fm maintains $\alpha_s(\mu)$ small at all distances, but for
$r\gg r_0$ makes the logarithm grow up, deteriorating the perturbative series' convergence. On the other hand, the \emph{canonical profile}
$\mu_s= \xi/r$ with $\xi$ of $\mathcal{O}(1)$ in natural units
makes the logarithm small at all distances, allowing to describe properly the short distance domain where the strong coupling is small as well for this profile. However, the nice perturbative behavior of the static energy in this case collapses for $r\gtrsim 1\,{\rm GeV}^{-1}\simeq 0.2\,$fm, since the renormalization scale crosses the Landau pole where $\alpha_s$ is ill-defined.

A solomonic solution merges the two previous approaches, that is, a renormalization scale that behaves as $\xi/r$ for small $r$ which freezes to a constant (perturbative) value as $r$ increases:
\begin{equation}
\mu_s=\left\{\begin{array}{ll}\dfrac{\xi}{r}&\mbox{ for }r\to 0\\
\mu_\infty &\mbox{ for }r\to \infty\end{array}\right..
\end{equation}
A global analytical function that encodes those two regimes is\footnote{Another choice reproducing the two regions at small and large $r$ is $\mu_s(r,\xi,\mu_\infty,\mu_\infty,0) = \frac{\xi}{r}+\mu_\infty$. We discard this possibility as the basis upon which we construct our profiles for small $r$ since, as opposed to Eq.~\eqref{eq:assimp}, it has $\mathcal{O}(r^0)$ corrections to the pure canonical behavior. Furthermore, at large distances it has $1/r$ positive corrections to the constant behavior, which are absent in Eq.~\eqref{eq:assimp}.}
\begin{align}\label{eq:assimp}
\mu_s(r,\xi,\mu_\infty,0,0)=\sqrt{\biggl(\frac{\xi}{r}\biggr)^{\!\!2}+\mu_\infty^2}
= \left\{\begin{array}{ll}\frac{\xi}{r}+\frac{\mu_\infty^2\,r}{2\xi}+\mathcal{O}(r^2)\\
\mu_\infty +\frac{\xi^2}{2\mu_\infty r^2}+\mathcal{O}(r^{-3})\end{array}\right.\!,
\end{align}
shown as a solid blue line in Fig.~\ref{fig:profile1}, together with the canonical profile (with $\xi=1.2$ in natural units, dashed purple) and the freeze-out scale $\mu_\infty=1$\,GeV (dashed black). It reduces to a pure canonical form for $\mu_\infty=0$, while for finite $\mu_\infty$ reproduces the canonical profile at small distances up to linear corrections in $r$. We choose $\xi=0.2+2^{\pm1}$ with default value $\xi=1.2$ instead of the usual $\xi=2^{\pm1}$ since numerical investigations reveal that, when fitting for the strong coupling, the standard choice produces
outliers, manifest as large values of $\alpha_s$ away from the bulk of results and caused by small renormalization scales. These are more abundant for datasets including mainly small distances. Any sensible method to estimate perturbative uncertainties must discard outliers, what in our case is equivalent to restricting $\xi\geq 0.7$. Our choice is however more conservative than simply throwing away anomalous results since includes values of $\xi$ as large as $2.2$, hence yielding larger perturbative uncertainties. This variation will be used for the renormalon subtraction renormalization scale $R$ as well.
\begin{table}[t!]
\centering
\begin{tabular}{|c c c|}
\hline
\textbf{Parameter} & \textbf{Default} & \textbf{Range} \\ \hline
$\xi$ & $1.2\times \hbar c$ & $[0.7,2.2]\times \hbar c$ \\
$\beta$ & $1.2\times \hbar c$ & $[0.7,2.2]\times \hbar c$ \\
$b\,$ & 0 & $[-0.3, 0.3]\times\hbar c$ \\
$\Delta\,$ & 0 & $[-0.6,0.6]\,$GeV \\
$\mu_\infty\,$ & $1\,$GeV &$[0.9,1.1]\,$GeV \\
$R_\infty\,$ & $1\,$GeV &$[0.9,1.1]\,$GeV \\
$\kappa$ & $1$ & $[0.8,1.2]$ \\
$\lambda$ & $1.8$& $[1.5,2.1]$ \\\hline
\end{tabular}\caption{Soft, ultrasoft, and renormalon-subtraction profile-function parameters and their variation ranges in the flat random scan. For completeness, we also include in the last row the R-evolution parameter $\lambda$ variation.\label{tab:profiles}}
\end{table}

\begin{figure}[t!]
\centering
\begin{subfigure}{0.455\textwidth}
\centering
\includegraphics[width=\textwidth]{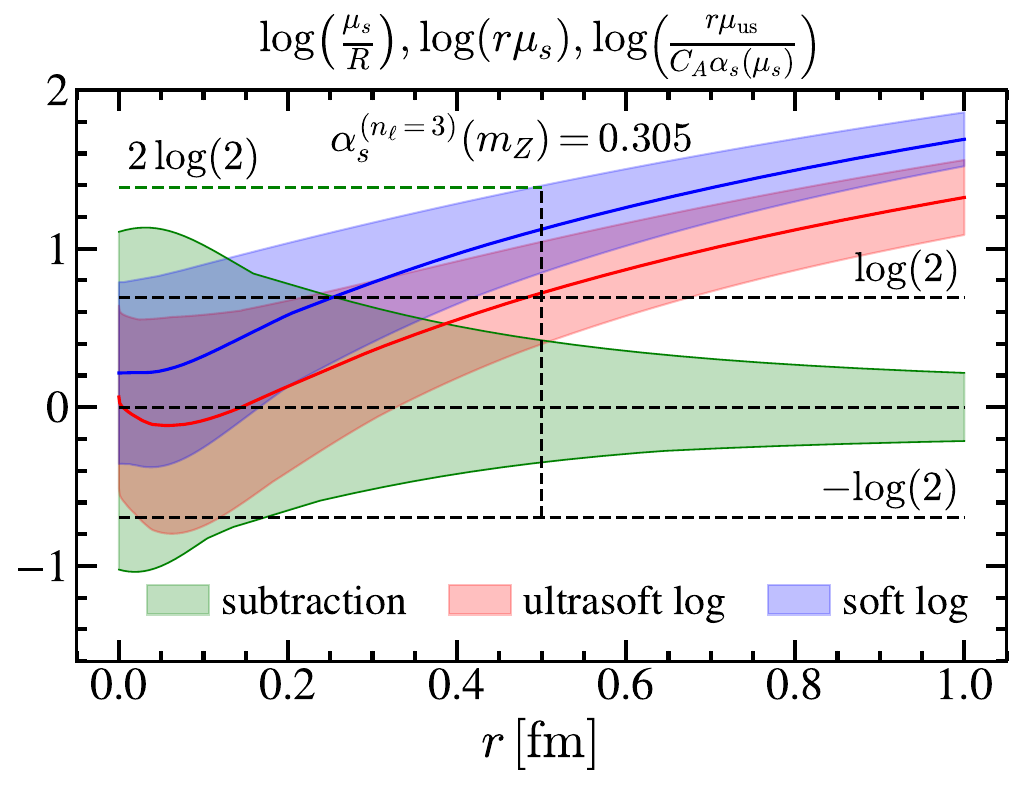}
\caption{}\label{fig:logs}
\end{subfigure}
\hfill
\begin{subfigure}{0.48\textwidth}
\centering
\includegraphics[width=\textwidth]{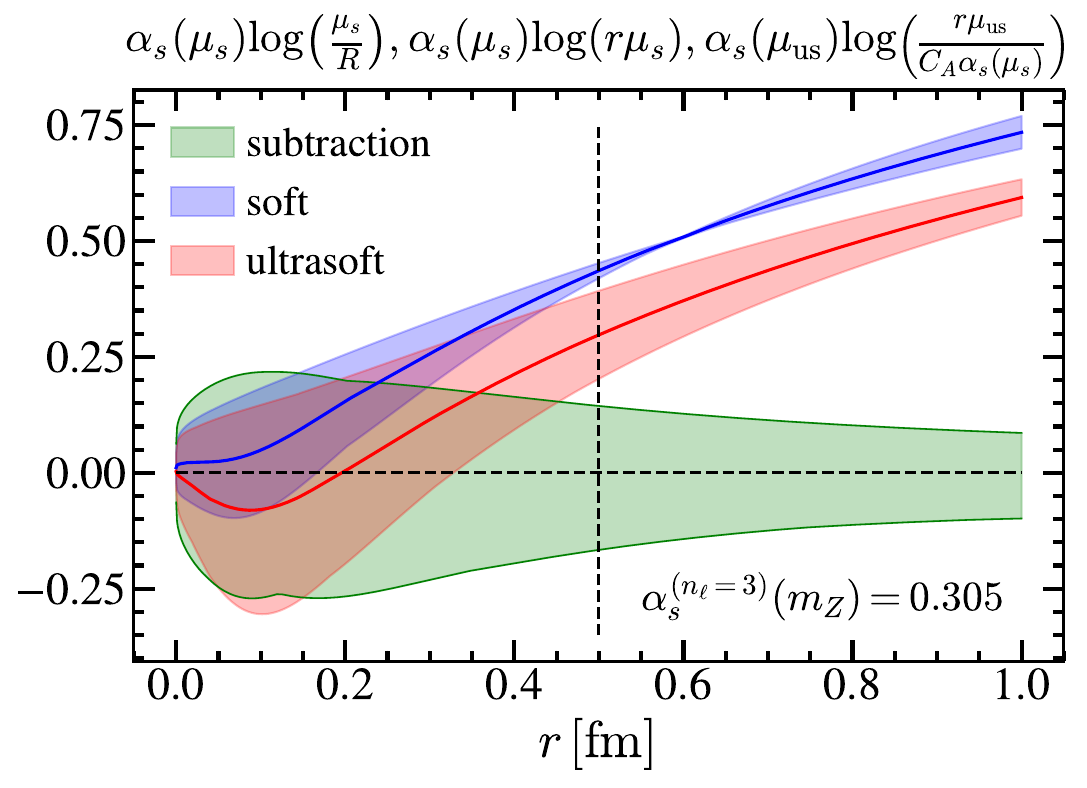}
\caption{}\label{fig:AlphaLog}
\end{subfigure}
\caption{\label{fig:ALogs}
Left panel: size of subtraction, soft and ultrasoft logarithms for the $R$, $\mu_s$ and $\mu_{\rm us}$ profile parameters varied within the ranges specified in Table~\ref{tab:profiles}. Dashed horizontal black lines mark the canonical size of logarithms under standard scale variation by factors of two, and the dashed green horizontal line marks $2\log(2)$.
Right panel: same as panel~(a), but multiplied by $\alpha_s(\mu_s)$, $\alpha_s(R)$ and $\alpha_s(\mu_{\rm us})$ the soft, subtraction and ultrasoft logarithms, respectively.
In both panels, the dashed black vertical line shows the maximal distance entering our fits.
Both plots are generated with $\alpha_s^{(3)}(m_\tau)=0.305$.}
\end{figure}

As already anticipated, with the aim of correctly estimating the theoretical uncertainty related with the choice of the renormalization scale, the parameter $\xi$
will be scaned over to implement ${}^{+\log(2.2)}_{-\log(1.43)}$ variations when $r$ is small. Since there are multiple functional forms that reproduce both the canonical and freeze-out behaviors, we account for this arbitrariness including two additional parameters and define
\begin{equation}\label{eq:profile}
\mu_s(r,\xi,\mu_\infty,\Delta,b) = \sqrt{\biggl(\frac{\xi}{r}\biggr)^{\!\!2}+\frac{b}{r}+(\mu_\infty-\Delta)^2}-\Delta
= \left\{\begin{array}{ll}\frac{\xi}{r}+\frac{b}{2\xi}-\Delta +\mathcal{O}(r)\\
\mu_\infty +\frac{b}{2(\Delta + \mu_\infty)r}+\mathcal{O}(r^{-2})\end{array}\right.\!,
\end{equation}
which has the asymptotic behavior show in Eq.~\eqref{eq:assimp} and reproduces its functional form for $b=\Delta=0$.
It reduces to a pure canonical scale if all parameters except for $\xi$ are set to zero. In the context of event shapes, such
renormalization scale parametrizations are known as {\it profile functions}, see e.g.\ Ref.~\cite{Abbate:2010xh,Hoang:2014wka} for piece-wise defined profiles.
Even though our profiles are not purely canonical in the entire spectrum, the soft logarithms appearing in the static potential remain $\mathcal{O}(1)$ and never become large, as can be seen in the blue band of Fig.~\ref{fig:logs}. In fact, the largest value the soft logarithm attains in the fit region is $2\log(2)\simeq 1.4$, which is definitely not large and should not alter perturbative convergence.

For the renormalon subtraction scale $R$ an identical functional dependence will be employed, with two different parameters $\beta$ and $R_\infty$ that play the role of $\xi$ and $\mu_\infty$ in the soft scale, respectively, and are (independently) varied in exactly the same ranges. They allow $\mu_s$ and $R$ to differ (we use, however, correlated values of $\Delta$ and $b$ in both scales):
\begin{equation}
R(r,\beta,R_\infty,\Delta,b) = \mu_s(r,\beta,R_\infty,\Delta,b)\,.
\end{equation}
With this functional form and the variation of the corresponding profile parameters we ensure that $\log(\mu_s/R)$ never becomes large, making sure our renormalon subtraction series is well behaved. This can be seen in the green translucent band of Fig.~\ref{fig:logs}, which is centered around zero and barely exceeds unity in absolute value.

Finally, the ultrasoft scale must have behave as
$\mu_{\rm us}\sim C_A\alpha_s(\mu_s)/r$ and satisfy the constraint $\mu_s\gg\mu_{\rm us}$
at small distances, but for large $r$ the hierarchy among scales
no longer holds and therefore we can simply identify $\mu_s\simeq \mu_{\rm us}$. In our parametrization, we smoothly merge the two renormalization scales to the values $\mu_s\sim\mu_{\rm us}\sim\mu_\infty$ when the radius becomes $\mathcal{O}(1\,{\rm fm})$, hence switching off ultrasoft resummation, while at short distances we allow for variations of $\mu_{\rm us}$ around its canonical value:
\begin{equation}\label{eq:usoftProf}
\mu_{\rm us}(r,\xi,\kappa,\mu_\infty,\Delta,b) = C_A\Bigl\{\mu_s(r,\kappa\xi,\mu_\infty,\Delta,b)\alpha_s\bigl[\mu_s(r,\kappa\xi,\mu_\infty,\Delta,b)\bigr]-\mu_\infty\alpha_s(\mu_\infty)\Bigr\} + \mu_\infty\,.
\end{equation}
The dependence on $\kappa\xi$ where $\kappa=1\pm 0.2$ ensures the condition $\mu_s>\mu_{\rm us}$ for small $r$ and avoids unnaturally small values of $\mu_{\rm us}$.
The ultrasoft scale is shown as a red solid line in Fig.~\ref{fig:profile1} for the default parameters $\xi=1.2$ (in natural units), $\kappa=1$, $b=\Delta=0$ and $\mu_\infty=1\,$GeV. Again, even if not fully canonical in the entire spectrum, this parametrization never yields large ultrasoft logarithms, as can be seen in the red band shown of Fig.~\ref{fig:logs}. The ultrasoft logarithm is smaller than the soft one, which was already argued to be $\mathcal{O}(1)$ in the fit range. We have also checked that the product of the strong coupling evaluated at the soft and ultrasoft scales times the corresponding logarithm, see Fig.~\ref{fig:AlphaLog}, is smaller than $0.5$ ($0.75$) for $r<0.5\,$fm ($r<1\,$fm), what ensures nice perturbative convergence.\footnote{We do not include the factor of $1/\pi$ accompanying $\alpha_s$ in our analysis that would make the product even smaller.} The same is true for the renormalon subtraction logarithm shown in green. One should keep in mind that the $n$-th power of the soft logarithm is multiplied by $[\alpha_s(\mu_s)]^{n+1}$ or higher (the same goes true for the renormalon-subtraction logarithm), and that the ultrasoft logarithm is also multiplied by three powers of $\alpha_s(\mu_s)$ and a single power of $\alpha_s(\mu_{\rm us})$.

In order to estimate the perturbative uncertainties of the static energy caused by missing higher orders we scan over the three renormalization scales. Since these are not Gaussian parameters, they should be varied within fixed ranges. Given the relatively large number of parameters describing the profile functions, some of them correlating the three scales, we perform a flat random scan on each parameter. Specifically, we randomly select $500$ different sets of profiles functions, where each parameter takes an arbitrary value within the ranges specified in Table~\ref{tab:profiles}. The default values around which the parameters are varied have been chosen to cancel the leading corrections to the canonical and freeze-out behaviors at small and large distances, respectively. Since the parameter $\lambda$ that specifies how R-evolution is carried out is also varied, and given that its variation is also related to perturbative uncertainties,
it is also varied within the same random scan.

\subsection{Perturbative Analysis}\label{sec:pert}
\begin{figure}[t!]
\centering
\begin{subfigure}{0.48\textwidth}
\centering
\includegraphics[width=\textwidth]{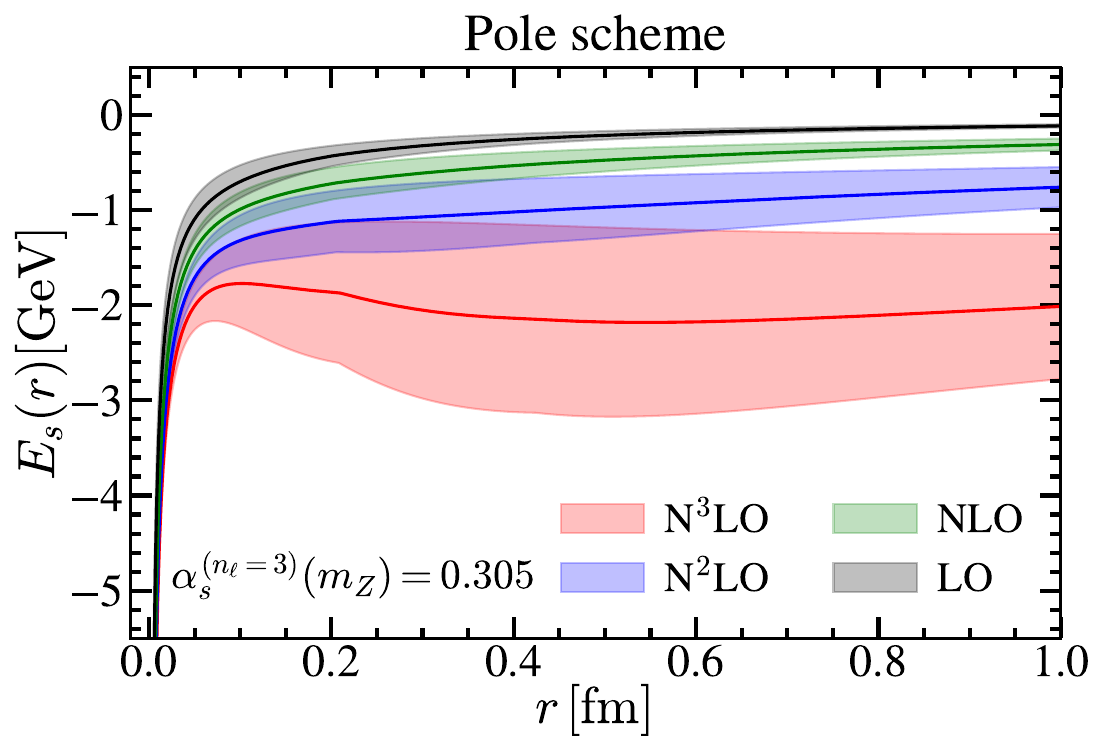}
\caption{}\label{fig:Pole}
\end{subfigure}
\hfill
\begin{subfigure}{0.48\textwidth}
\centering
\includegraphics[width=\textwidth]{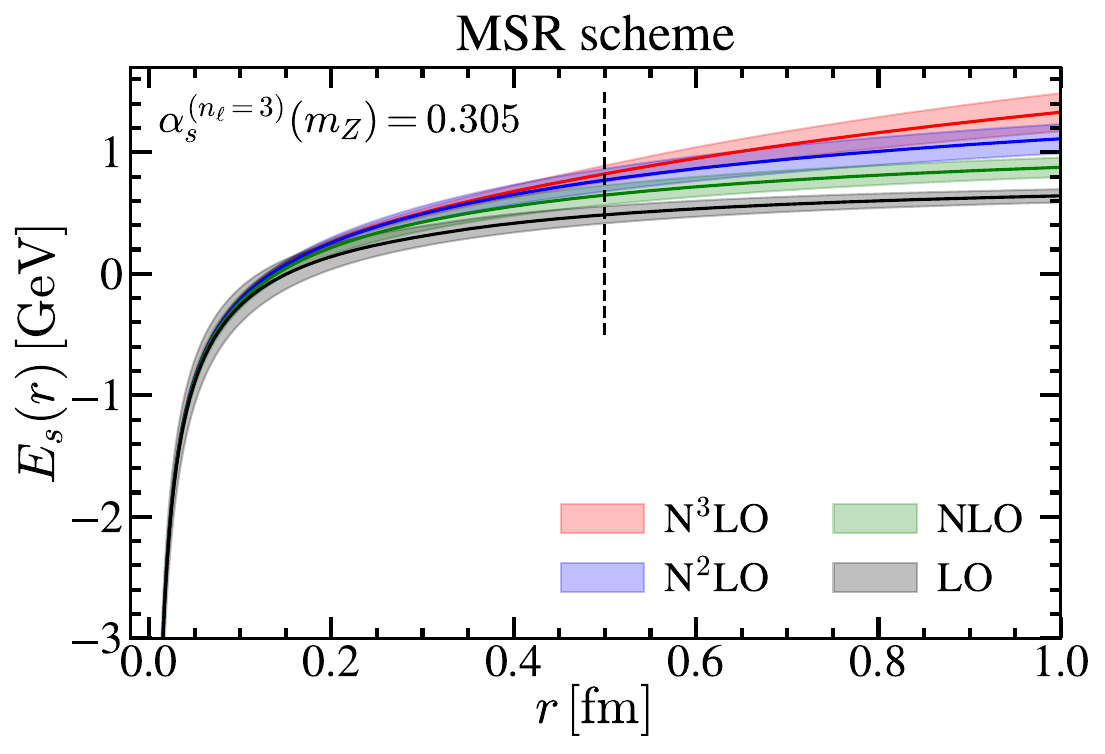}
\caption{}\label{fig:MSR}
\end{subfigure}
\caption{\label{fig:Orders}
Static energy at LO (gray), NLO (green), N$^2$LO (blue) and N$^3$LO (red), with uncertainty bands generated by varying the profile parameters within the ranges specified in Table~\ref{tab:profiles}. The N$^2$LO (N$^3$LO) prediction includes N$^2$LL (N$^3$LL) ultrasoft resummation. Left and right panels show the static energy in the pole and MSR schemes, respectively. Both panels use $\alpha_s^{(3)}(m_\tau)=0.305$ and $R_0=2\,$GeV.}
\end{figure}

In this section we perform a brief numerical analysis of the static energy's perturbative behavior before delving into our strong coupling fits. Unless otherwise stated, all figures and investigations are based on the MSR scheme with $R_0=2\,$GeV and the boundary condition $\alpha_s^{(n_\ell=3)}(m_\tau)=0.305$. This value agrees within errors with the most recent determination from hadronic $\tau$ decays in Ref.~\cite{Boito:2025pwg}, namely $\alpha_s^{(n_\ell=3)}(m_\tau)=0.2983\pm0.0101$. The N$^k$LO prediction is obtained implementing the potential and subtraction series at $\mathcal{O}(\alpha_s^{k+1})$. While truncating those
at the same order is fundamental in order to cancel the renormalon, there is some freedom in RG evolution. We choose to evolve the strong coupling with the \mbox{five-loop} beta function and use R-evolution at the highest available order N$^3$LL in all cases, independent of the order at which the static energy is truncated, since we have checked this choice yields a more conservative estimate of perturbative errors and makes varying $\lambda$ mandatory at every order, so that the treatment is identical.
Other equally valid choices are to evolve the MSR mass and/or the strong coupling with less accuracy at lower orders.

\begin{figure}[t!]
\centering
\begin{subfigure}{0.48\textwidth}
\centering
\includegraphics[width=\textwidth]{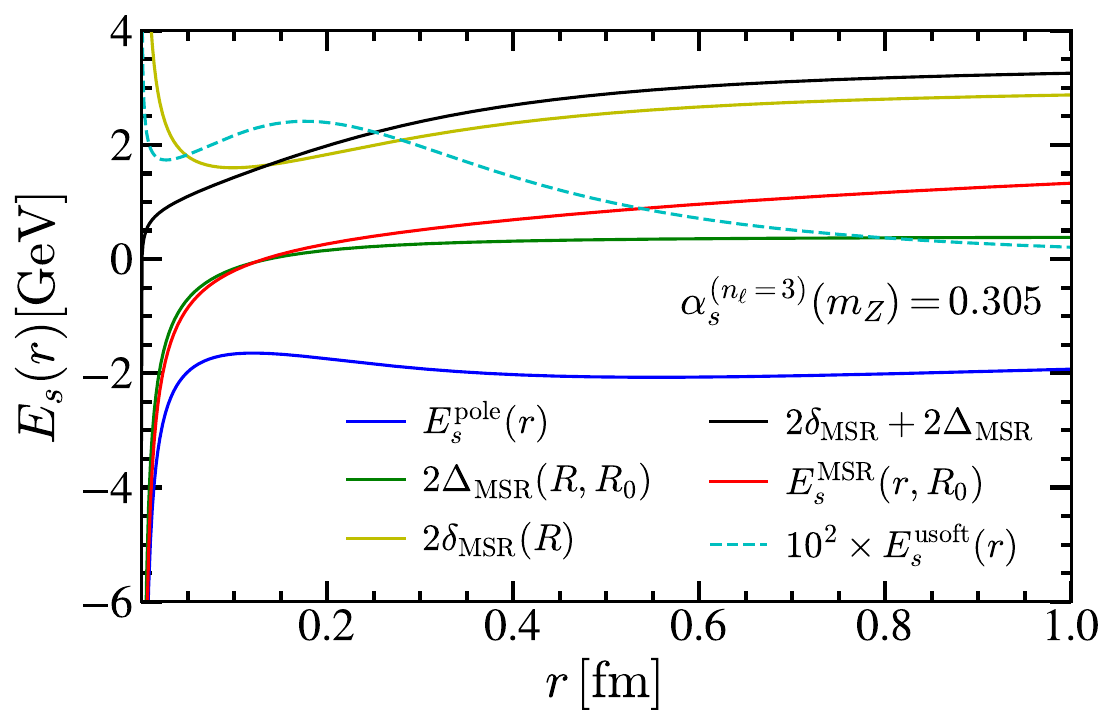}
\caption{}\label{fig:Components}
\end{subfigure}
\hfill
\begin{subfigure}{0.491\textwidth}
\centering
\includegraphics[width=\textwidth]{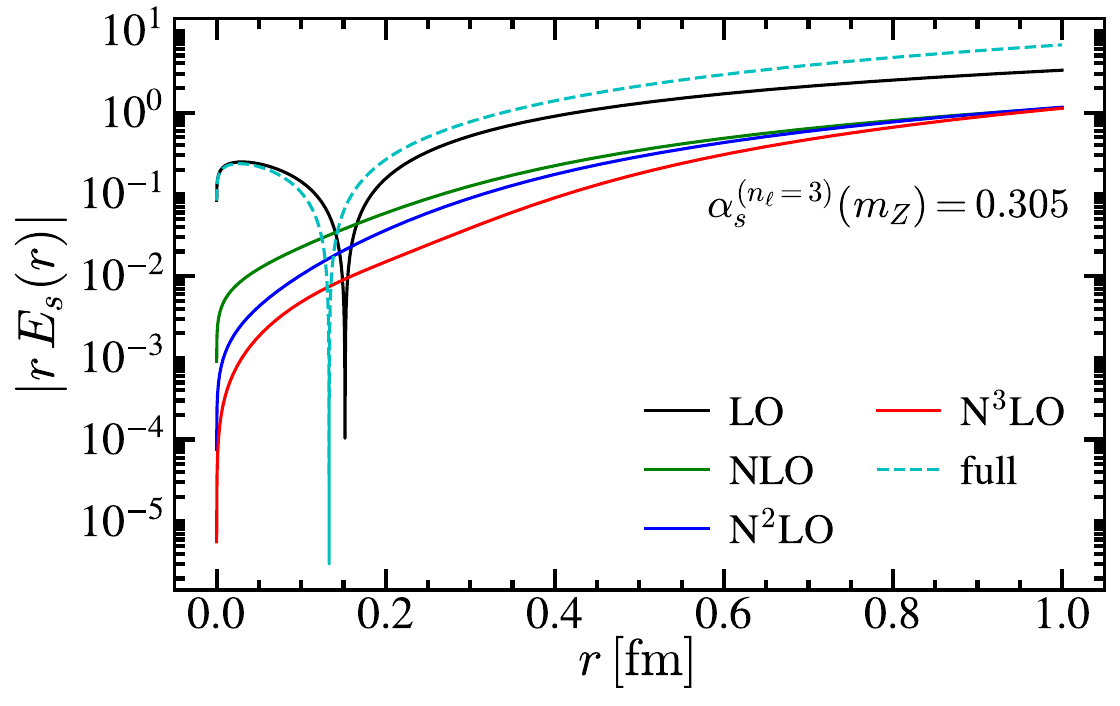}
\caption{}\label{fig:Ords}
\end{subfigure}
\caption{\label{fig:Sens}
Left panel: Various components of the static energy: solid blue shows the pole-scheme contribution (with ultrasoft resummation), yellow and green correspond to the MSR subtractions and R-evolution kernel, respectively, where the sum is shown as a solid black line. The sum of black and blue, i.e., the MSR-scheme static energy, is represented in solid red. The resummed ultrasoft contribution multiplied by a factor of 100 appears as a dashed cyan line. Right panel: Leading-order static energy (black line) along with the NLO, N$^2$LO, and N$^3$LO corrections shown in green, blue, and red, respectively. In all curves we plot the absolute value multiplied by $r$
to yield a positive dimensionless quantity. The full N$^3$LO prediction is shown as a dashed cyan line. Both panels use canonical profiles with $\alpha_s^{(3)}(m_\tau)=0.305$ and $R_0=2\,$GeV.}
\end{figure}
We start discussing the order-by-order convergence of the static energy, comparing the pole and MSR-improved setups in Fig.~\ref{fig:Orders}. Perhaps the most prominent feature of the pole scheme, shown in Fig.~\ref{fig:Pole}, is that uncertainty bands are larger at higher than lower orders already for $r>0.02\,$fm, the opposite of what should be expected. On the other hand, in the MSR scheme shown in Fig.~\ref{fig:MSR}, below $0.5\,$fm the N$^3$LO uncertainty band is the smallest, and below $0.35\,$fm the rest of bands are ordered according to the perturbative precision. As a benchmark, we quote the uncertainties at $r=0.3\,$fm at LO, NLO, N$^2$LO and N$^3$LO: $[0.16, 0.31, 0.71, 1.86]\,$GeV in the pole scheme, and $[0.17, 0.15, 0.13, 0.07]\,$GeV in the MSR scheme. As for central values, at $r=0.5\,$fm the pole-scheme prediction jumps $-0.26\,$GeV, $-0.49\,$GeV and $-1.21\,$GeV from LO to NLO, N$^2$LO and N$^3$LO, respectively. The respective shifts in the MSR scheme are $0.16\,$GeV, $0.12\,$GeV and $0.05\,$GeV, decreasing at each order, as expected. We conclude then that the pole scheme is not adequate for a high-precision analysis and that our uncertainty estimates in the MSR scheme within the fit range encode all the nice features one expect in well behaved series: order-by-order convergence, decreasing uncertainty bands and shifts that progressively decrease as higher-order corrections are included. Hence, for $r \lesssim 0.45\,$fm we see no evidence of the $u=3/2$ renormalon effect. On the other hand, for larger distances subtracting such renormalon, whose ambiguity is proportional to $r^2$, could lead to better order-by-order convergence, although this implies including the corresponding non-perturbative correction in the form of a non-local condensate.

Next we discuss the importance of each contribution to the MSR-improved static energy, illustrated in Fig.~\ref{fig:Components}. For this analysis we consider the highest order with default profiles. The solid blue line shows the pole-scheme contribution, which is always negative and becomes very flat already at $r\simeq 0.2\,$fm, not displaying the Cornell-like linear rising potential responsible for confinement. The N$^3$LL resummed ultrasoft terms are very small and to make them visible they are magnified by a factor of 100, shown as a dashed cyan line. The subtraction series $\delta_{\rm MSR}$ is represented in solid yellow and, as expected, diverges as $1/r$ and is always positive. The R-evolution kernel $\Delta_{\rm MSR}$ is depicted in green solid and attains both positive and negative values. The sum of these MSR-related contributions is shown in solid black, and the addition of all four contributions, making up our final result, in solid red. It is also instructive to display the size of each perturbative correction. To that end, in Fig.~\ref{fig:Ords} we show for the default profiles that higher-order corrections become increasingly smaller if the renormalon is subtracted, see caption of the figure for more details. The results are displayed in absolute value and on a logarithmic scale, since they would otherwise be hardly visible.

\begin{figure}[t!]
\centering
\begin{subfigure}{0.49\textwidth}
\centering
\includegraphics[width=\textwidth]{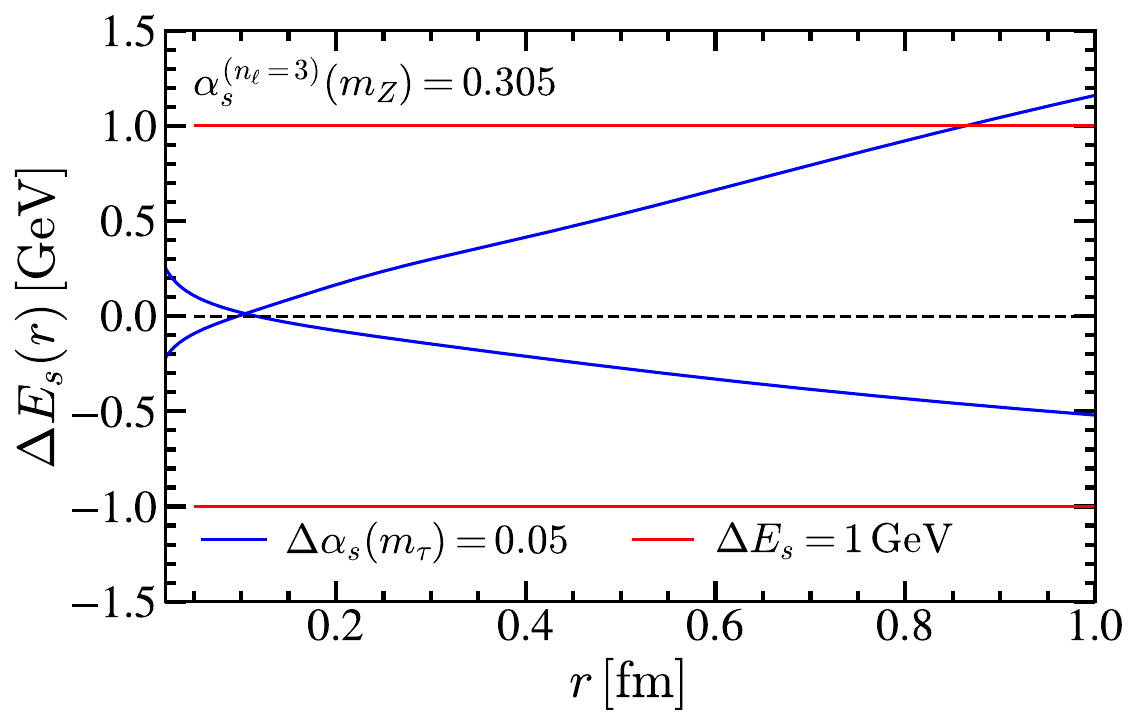}
\caption{}\label{fig:DeltaV}
\end{subfigure}
\hfill
\begin{subfigure}{0.48\textwidth}
\centering
\includegraphics[width=\textwidth]{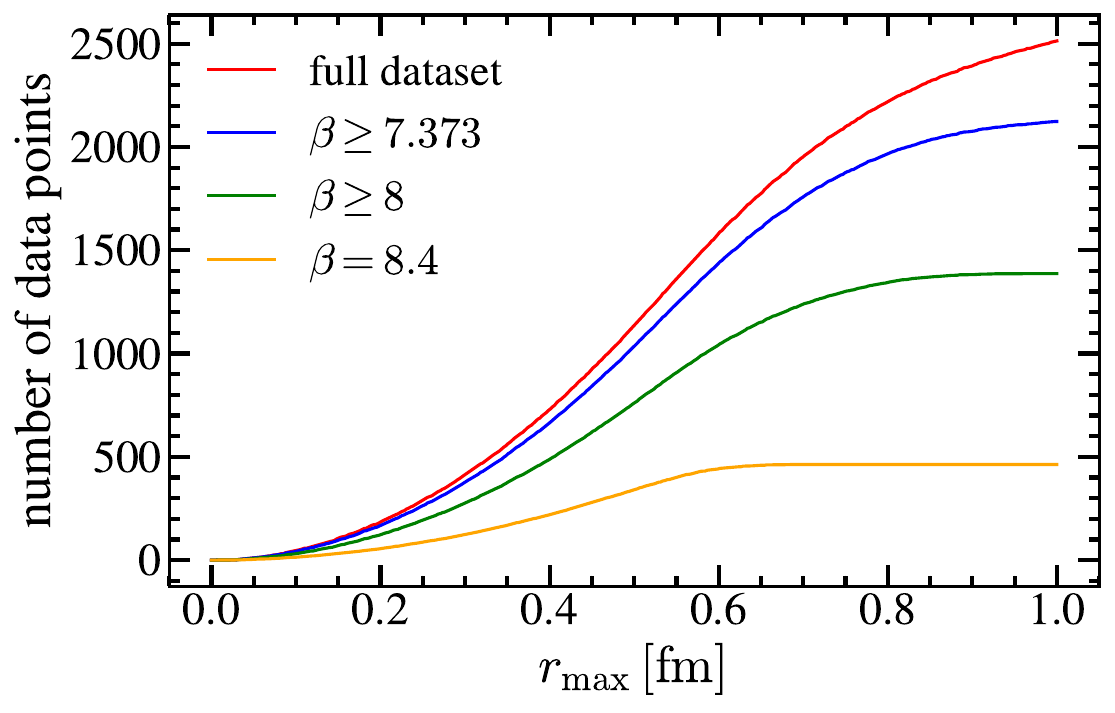}
\caption{}\label{fig:DOF}
\end{subfigure}

\caption{\label{fig:DeltaVDOF} Left panel: effect caused in the static energy by varying $\alpha_s^{(3)}(m_\tau)$ up and down by $0.05$, and by shifting the potential $1\,$GeV upwards or downwards for canonical profiles,
$\alpha_s^{(3)}(m_\tau)=0.305$ and $R_0=2\,$GeV. Right panel: number of data points contained in the HotQCD dataset as a function of the maximal distance $r_{\rm max}$ considered for various selections of lattice ensembles as indicated in the caption.}
\end{figure}

In order to carry two-parameter fits, it is important to make sure there are no strong correlations between these parameters that could cause flat directions in the $\chi^2$. In Fig.~\ref{fig:DeltaV} we compare the effects of varying the offset (solid red lines) by $1\,$GeV and varying $\alpha_s^{(n_\ell=3)}(m_\tau)$ by $0.05$ (dashed blue line).
These two variations cause very distinct modifications in the static energy even for small ranges of $r$, hence it appears fitting both is warranted. Fits which include larger ranges of $r$ should display smaller correlations and hence are favored.

\section{Lattice Data}\label{sec:data}
In this section we briefly describe the results from lattice QCD simulations on the static energy that will be used to determine the strong coupling through a comparison with our theoretical perturbative QCD (pQCD for short) prediction. We also introduce a method to statistically combine all available data into a single dataset containing significantly less points than the original set of simulations.

\subsection{HotQCD Lattice Simulations}
\begin{figure}[t!]
\centering
\begin{subfigure}{0.473\textwidth}
\centering
\includegraphics[width=\textwidth]{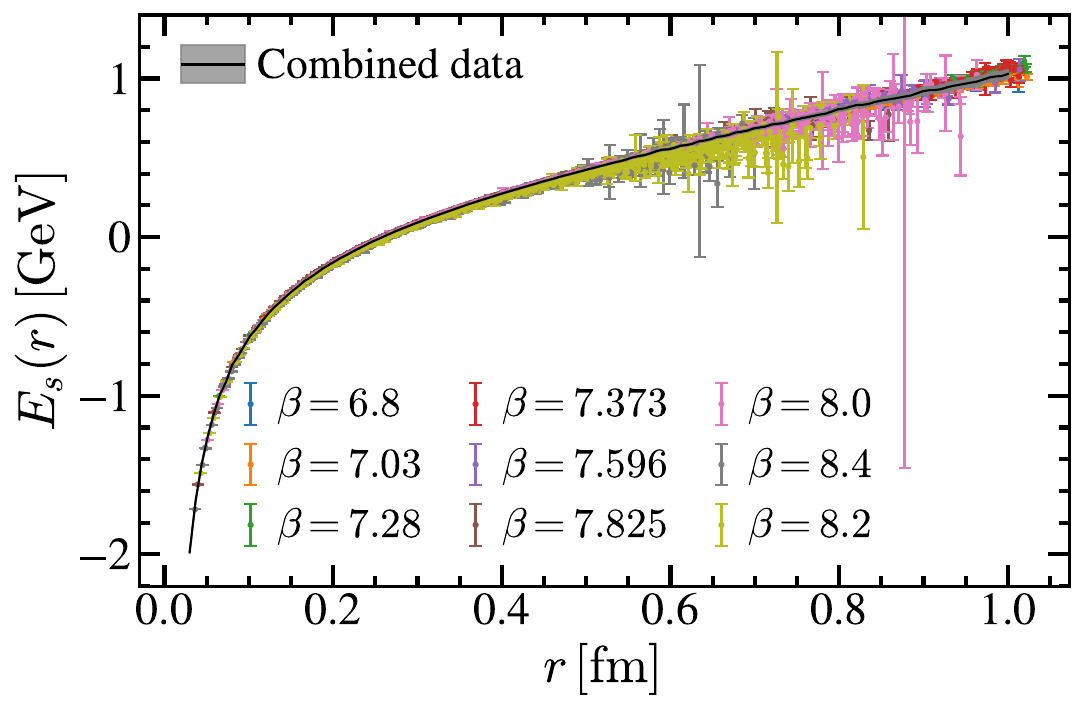}
\caption{}\label{fig:Cluster}
\end{subfigure}
\hfill
\begin{subfigure}{0.499\textwidth}
\centering
\includegraphics[width=\textwidth]{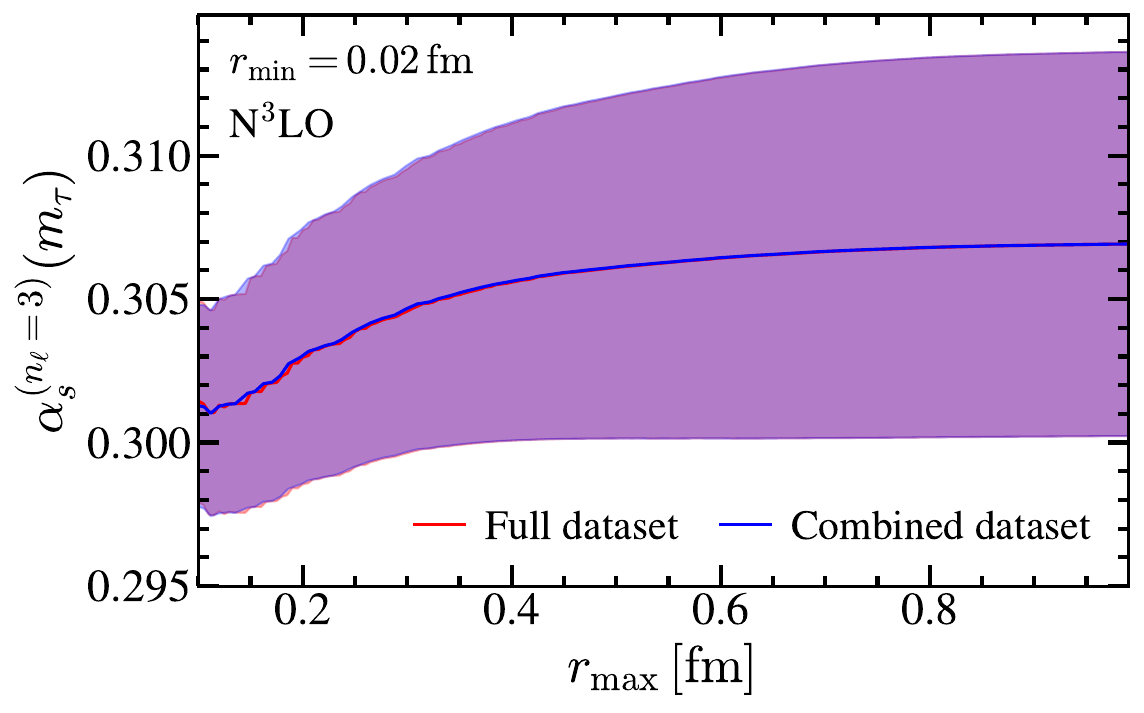}
\caption{}\label{fig:ClusterFit}
\end{subfigure}
\caption{\label{fig:FitCluster} Left panel: Lattice data on the static energy used in our analysis displayed in different colors according to the various $\beta$ values, confronted with the recombination shown as a black solid line.
The uncertainty band on the reconstructed static energy is too small to be seen in the plot. Right panel: results for the strong coupling from N$^3$LO fits with a single offset to the full (red) and combined (blue) datasets.}
\end{figure}

We will compare to lattice QCD data from the HotQCD Collaboration published in Refs.~\cite{Bazavov:2014pvz,Bazavov:2016uvm,Bazavov:2017dsy}. The
data is obtained from a (2+1)-flavor QCD simulation using the highly-improved staggered quark (HISQ) action with tree-level improvement. The strange quark
mass $m_s$ is set
close to its physical value, while that of the two lightest quarks \mbox{$m_\ell\equiv m_u=m_d$} is fixed to $m_s/20$ or $m_s/5$, which gives a pion mass of
about $160$\,MeV or $320$\,MeV in the continuum limit~\cite{Bazavov:2014pvz}.
We use nine different sets with distinct values for the bare lattice gauge coupling $\beta=\{6.8, 7.03, 7.28, 7.373, 7.596, 7.825, 8.0, 8.2, 8.4\}$, including three additional ensembles with smaller $\beta$ compared to those used in Ref.~\cite{Bazavov:2019qoo}. These correspond to the lattice spacing $a=\{0.104, 0.084, 0.067, 0.061, 0.05, 0.041, 0.035, 0.03, 0.025\}\,$fm. In Fig.~\ref{fig:Cluster} lattice data are displayed in various colors to distinguish the different ensembles. Even though the compatibility among different sets is ensured, a small difference at short distances may be present, though, due to discretization lattice artifacts.
The static energy calculated on the lattice has a linearly divergent part that needs to be removed by additive renormalization, as studied in Ref.~\cite{Bazavov:2016uvm}.
Datasets with distinct values of $\beta$ are uncorrelated, but data points with different $r$ and a common $\beta$ are correlated.

As can be inferred by reading the various articles mentioned earlier in this section, different datasets have origins of the static energy compatible within lattice uncertainties, and accordingly one could consider there is a common offset with respect to the pQCD prediction. However, their central values are not identical and, given the high accuracy of our theoretical prediction, this can diminish the fit's quality. Therefore, in one of our approaches we consider different lattice ensembles have independent origins of energy and regard the various offsets as individual fit parameters. Since offsets are statistically compatible, in our second approach we perform fits assuming all offsets are the same (hence fitting for $\alpha_s$ and a unique offset). Both approaches shall also be followed in Sec.~\ref{sec:cluster}.

In order to convert from lattice units (given as multiples of the distance $r_1$) to physical units, previous studies have used $r_1=0.3106(17)$ fm obtained from $f_\pi$ as given in Ref~\cite{Bazavov:2014pvz}.
In Ref.~\cite{Leino:2025pvl} this value was updated to $r_1=0.3037(25)$fm as determined from $a_{f_{p4s}}$. The new result is $2.2\%$ lower (more than 2-$\sigma$ away, hence statistically only marginally compatible) and slightly less precise ($0.82\%$ vs $0.55\%$). This uncertainty can be interpreted as a redefinition of $\Lambda_{\rm{QCD}}$ or, even simpler, of the energy at which the strong coupling takes its boundary value that, in our case, is the $\tau$-lepton mass. A $1\%$ positive shift in $r_1$ translates into a $0.5\%$ increase for $\alpha_s^{(n_\ell=3)}(m_\tau)$, hence four times smaller than our total uncertainty. When matching at the charm and bottom thresholds, and after evolving the strong coupling to the $Z$-pole, we find that this shift makes $\alpha_s^{(n_f=5)}(m_Z)$ only $0.2\%$ higher.

\subsection{Lattice QCD Data Clustering}\label{sec:cluster}
The complete lattice dataset, which includes distances up to $r=1\,$fm, contains 2682 points. The dataset's size dependence with the maximal distance considered for various selections of lattice ensembles can be seen in Fig.~\ref{fig:DOF}. This makes computing and minimizing the $\chi^2$ functions computationally very expensive, what is particularly inefficient for studies carried out while gearing up for the final fits. Therefore, with the purpose of reducing the large number of
lattice data points on the static energy, we have implemented a method to combine all the lattice QCD ensembles into a single one. This procedure permits
working with only one dataset and, at the same time, extract the correlation and offsets between different ensembles. It is based on previous algorithms employed to combine data on the so called $e^+e^-\to\,$hadrons R-ratio, which can be used to determine the vacuum polarization contribution to the $g-2$ of the muon using experimental data~\cite{Hagiwara:2003da}. It has also been used to obtain moments of the charm correlator in the context of QCD sum rules~\cite{Dehnadi:2011gc,Dehnadi:2015fra}.

The combination of $N_s$ different lattice QCD ensembles does not consider any
assumption (in particular, it is not biased by theoretical expectations such as pQCD) other than the statistical compatibility of all the static energy data among different ensembles labeled by $1\leq k\leq N_s$, modulo $N_s$ global shifts that account for the arbitrariness of the potential energy's origin, which differs among datasets. We consider a distance range $r_{\rm min}\leq r\leq r_{\rm max}$ divided in $N$ bins which in general are not
equally long, whose $N+1$ borders are denoted by $r_i$ with $0\leq i\leq N$ that satisfy $r_i>r_j$ if $i>j$. The procedure yields a new value for the distance and static energy for each bin, along with a unique covariance matrix which correlates all the static energy predictions. The new value for $r$ is given by the weighted average of all the lattice data for the distance within that bin:
\begin{align}
\tilde r_i = \frac{\sum_{k=1}^{N_s}\sum_{j=1}^{n^k_i}\frac{r^k_{ij}}{(\sigma^k_{ij})^2}}{\sum_{k=1}^K\sum_{j=1}^{n^k_i}\frac{1}{(\sigma^k_{ij})^2}}\,,
\end{align}
satisfying $r_{i-1}\leq \tilde r_i\leq r_{i}$, where $i$ denotes the $i$-th bin, $r^k_{ij}$ is distance of a lattice data point in the $k$-th ensemble and $i$-th bin, i.e., satisfying $r_{i-1}\leq r^k_{ij}<r_{i}$. We label the distances such that $r^k_{ij}>r^k_{i\ell}$ if $j>\ell$, and of course $r^k_{ij}>r^n_{m\ell}$ if $i>m$ for any $k,n,j,\ell$, what ensures that $\tilde r_i>\tilde r_j$ for $i>j$.
Finally, $n_i^k$ is the number of lattice QCD data points from the $k$-th ensemble satisfying $r_{i-1}\leq r< r_{i}$, which defines the total number of points in the bin $n_i=\sum_{k=1}^{N_s} n^k_i$,
and $\sigma^k_{ij}$ is the static energy prediction's uncertainty at $r^k_{ij}$. This choice is not unique and other possibilities such as the regular mean
$\hat r_i=\sum_{k=1}^{N_s}\sum_{j=1}^{n^k_i}r^k_{ij}/n_i$ or the bin center $\bar r_i=(r_{i-1}+ r_i)/2$ are also possible, but we consider the weighted average less biased. In any case, upon the fitting procedure that shall be described next, the specific choice's impact should be minimized such that all predictions for the static energy are very similar. We include only statistically uncertainties, as these are the only ones currently available. Furthermore, we do not consider statistical correlations among lattice data on the same ensemble since this information is not public.

The value of the static energy at each distance $\tilde r_i$ is calculated minimizing the following $\chi^2$ function:
\begin{align}\label{eq:Chi2Clus}
\chi_n^2(\{\tilde V_i\},\{A_k\}) = \sum_{i=1}^{N}\sum_{k=1}^K\sum_{j=1}^{n_i^k} \Biggl[\frac{V^k_{ij}-V^n_s(\{\tilde V_i\},r^k_{ij}) + A_k}{\sigma^k_{ij}}\Biggr]^2\,,
\end{align}
with respect to $A_k$ and $\tilde V_i$, where $V^k_{ij}$ is the static-energy lattice prediction at $r^k_{ij}$, so that $E_s(r^k_{ij})=V^k_{ij}\pm \sigma^k_{ij}$,
$\tilde V_i$ is the recombined potential at $\tilde r_i$, that is $E_s(\tilde r_i)=\tilde V_i\pm \sigma_i$, where $\sigma_i$ is the fit uncertainty of $\tilde V_i$ to be discussed next, and $A_k$ is the $k$-th ensemble's offset. Finally, $V^n_s(\{\tilde V_i\},r)$ is an interpolation function that goes through the $N$ points $\{\tilde r_i,\tilde V_i\}$ with \mbox{order-$n$} splines. For $n=0$ one has $V^0_s(\{\tilde V_i\},r^k_{ij}) =\tilde V_i$, which does not induce correlations among bins, while higher values of $n$ imply smoother functional forms that can better adapt to the strongly varying static energy at small radii, but imply stronger correlations among bins. We have investigated various values of $n$ and concluded that $n=3$ is a nice compromise between a flexible functional form and a moderate bin-by-bin correlation, but
using larger values leads to almost identical results. The interpolations are computed using the \texttt{Python} function \texttt{interp1d} included in \texttt{scipy.interpolate}. Even for 0-th order interpolations, the fit parameters $A_k$ will yield correlations among the various bins. Regardless of the interpolation order, the $\chi^2$ function in Eq.~\eqref{eq:Chi2Clus} is quadratic in all its fit parameters.

The unknown (fit) parameters are the $N$ static energies $\tilde V_i$ and the $N_s-1$ offsets $A_k$, since without loss of generality we set $A_1=0$, that is, we `force' the predictions of all simulations to agree with the first ensemble. This is mandatory as the fit cannot choose a particular value for the overall shift which is anyway arbitrary. This leaves a global offset (which can be identified with $A_1$) with respect to the pQCD prediction that will be fit along with the strong coupling. Performing a minimization of
$\chi^2_n$, function, which can be done analytically using linear algebra and involves only a matrix inversion, we obtain a list of $\big\{\tilde r_i,\tilde V_i\big\}$ combining all the lattice QCD ensembles.
The covariance matrix accounting for the uncertainties and correlations of the $\tilde V_i$ is obtained inverting the hessian and discarding the rows and columns corresponding to $A_k$, which play no role. Given the Gaussian nature of $\chi_n^2$, this procedure is exact and tantamount to looking into $\Delta \chi^2=1$ contours. Since we find $\mathcal{O}(10)$ reduced minimal $\chi^2$ values, fit uncertainties are rescaled by $\sqrt{\chi^2_{\rm min}/{\rm d.o.f.}}$ to be conservative. It is through the hessian's inversion that correlations among
$\tilde V_i$ are generated even for flat splines.

We also consider the clustering procedure under the assumption that all lattice ensembles are defined with respect to a common energy offset. In this case we simply set $A_k=0$ for all $1\leq k\leq K$ and minimize $\chi_n^2(\{\tilde V_i\},\{0\})$ with respect to $\tilde V_i$ for $1\leq i\leq N$. This yields weaker correlations among the $\tilde V_i$ since there is no common floating parameter that affects them. On the other hand, the minimal value of $\chi^2_n$ will be higher.

In order to have a proper compatibility between all the ensembles at low distances ($r<0.2$ fm), it is better to select a small bin size for that region, while a coarser
spacing can be used for medium and larger distances, where lattice QCD ensembles are more widely spread and uncertainties are larger. We will consider a bin size of $d_1=0.0083\,$fm for $r \le 0.25\,$fm and $d_2=0.0107\,$fm for $r>0.25\,$fm, but have checked that other choices yield very similar outcomes. The result of this recombination is shown as a solid black line in Fig.~\ref{fig:Cluster} with uncertainties visible as a gray band.

\section{Fitting Procedure}\label{sec:procedure}
In this section we lay out our method to perform strong-coupling fits for either one or several offsets since in our numerical studies we will consider both kind of analyses and compare their outcome. Even though one could in principle numerically minimize the $\chi^2$ function with respect to all fit parameters, since the dependence on $A_k$ is quadratic,
we analytically marginalize with respect to the offsets obtaining a one-parameter function $\tilde\chi^2(\alpha_s)$. The resulting marginalized function is then numerically minimized. Details on the procedure shall be given below.

\subsection
{Fits with a Single Offset}
We start discussing a scenario that will take place either if there is a single lattice ensemble to be fit, if we assume a common offset for all lattice ensembles, or if the recombined dataset is used. To be as general as possible, we consider
non-trivial correlations among data. For a given value of $\alpha_s^{(n_\ell=3)}(m_\tau)$
the pQCD theoretical prediction for the static energy $E_s$ can be written as the sum of an $r$-dependent piece and a constant term \mbox{$E_s(\alpha_s,r_i)=V_i(\alpha_s)+A$}, where $A$ is the unphysical offset common to all $r_i$ that depends on the
lattice simulation. Thus, the $\chi^2$ function is

\begin{equation}\label{eq:chi2}
\chi^2(\alpha_s,A) = \sum_{i,j=1}^{n_{\rm data}} \big[V_i(\alpha_s)+A-V_i^{\rm exp}\big][\sigma^{-1}]_{ij}\big[V_j(\alpha_s)+A-V_j^{\rm exp}\big] \,,
\end{equation}
where $\sigma^{-1}$ is the inverse covariance matrix,
$V_i^{\rm exp}$ is the corresponding lattice (or recombination) result, and $n_{\rm data}$ is the dataset's size. In order to obtain the best-fit $\alpha_s$, we also have to minimize with respect to $A$. Since the offset $A$ is common to all radii,
one can expand the $\chi^2$ function as a second-order polynomial in $A$.
Exploiting the quadratic dependence on the offset, which guaranties the minimum of $\chi^2$ for any fixed value of $\alpha_s$ is unique, we can analytically minimize the $\chi^2$ function with respect to $A$ for a given value of the strong coupling
\begin{align}\label{eq:marginA}
\tilde A(\alpha_s) = \frac{\sum_{i,j=1}^{n_{\rm data}}[\sigma^{-1}]_{ij}\big[V_j^{\rm exp} - V_j(\alpha_s)\big]}{\sum_{i,j=1}^{n_{\rm data}}[\sigma^{-1}]_{ij}}\,,
\end{align}
where we have used $[\sigma^{-1}]_{ij}=[\sigma^{-1}]_{ji}$. Of course, if $V_j(\alpha_s)=V_j^{\rm exp}$ for all $1\leq j\leq n_{\rm data}$, one obtains $\tilde A(\alpha_s)=0$. As shall be justified later, this expression can be interpreted as the weighted average of the displacements taking into account lattice correlations. With this method, we define the so-called \emph{marginalized} $\chi^2$-function, denoted as $\tilde \chi^2(\alpha_s)$, by setting the offset to its $\alpha_s$-dependent best-fit value. The strong coupling can be determined numerically minimizing $\tilde \chi^2(\alpha_s)$, which is given by
\begin{align}\label{eq:marginchi2}
\tilde \chi^2(\alpha_s) \equiv \chi^2(\alpha_s,\tilde A(\alpha_s)) = \chi^2(\alpha_s,0)
- \frac{\Bigl\{\sum_{i,j=1}^{n_{\rm data}}[\sigma^{-1}]_{ij}\big[V_j(\alpha_s)-V_j^{\rm exp}\big]\Bigr\}^{\!2}}{\sum_{i,j=1}^{n_{\rm data}}[\sigma^{-1}]_{ij}}.\!\!
\end{align}
The two subtracted terms are definite positive, but given that $\tilde \chi^2$ stems from a particular choice of $A$ in the $\chi^2$ function defined in Eq.~\eqref{eq:chi2}, which is definite positive for any value of its two arguments, we trivially conclude that the first term in Eq.~\eqref{eq:marginchi2} is larger than the second.
Hence, $\chi^2(\alpha_s,0)> \tilde \chi^2(\alpha_s)$, indicating that having a second fit parameter necessarily decreases $\chi^2_{\rm min}$. The expression above is invariant under $V_j(\alpha_s)\to V_j(\alpha_s)+b(\alpha_s)$ for \mbox{$r$-independent} $b(\alpha_s)$.
If we assume there is no correlation between different $V_i^{\rm exp}$, the inverse covariance matrix takes the form $[\sigma^{-1}]_{ij}\equiv\delta_{ij}/\sigma_i^2$, and $\tilde \chi^2(\alpha_s)$ simplifies to
\begin{align}\label{eq:uncor}
\tilde \chi^2(\alpha_s) =& \sum_{i=1}^{n_{\rm data}} \frac{\big[V_i(\alpha_s)-V_i^{\rm exp}\big]^2}{\sigma_i^2}-\frac{\left[\sum_{i=1}^{n_{\rm data}} \frac{V_i(\alpha_s)-V_i^{\rm exp}}{\sigma_i^2}\right]^2}{\sum_{i=1}^{n_{\rm data}} \frac{1}{\sigma_i^2}}\\
\equiv{} &\Bigl\{\bigl\langle [V^{\rm exp} - V(\alpha_s)]^2\bigr\rangle-\bigl[\langle V^{\rm exp} - V(\alpha_s)\rangle\bigr]^2\Bigr\} \sum_{i=1}^{n_{\rm data}} \frac{1}{\sigma_i^2}\,,\nonumber\\
\tilde A(\alpha_s) ={} &\frac{\sum_{i=1}^{n_{\rm data}}\frac{V_i^{\rm exp}-V_i(\alpha_s)}{\sigma_i^2}}{\sum_{i=1}^{n_{\rm data}} \frac{1}{\sigma_i^2}}\equiv\langle V^{\rm exp} - V(\alpha_s)\rangle\,,\nonumber
\end{align}
where we have introduced the notation $\langle B\rangle$ to denote $B$'s weighted average. Now $\tilde A(\alpha_s)$ corresponds to the weighted average of the difference between theory and experiment. Incidentally, the middle line makes clear that $\tilde \chi^2(\alpha_s)$ is positive definite.

By minimizing the marginalized $\tilde \chi^2(\alpha_s)$ we obtain the best-fit value for the strong coupling $\alpha_s^{\rm BF}$ that verifies $\tilde \chi^2(\alpha_s^{\rm BF})=\chi^2(\alpha_s^{\rm BF},A^{\rm BF})=\chi^2_{\rm min}$, and for the offset \mbox{$A^{\rm BF}=\tilde A(\alpha_s^{\rm BF})$} at the same time. We also obtain the strong coupling 1-$\sigma$ fit uncertainty $\Delta_{\rm fit}\alpha_s$ requiring \mbox{$\tilde\chi^2(\alpha_s^{\rm BF}\pm\Delta_{\rm fit}\alpha_s)=\chi^2_{\rm min}+1$}.\footnote{One can also obtain the strong-coupling fit uncertainty assuming a Gaussian approximation with the hessian matrix
$\Delta_{\rm fit}\alpha_s=\sqrt{2/\Bigl[{\rm d}^2\tilde\chi^2(\alpha_s)/{\rm d}\alpha_s^2\Bigr]_{\alpha_s^{\rm BF}}}$.
We have checked this yields slightly smaller uncertainties, and hence stick to the more conservative procedure outlined in the main text.}
In the Gaussian approximation one gets the $n$-$\sigma$ interval by the condition $\tilde\chi^2(\alpha_s^{\rm BF}\pm n\Delta_{\rm fit}\alpha_s)=\chi^2_{\rm min}+n^2$ corresponding to a confidence level \mbox{$p={\rm erf}(n/\sqrt{2})$}, expression involving the error function.
Moreover, the offset uncertainty due to the incertitude on $\alpha_s$ is obtained as
\begin{equation}\label{eq:difference}
\Delta_{\alpha_s} A=\frac{1}{2}\bigl[\tilde A(\alpha_s^{\rm BF}+\Delta_{\rm fit}\alpha_s) - \tilde A(\alpha_s^{\rm BF}-\Delta_{\rm fit}\alpha_s)\bigr]\,.
\end{equation}
It is also interesting to define the one-sided minimized $\chi^2$ function as $\hat\chi^2(\alpha_s)=\chi^2(\alpha_s,A^{\rm BF})$ that has its minimum at $\alpha_s=\alpha_s^{\rm BF}$, satisfies $\hat\chi^2(\alpha_s^{\rm BF})=\chi^2_{\rm min}$ and can be used to decompose the fit uncertainty $(\Delta_{\rm fit}\alpha_s)^2=(\Delta_{\rm lat}\alpha_s)^2+(\Delta_A\alpha_s)^2$ into a pure lattice error $\Delta_{\rm lat}\alpha_s$ and an incertitude caused
by $\Delta_{\rm fit}A$, which will be denoted $\Delta_A\alpha_s$. The former is obtained by the condition \mbox{$\hat\chi^2(\alpha_s^{\rm BF}\pm\Delta_{\rm lat}\alpha_s)=\chi^2_{\rm min}+1$} and the latter from the error decomposition.
Fixing the strong coupling to $\alpha_s^{\rm BF}$ defines the $A$-dependent $\chi^2$ function \mbox{$\bar\chi^2(A)=\chi^2(\alpha_s^{\rm BF},A)$} ---\,attaining its minimum at $\bar\chi^2(A^{\rm BF}) = \chi^2_{\rm min}$\,---
that can be used to obtain the pure lattice uncertainty on $A$ by the condition $\bar\chi^2(A^{\rm BF}\pm\Delta_{\rm lat}A)=\chi^2_{\rm min}+1$. The total fit incertitude is obtained as $(\Delta_{\rm fit}A)^2=(\Delta_{\rm lat}A)^2+(\Delta_{\alpha_s}A)^2$. The inequality $\Delta_{\rm fit}\alpha_s >\Delta_{\rm lat}\alpha_s$ implies $A^{\rm BF}$ and $\alpha_s^{\rm BF}$ have a correlation that is obtained as
\begin{equation}\label{eq:corr}
|\rho|=\frac{\Delta_A\alpha_s}{\Delta_{\rm fit}\alpha_s} = \frac{\Delta_{\alpha_s}A}{ \Delta_{\rm fit}A}\,,
\end{equation}
where the last equality follows from the uniqueness of $\rho$. The sign of $\rho$ is opposite to that of the slope of $\tilde A(\alpha_s)$ at $\alpha_s=\alpha_s^{\rm BF}$.
To justify these relations we expand $\chi^2$ near the minimum up to quadratic terms:

\begin{align}\label{eq:quadratic}
\!\!\!\! \chi^2(\alpha_s,A) \simeq \chi^2_{\rm min} + \!\frac{1}{1-\rho^2}\!\biggl[\!\biggl(\frac{\alpha_s-\alpha_s^{\rm BF}}{\Delta_{\rm fit}\alpha_s}\biggr)^{\!\!2}
\!+\biggl(\frac{A-A^{\rm BF}}{\Delta_{\rm fit}A}\biggr)^{\!\!2}\!
+\frac{2\rho(\alpha_s-\alpha_s^{\rm BF})(A-A^{\rm BF})}{(\Delta_{\rm fit}\alpha_s)(\Delta_{\rm fit}A)}\biggr].\!
\end{align}
One trivially obtains $\Delta_{\rm latt}\alpha_s = \sqrt{1-\rho^2}\Delta_{\rm fit}\alpha_s$ and $\Delta_{\rm latt}A = \sqrt{1-\rho^2}\Delta_{\rm fit}A$,
what justifies our method to obtain these uncertainties
based on $\hat \chi^2(\alpha_s)$ and $\bar\chi^2(A)$, respectively, since near the minimum
\begin{equation}\label{eq:nearMin}
\hat \chi^2(\alpha_s) \simeq \chi^2_{\rm min} + \frac{1}{1-\rho^2}\biggl(\frac{\alpha_s-\alpha_s^{\rm BF}}{\Delta_{\rm fit}\alpha_s}\biggr)^{\!\!2}\,,
\qquad
\bar \chi^2(A) \simeq \chi^2_{\rm min} +\frac{1}{1-\rho^2}\biggl(\frac{A-A^{\rm BF}}{\Delta_{\rm fit}A}\biggr)^{\!\!2}\,.
\end{equation}
Since trivially $\Delta_A\alpha_s = |\rho|\Delta_{\rm fit}\alpha_s$ and $\Delta_{\alpha_s} A = |\rho|\Delta_{\rm fit}A$, this justifies both equalities in Eq.~\eqref{eq:corr}. Marginalizing Eq.~\eqref{eq:quadratic} we obtain
\begin{equation}\label{eq:correlationAalpha}
\tilde A(\alpha_s) = A^{\rm BF}-\rho\frac{\Delta_{\rm fit}A}{\Delta_{\rm fit}\alpha_s}\big(\alpha_s-\alpha_s^{\rm BF}\big)\,,\qquad
\tilde \chi^2(\alpha_s) = \chi^2_{\rm min} + \biggl(\frac{\alpha_s-\alpha_s^{\rm BF}}{\Delta_{\rm fit}\alpha_s}\biggr)^{\!\!2}\,,
\end{equation}
relations that
justify our numerical methods to obtain $\Delta_{\rm latt}\alpha_s$ and $\Delta_{\alpha_s}A$, c.f.\ Eq.~\eqref{eq:difference} and the relation three lines above it, as well as our procedure to figure out the sign of $\rho$. The two-dimensional ellipse representing a confidence level $p$ is given by
\begin{equation}\label{eq:confidence}
\chi^2(\alpha_s,A) = \chi^2_{\rm min} -2\log(1-p)\,.
\end{equation}
This confidence level can be interpreted as an $n$-$\sigma$ interval with the relation $n=\sqrt{2}\, \text{erf}^{\,-1}(p)$, involving the error function's inverse.

\subsection[$\chi^2$ with Multiple Offsets]{\boldmath $\chi^2$ with multiple offsets}\label{sec:Multiple}
Let us assume now we have $N_s$ lattice sets labeled by the index $1\leq k\leq N_s$, each one with a different offset $A_k$. In general, the offsets for the different ensembles are arbitrary and uncorrelated. The parameters we need to minimize are the various $A_k$ and the global $\alpha_s$. We will assume there are no correlations among different sets, therefore will consider $N_s$ covariance matrices denoted by $\sigma_k$.\footnote{The more general case of dataset-to-dataset correlation can be easily worked out but we focus on the case of interest for our analysis.} In this case, the $\chi^2$-function can be expressed as the sum of $N_s$ individual functions each looking like that defined in Eq.~\eqref{eq:chi2}:
\begin{align}\label{eq:ChiSum}
\chi^2(\alpha_s,\{A_k\}) =& \sum_{k=1}^{N_s} \chi^2_k(\alpha_s,A_k)\,,\\
\chi^2_k(\alpha_s,A_k) =& \sum_{i,j=1}^{n^k_{\rm data}} \big[V_{ik}(\alpha_s)+A_k-V_{ik}^{\rm exp}\big][\sigma_k^{-1}]_{ij}\big[V_{jk}(\alpha_s)+A_k-V_{jk}^{\rm exp}\big] \,,\nonumber
\end{align}
where $V_{ik}^{\rm exp}$ is the $k$-th lattice set prediction for the static energy at the $i$-th distance $r=r_i^k$ with $r_i^k>r_j^k$ if $i>j$, $[\sigma_k^{-1}]_{ij}$ is the
matrix element obtained after inverting $\sigma_k$, and $n_{\rm data}^k$ is the number of data points belonging to the $k$-th set such that $n_{\rm data}=\sum_{k=1}^{N_s}n_{\rm data}^k$. The theoretical prediction is given by $E^{\rm pQCD}_s(r^k_i)=V_{ik}(\alpha_s)$. This functional form ensures the hessian will have zeros in the entries related to different offsets. However, upon inverting this matrix, correlations among the various offsets will be generated, which is just to be expected as they are determined along with $\alpha_s$, creating these correlations.
Given the form of Eq.~\eqref{eq:ChiSum}, one can marginalize the $\chi^2$ function by doing so for each of its terms:
\begin{align}\label{eq:multiple}
\tilde \chi^2(\alpha_s) ={}& \!\sum_{k=1}^{N_s} \tilde \chi^2_k(\alpha_s)\,,\\
\tilde A_k(\alpha_s) ={} & \frac{\sum_{i,j=1}^{n^k_{\rm data}}[\sigma_k^{-1}]_{ij}\big[V_{jk}^{\rm exp} - V_{jk}(\alpha_s)\big]}{\sum_{i,j=1}^{n^k_{\rm data}}[\sigma_k^{-1}]_{ij}}\,,\nonumber\\[0.1cm]
\tilde \chi^2_k(\alpha_s) ={}& \chi^2_k(\alpha_s,\tilde A_k(\alpha_s))
- \frac{\Bigl\{\sum_{i,j=1}^{n^k_{\rm data}}[\sigma_k^{-1}]_{ij}\big[V_{jk}(\alpha_s)-V_{jk}^{\rm exp}\big]\Bigr\}^{\!2}}{\sum_{i,j=1}^{n^k_{\rm data}}[\sigma_k^{-1}]_{ij}}. \nonumber
\end{align}
From this point on, we proceed as in the previous section, namely, numerically minimizing $\tilde \chi^2(\alpha_s)$ with respect to $\alpha_s$ in order to obtain $\alpha_s^{\rm BF}$,
which is then introduced in $\tilde A_k$ to get $A_k^{\rm BF}$.
To determine the fit uncertainty we again solve $\tilde\chi^2(\alpha_s^{\rm BF}+\Delta_{\rm fit}\alpha_s)=\chi^2_{\rm min}+1$, while the pure lattice uncertainty is obtained solving an analogous equation for
$\hat\chi^2(\alpha_s)= \chi^2(\alpha_s,\{A_i^{\rm BF}\})$. From the quadratic difference, one gets the uncertainty that globally accounts for the multiple offsets' incertitudes. When dealing with various datasets we will not be concerned with offset uncertainties and correlation coefficients, hence we do not discuss how these can be obtained. It is instructive to inspect the $\chi^2$ function around the minimum in terms of the hessian $M$, which is symmetric and positive definite
\begin{align}\label{eq:MultipleA}
\chi^2(\alpha_s,A_1,\cdots, A_{N_s})\simeq{}& \chi^2_{\rm min} + M_{11}(\alpha_s-\alpha^{\rm BF}_s)^2+\sum_{k=1}^{N_s}(A_k-A_k^{\rm BF})^2M_{k+1,k+1}\\[-0.3cm]
&+2(\alpha_s-\alpha^{\rm BF}_s)\sum_{k=1}^{N_s} (A_k-A_k^{\rm BF})M_{1,k+1}\,,\nonumber\\
\tilde A_k(\alpha_s) \simeq{}& A_k^{\rm BF}-\big(\alpha_s-\alpha_s^{\rm BF}\big)\frac{M_{1,k+1}}{M_{k+1,k+1}}\,,\nonumber\\
\tilde \chi^2(\alpha_s) \simeq{}& \chi^2_{\rm min} + (\alpha_s-\alpha^{\rm BF}_s)^2\Biggr(M_{11}-\sum_{k=1}^{N_s}\frac{M_{1,k+1}^2}{M_{k+1,k+1}}\Biggr).\nonumber
\end{align}
Once more, one trivially obtains $M_{11}=[\Delta_{\rm latl}\alpha_s]^{-2}$ justifying our numerical procedure to obtain this uncertainty since near the minimum $\hat \chi^2(\alpha_s) \simeq \chi^2_{\rm min} + (\alpha_s-\alpha_s^{\rm BF})^2M_{11}$. Given the large amount of zeros in the hessian, we can compute the fit error in a generic way:
\begin{equation}
\Delta_{\rm fit}\alpha_s = \sqrt{\frac{\prod_{i=2}^{N_s+1}M_{ii}}{{\rm det}(M)}}\,.
\end{equation}
While not evident at first sight, by inverting and squaring the previous relation one obtains the coefficient of $(\alpha_s-\alpha_s^{\rm BF})^2$ in the last line of Eq.~\eqref{eq:MultipleA}.
With this we finish the justification of our numerical procedure.

\begin{table}[t!]
\centering
\begin{tabular}{ c ccccc } \toprule $N_{1/2}$ & MSRn & MSRp & Force & Potential & PS \\\midrule
$\lambda=1$ & $0.5371$ & $0.5373$ & $0.5509$ & $0.5440$ & $0.5721$ \\[0.1cm]
$\lambda=2^{\pm 1}$ & $0.526\pm 0.012$ & $0.522 \pm 0.017$ & $0.540 \pm0.012$ & $0.489\pm 0.062$ & $0.559\pm 0.043$\\\bottomrule
\end{tabular}
\caption{Estimate of $N_{1/2}$ using the $\mathcal{O}(\Lambda_{\rm QCD})$ renormalon sum rule on different subtraction schemes for $\lambda=1$ (middle row) and $\lambda=2^{\pm1}$ (bottom row). \label{tab:N12}}
\end{table}

\section{Study of renormalon subtractions}\label{sec:RSlam}
\begin{figure}[t!]
\centering
\begin{subfigure}{0.488\textwidth}
\centering
\includegraphics[width=\textwidth]{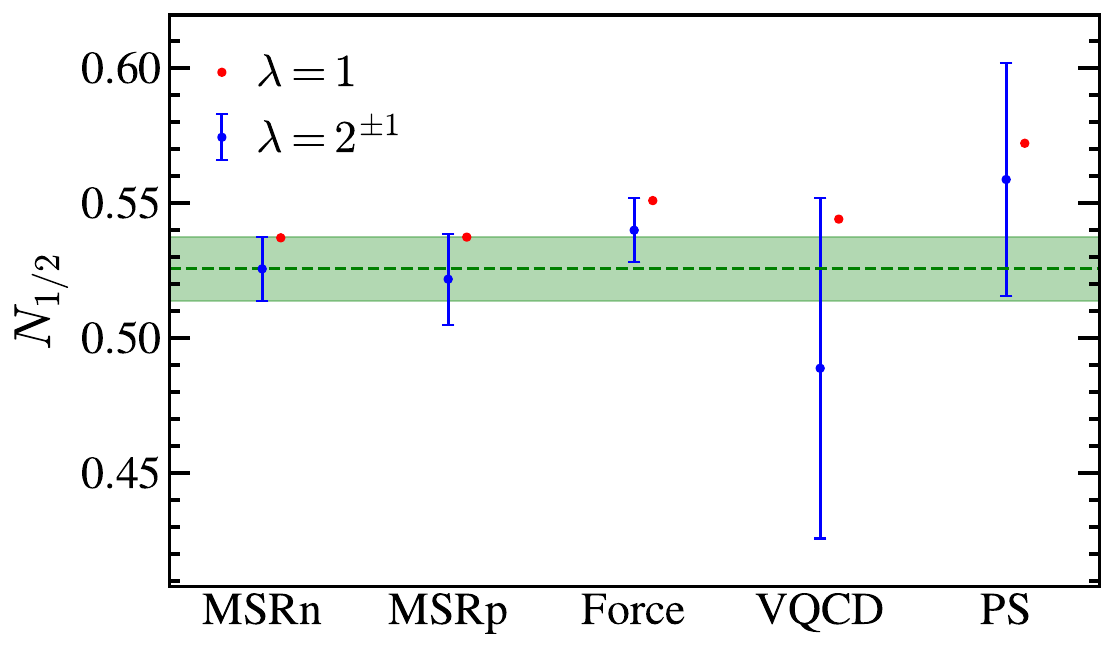}
\caption{}\label{fig:N12lambda}
\end{subfigure}
\hfill
\begin{subfigure}{0.475\textwidth}
\centering
\includegraphics[width=\textwidth]{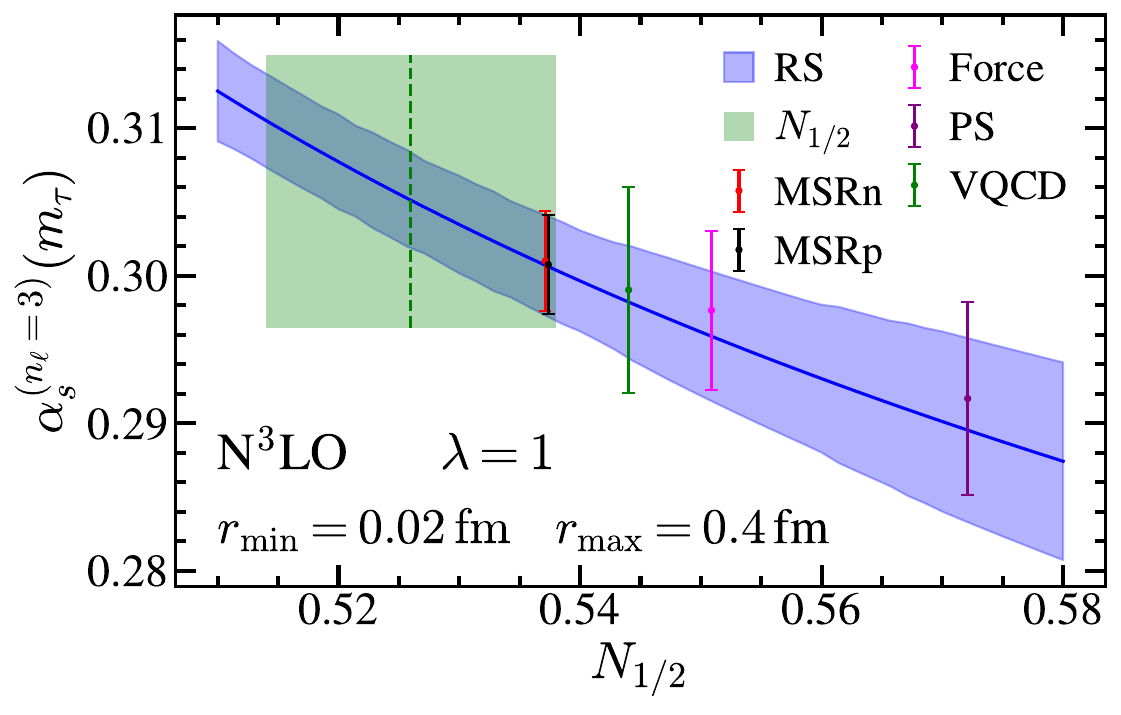}\vspace*{-0.25cm}
\caption{}\label{fig:RS}
\end{subfigure}
\caption{\label{fig:Renormalon}
Left panel: Estimate of $N_{1/2}$ using the $\mathcal{O}(\Lambda_{\rm QCD})$ renormalon sum rule on various renormalon subtraction schemes setting $\lambda=1$ (red dots) and $\lambda=[0.5,2]$ (blue error bars). Right panel: strong-coupling fits in the RS scheme as a function of $N_{1/2}$ (blue band) compared to various renormalization schemes with $\lambda=1$ (colored dots), where the error bars have been generated by profile-parameter variation. The green translucent band represents our best estimate for $N_{1/2}$, including its uncertainty, from the MSRn scheme.}
\end{figure}

An important aspect one needs to take care off before starting high-precision strong-coupling fits is the renormalon subtraction scheme choice. Several such schemes have been presented in Sec.~\ref{sec:renormalon}, and we have anticipated that some of them are not adequate due to their sensitivity to ultrasoft physics. In this section we make this statement more quantitative by studying the impact of this choice on the extraction of $\alpha_s$, and draw important conclusions that can be applied to other scenarios. We start with the basic assumption that the RS mass, provided a good estimate for $N_{1/2}$ is used, perfectly subtracts the static potential's renormalon and has no sensitivity to low scales since it reproduces the high-order asymptotic behavior and nothing else. Estimates for $N_{1/2}$ using the $\mathcal{O}(\Lambda_{\rm QCD})$ renormalon sum rule on the various schemes considered in our work are collected in Table~\ref{tab:N12} and Fig.~\ref{fig:N12lambda} where one can observe that, even if compatible within uncertainties, error estimates strongly depend on the scheme while central values can vary by a few sigma. Having the $\lambda=1$ prediction very close to the upper uncertainty limit is observed for all schemes except for the PS mass. These $\lambda=1$ results can also differ by a few sigma from our best estimate. The last important observation is that the strong coupling's best-fit value within a given renormalon-subtraction scheme is strongly correlated with the value of $N_{1/2}$ predicted by that scheme, as is clearly seen in Fig.~\ref{fig:RS}. Here we compare N$^3$LO fits in the RS scheme varying $N_{1/2}$ between 0.51 and 0.58 with results obtained from the schemes studied in this article using $\lambda=1$. The latter are placed in the plot according to their $N_{1/2}(\lambda=1)$ prediction (red dots of Fig.~\ref{fig:N12lambda}). The correlation appears evident and suggests that schemes yielding an inaccurate prediction for $N_{1/2}$ should be avoided for high-precision analyses since otherwise the central value can be biased and the uncertainty inflated. We also believe the RS scheme should be discarded since it forces including an uncertainty accounting for that of $N_{1/2}$ which amounts to $\Delta_{N_{1/2}}\alpha_s=0.0052$, affecting the accuracy $\alpha_s$ can be extracted with, as can be seen by the green translucent square in Fig.~\ref{fig:RS}. Moreover, we have observed that in the RS scheme outliers are generated for small values of $N_{1/2}$ ---\,which have been removed to generate the band in Fig.~\ref{fig:RS}\,--- even if using the profiles discussed in Sec.~\ref{sec:profiles}. We believe the MSRn scheme (from now on referred simply as MSR) is the optimal choice since a)~it does not necessitate an explicit prediction for $N_{1/2}$ that can affect $\alpha_s$ uncertainties, and b)~optimally predicts $N_{1/2}$ as long as $\lambda$ is varied in the range advocated for at the end of Sec.~\ref{sec:renormalon}. This is the scheme and setup that shall be used in the rest of this article.

\section{Fits for the Strong Coupling}\label{sec:fit}
After having set up the theoretical formalism to predict the static energy within pQCD, we are ready to determine the strong coupling by fitting the results obtained in lattice QCD simulations. With our improved \texttt{REvolver} code, we directly fit for $\Lambda_{\rm{QCD}}^{(n_\ell=3)}$, and later translate this value to the strong coupling. However, for the sake of clarity, in this section we will assume the actual fit parameter is the strong coupling at a given reference scale. Specifically, in order to ease comparisons with determinations coming from hadronic tau decays, we will take $\alpha_s^{(n_\ell=3)}(m_\tau)$ as our fit parameter, which will be referred to as simply $\alpha_s$ in this section. As a sanity check, MSR-scheme fits are performed directly with our \texttt{C++} code and also independently via the \texttt{Python} code using interpolations. Results agree within 7 decimal places. All numerical explorations discussed in what follows, along with our final results, use the full lattice dataset but, as seen in Fig.~\ref{fig:ClusterFit}, fitting the recombined dataset yields an almost identical result and is hundreds of times faster.
\begin{figure}[t!]
\centering
\begin{subfigure}{0.499\textwidth}
\centering
\includegraphics[width=\textwidth]{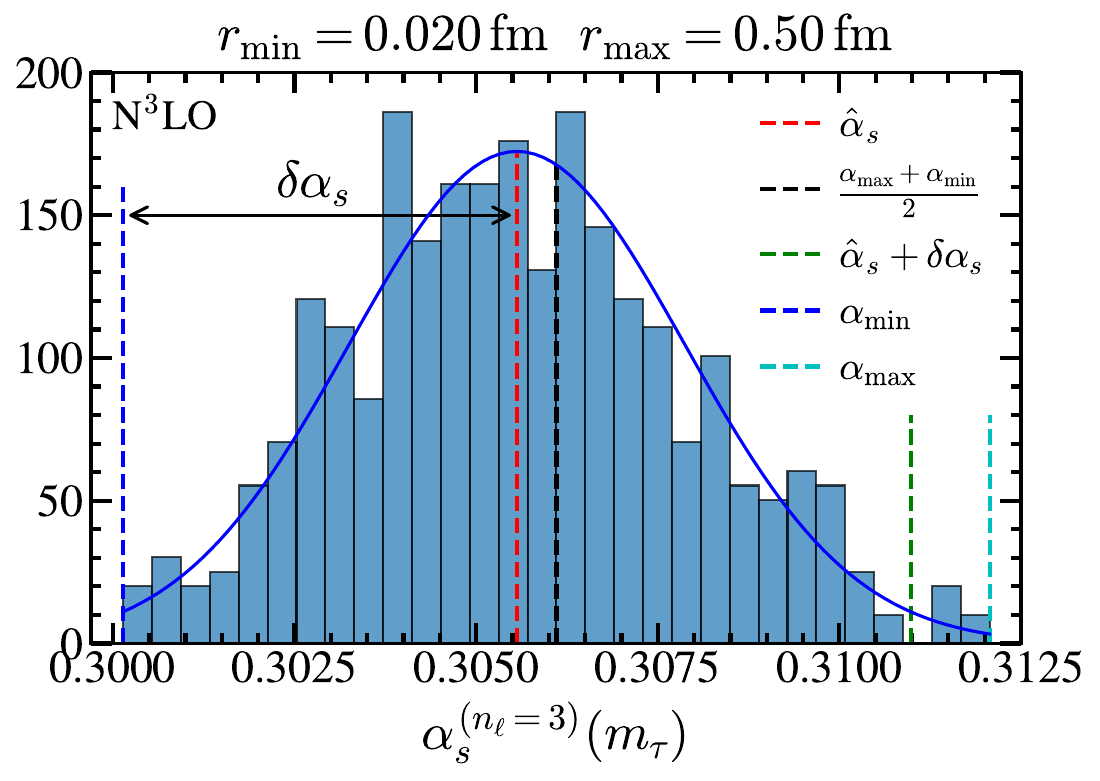}
\caption{}\label{fig:discard}
\end{subfigure}
\hfill
\begin{subfigure}{0.488\textwidth}
\centering
\includegraphics[width=\textwidth]{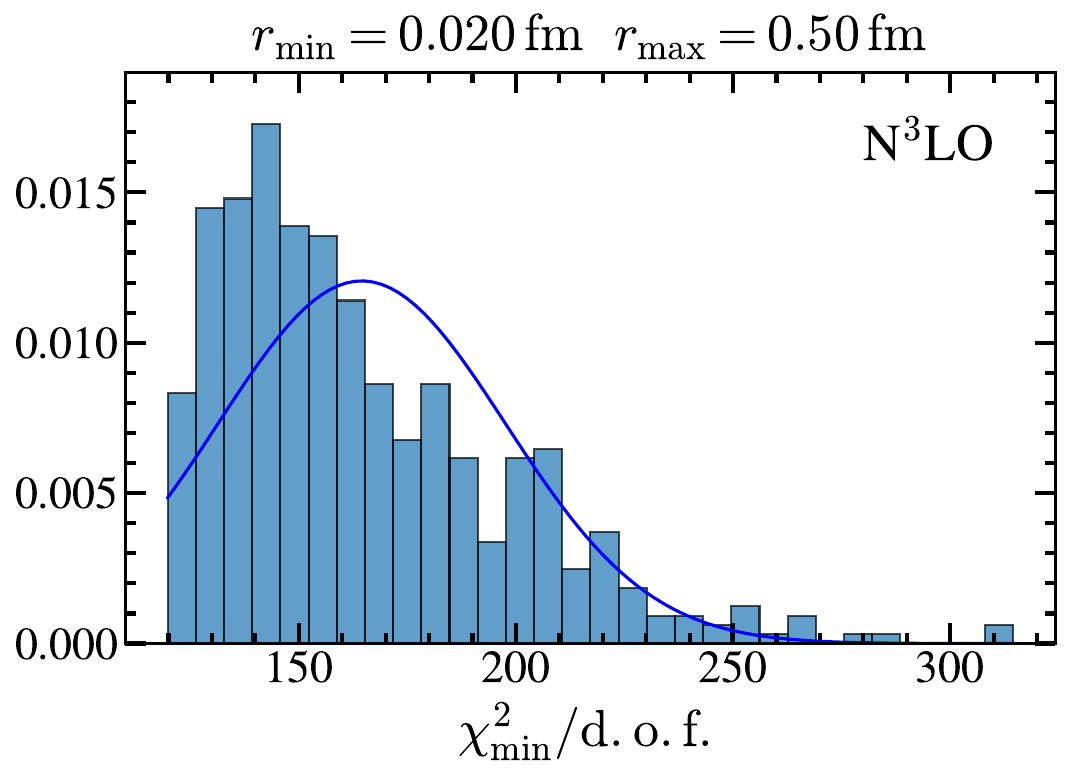}
\caption{}\label{fig:chi2}
\end{subfigure}
\caption{\label{fig:Histos}
Left panel: Histogram representing the frequency of $\alpha_s$ best-fit values in the 500 random scan for $r_{\rm min}=0.02\,$fm and $r_{\rm max}=0.5\,$fm at N$^3$LO and considering a single offset. The blue and cyan vertical dashed lines mark the minimum and maximum best-fit values of the strong coupling from the scan, respectively.
The black and red lines indicate the average of these two values and of the 500 $\alpha_s$ results (marked as $\hat \alpha_s$ in the legend), respectively.
The green dashed vertical line is obtained by adding $\hat \alpha_s$ and $\delta \alpha_s = \hat\alpha_s-\alpha_{\rm min}$. The blue solid curve corresponds to a Gaussian distribution with mean and variance computed from the set of $\alpha_s$ best-fit values. Right panel: same for the distribution of $\chi^2_{\rm min}/{\rm d.o.f.}$}
\end{figure}
\subsection{Estimate of Theoretical and Lattice Uncertainties}\label{sec:errors}
For a given dataset, there are two distinct kinds of uncertainties affecting the strong coupling, namely perturbative and that coming from lattice simulations. In our case, only statistical lattice errors are known and, given those are very small, theoretical uncertainties overly dominate. It would be desirable to include both theory and lattice uncertainties in the $\chi^2$ function but, as of today, there is no standard and universally accepted method to incorporate perturbative uncertainties in $\chi^2$ functions. One alternative, presented in Ref.~\cite{Benitez:2025vsp}, is to use renormalization scale variation by randomly varying the parameters specifying the profile functions in finite ranges. With this flat parameter scan one can compute the full covariance matrix that, with some caveats, can be interpreted as having the usual Gaussian meaning. We have checked that this procedure yields very strong correlations making the covariance matrix almost singular, being then severely affected by the d'Agostini bias~\cite{DAgostini:1993arp}. In Ref.~\cite{Tackmann:2024kci} the use of theory nuisance parameters has been advocated. Since such approach necessitates a dedicated study, and has never been applied to series suffering from $u=1/2$ renormalons, we believe it is beyond the scope of this work and leave it for future investigations. The approach we follow is the same used in Refs.~\cite{Benitez:2024nav,Abbate:2010xh,Abbate:2012jh,Hoang:2015hka}, that is, including only lattice uncertainties in the $\chi^2$ function and carrying out a so-called ``random scan''. Even though this procedure has been extensively explained in articles related to event shapes, we believe streamlining the method here is warranted. A sample of $500$ sets of profile parameters (which includes also $\lambda$) is produced, where each parameter is chosen from a flat random distribution within the limits specified in Table~\ref{tab:profiles}. For each of the sets, a $\chi^2$ function which includes only lattice uncertainties and uses theory predictions for the profiles defined by the set is minimized to obtain the best-fit value $\alpha_s^{\rm BF}$ and fit uncertainty $\Delta_{\rm fit}\alpha_s$
as explained in Sec.~\ref{sec:procedure}.
Taking a conservative approach, we inflate the lattice error by multiplying it by $\sqrt{\chi^2_{\rm min}/{\rm d.o.f.}}$ ---\,being ${\rm d.o.f.}$ the degrees of freedom\,--- so that the minimal reduced $\chi^2$ equals unity. This yields 500 tuples $\{\alpha_s^{\rm BF},\Delta_{\rm fit}\alpha_s,\chi^2_{\rm min}\}$ for each dataset which, as shall be discussed, is characterized by the minimal ($r_{\rm min}$) and maximal ($r_{\rm max}$) value of the distance. With our choice of profile-parameter variation no outliers are generated.
This is clearly seen in Fig.~\ref{fig:discard}, where we show a histogram of the 500 points in our random scan for $r_{\rm min}=0.02\,$fm and $r_{\rm max}=0.5\,$fm, where the binned distribution agrees reasonably well with a Gaussian whose parameters are obtained from the individual points. That the average of all 500 best-fit results $\hat \alpha_s$ is very close to
that of the maximal ($\alpha_{\rm max}$) and minimal ($\alpha_{\rm min}$) values $\bar\alpha_s=(\alpha_{\rm max} + \alpha_{\rm min})/2$,
confirms the absence of outliers that could bias the central value and overestimate the perturbative uncertainty.\footnote{We find an identical behavior for the distribution of offsets. To avoid cluttering, we do not show the corresponding histogram.} One could discard best-fit values of $\alpha_s$ larger than $\hat \alpha_s+\delta\alpha_s$ being $\delta \alpha_s=\hat \alpha_s - \alpha_{\rm min}$,
but this has a very small effect and hence we do not adopt this prescription.
The final central value corresponds to $\bar\alpha_s$ and the fit uncertainty is obtained averaging the corresponding 500 errors.
The perturbative uncertainty is computed as $\Delta_{\rm pert}\alpha_s=(\alpha_{\rm max} - \alpha_{\rm min})/2$.
From our point of view, this procedure is the most conservative among the ones used in the literature concerning $\alpha_s$ determinations from the static energy.
\begin{figure}[t!]
\centering
\begin{subfigure}{0.47\textwidth}
\centering
\includegraphics[width=\textwidth]{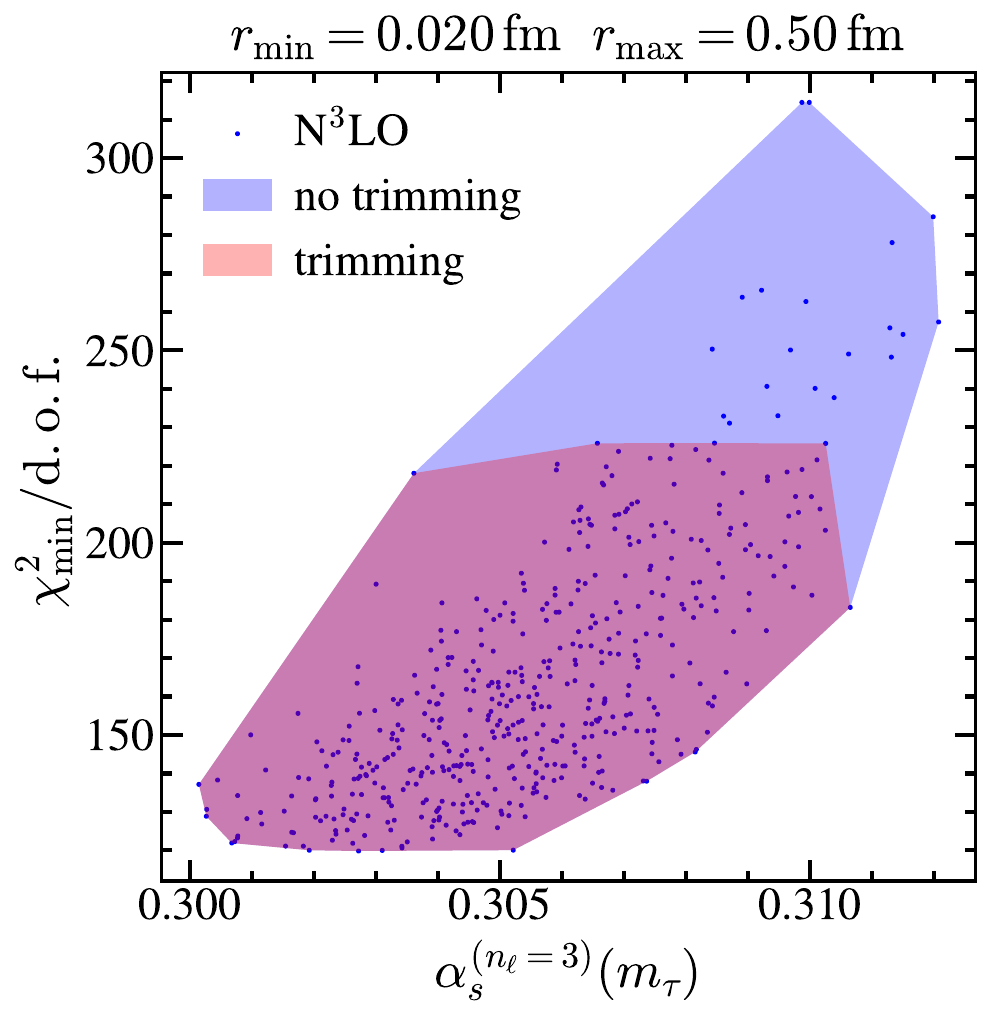}
\caption{}\label{fig:Chi2Trim}
\end{subfigure}
\hfill
\begin{subfigure}{0.47\textwidth}
\centering
\includegraphics[width=\textwidth]{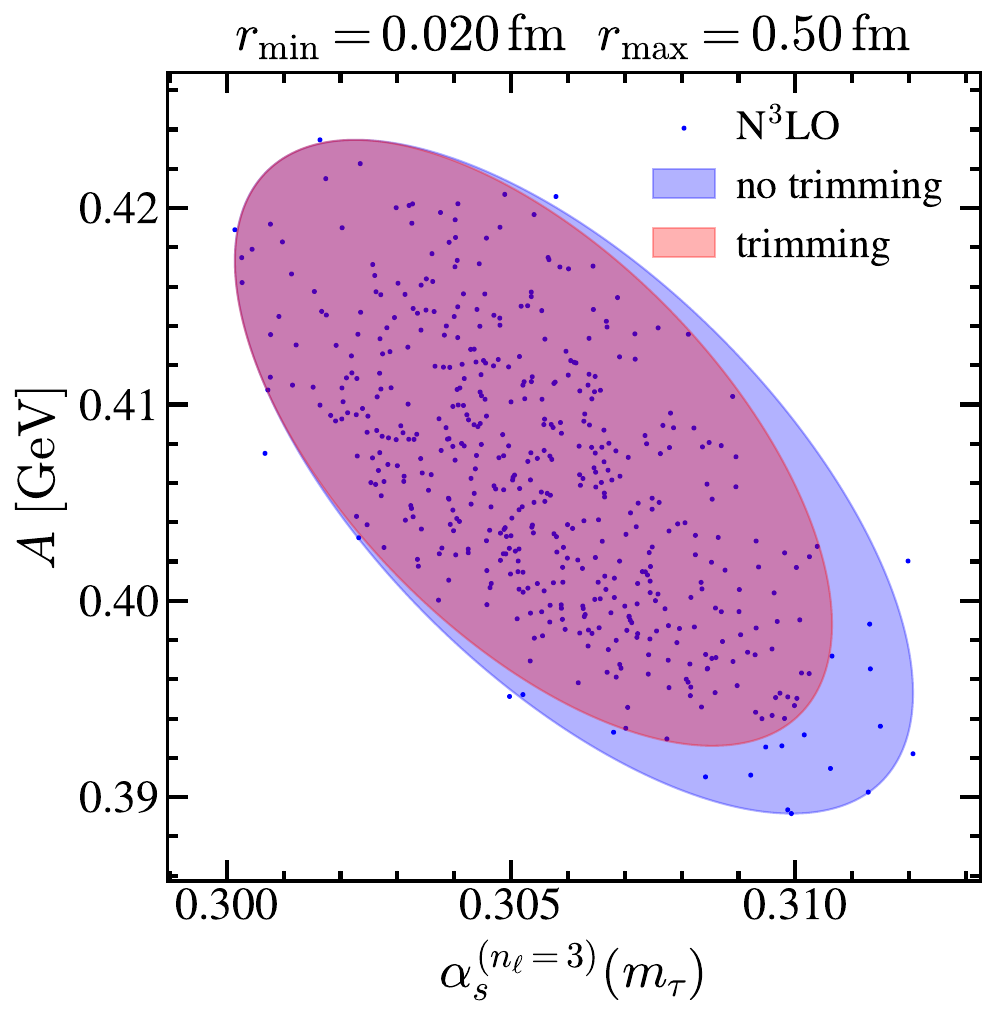}
\caption{}\label{fig:Ellipse}
\end{subfigure}
\caption{ \label{fig:Trimming}
Results for the 500 N$^3$LO fits corresponding to our random scan for \mbox{$r_{\rm min}=0.02\,$fm} and $r_{\rm max}=0.5\,$fm, considering a single offset $A$.
All results are shown in blue, while those surviving the $\chi^2$-trimming are in red.
The left and right panels show the projections in the $\alpha_s$-$(\chi^2_{\rm min}/{\rm d.o.f.})$ and $\alpha_s$-$A$ planes, respectively.}
\end{figure}

There is a minor point to be discussed before we finish this section. Looking at the distribution of minimal $\chi^2$ values displayed as a histogram, see Fig.~\ref{fig:chi2}, one finds that results do not follow a Gaussian and that there are few fits with values much higher than the bulk. One could be tempted to call those ``outliers'' and trim them away. A simple criterion that effectively but not too aggressively removes those outliers is requiring that $\chi^2_{\rm min}$ should be smaller than the average of values plus one and a half standard deviations. The result of that procedure is shown in Fig.~\ref{fig:Trimming}. Since discarding points which do not satisfy this criterion does not significantly alter central value and perturbative uncertainty for the strong coupling, we prefer to stay conservative and keep them in our analyses.

\begin{figure}[t!]
\centering
\begin{subfigure}{0.47\textwidth}
\centering
\includegraphics[width=\textwidth]{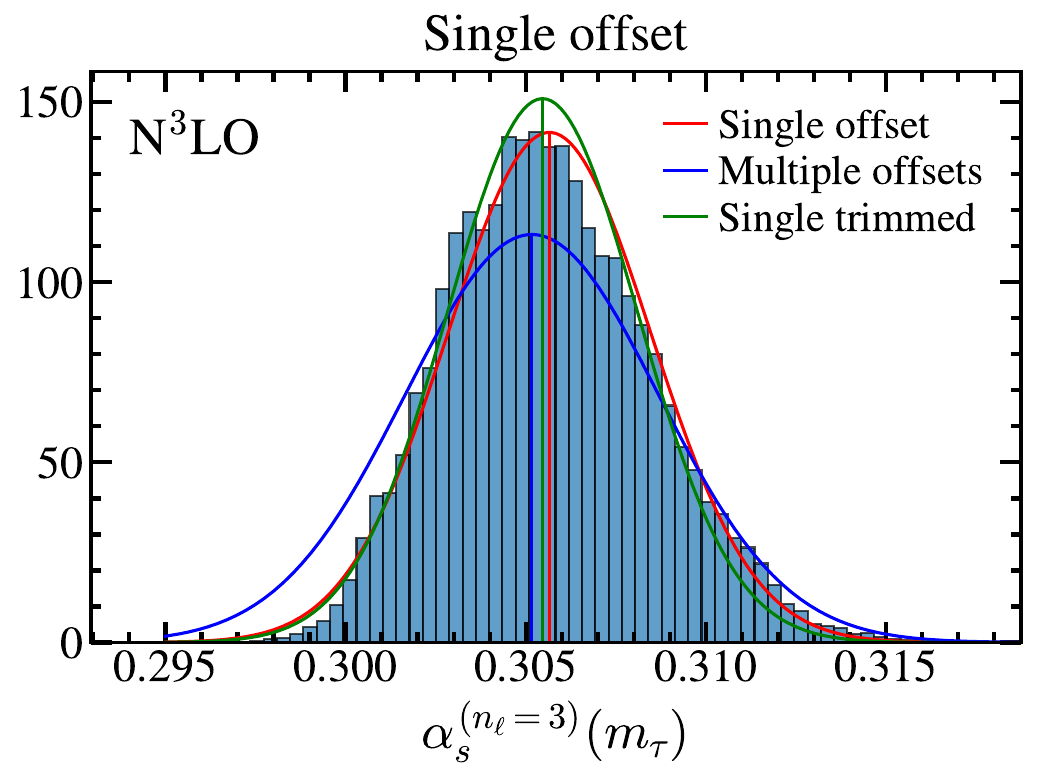}
\caption{}\label{fig:Full}
\end{subfigure}
\hfill
\begin{subfigure}{0.47\textwidth}
\centering
\includegraphics[width=\textwidth]{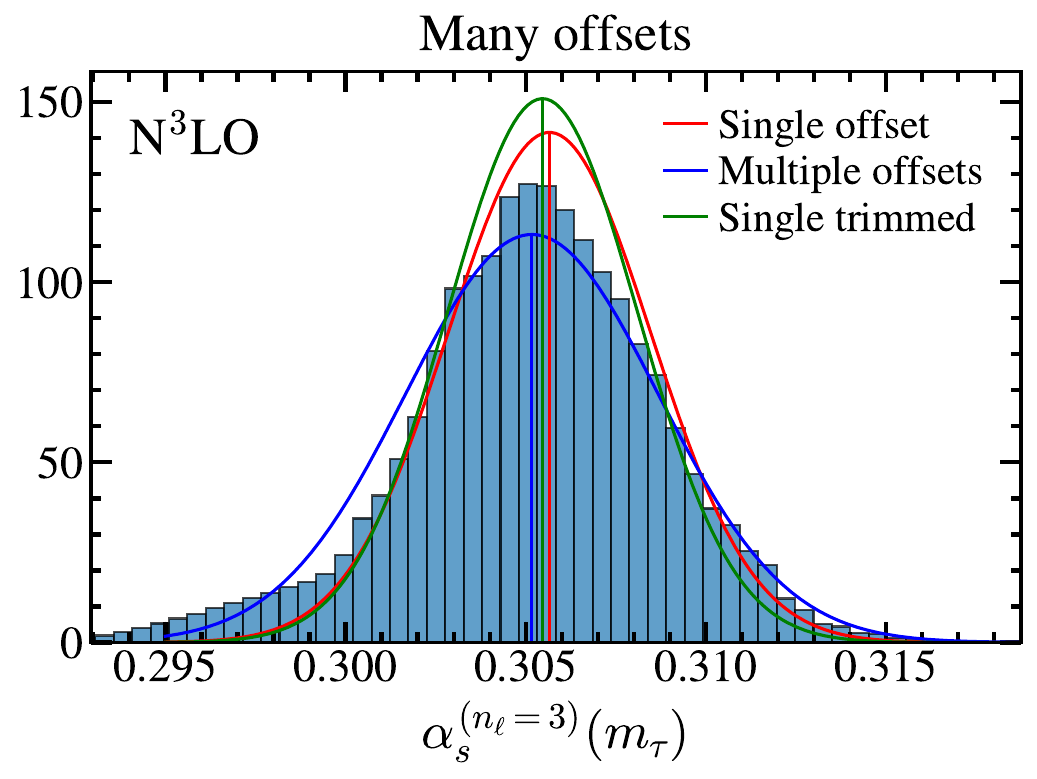}
\caption{}\label{fig:HistoMany}
\end{subfigure}
\vskip 0.2cm
\begin{subfigure}{0.47\textwidth}
\centering
\includegraphics[width=\textwidth]{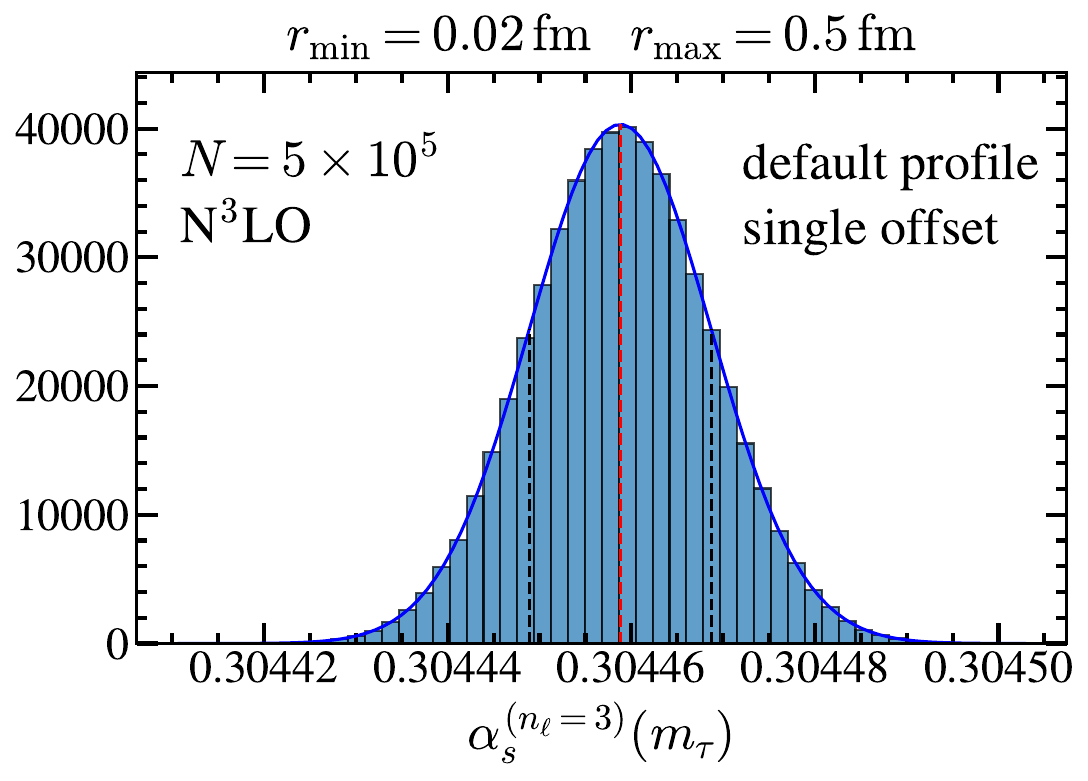}
\caption{}\label{fig:Replica}
\end{subfigure}
\hfill
\begin{subfigure}{0.47\textwidth}
\centering
\includegraphics[width=\textwidth]{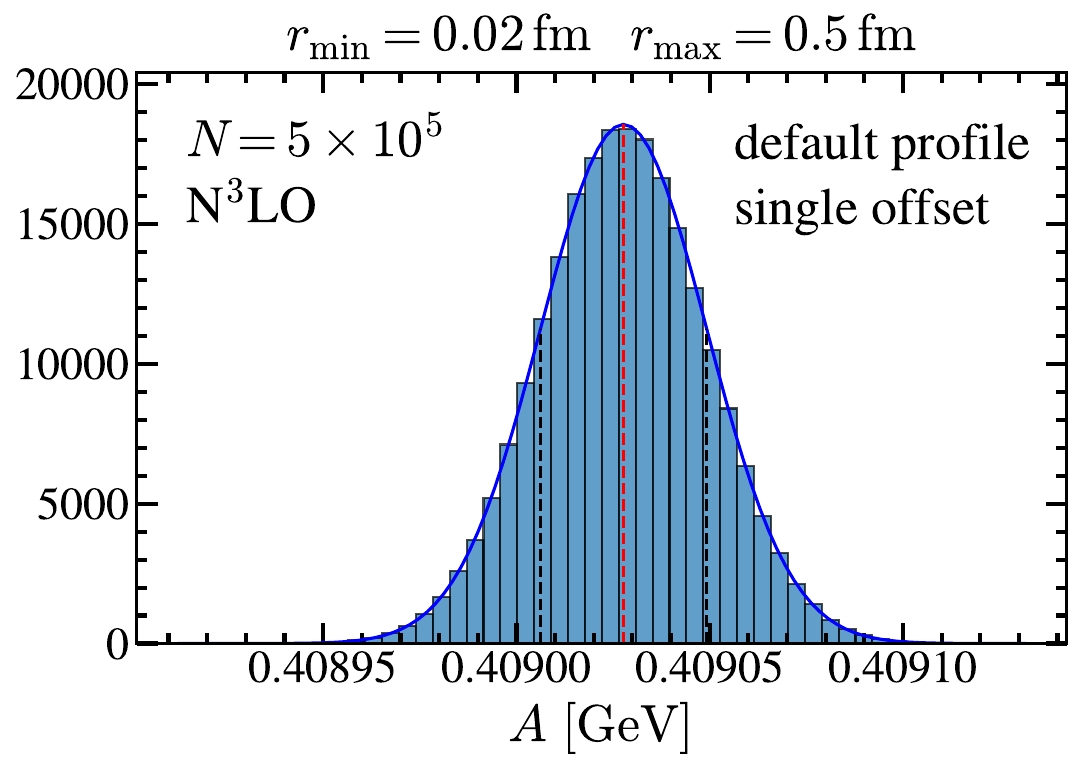}
\caption{}\label{fig:A-Replica}
\end{subfigure}
\caption{\label{fig:Histogram}
Upper panels: Distribution of all 543\,000 best-fit values for the strong coupling obtained from performing 500 fits for each of the 1086 fit ranges considered, shown in 50 bins for a single (a) or multiple (b) offsets. The red and blue solid lines show Gaussian distributions with parameters obtained from the full dataset for one or multiple offsets, respectively. The green line corresponds to one-parameter fits after applying the $\chi^2$ trimming procedure, leaving 518\,445 results. Lower panels: Distribution of the $5\times 10^5$ best-fit values for $\alpha_s$ (c) and $A$ (d) as obtained with the replica method, displayed in 50 bins for our default profile parameters. The blue and red dashed vertical lines indicate the central value and 1-$\sigma$ confidence-level region, respectively. The blue line corresponds to a Gaussian distribution obtained with the mean and variance of the half a million results.}
\end{figure}

\subsection{Numerical Analysis}\label{sec:RangeFits}
In this section we explore various aspects that can affect the fit results.
We vary the dataset
considering six values for $r_{\rm min}$ between $0.020\,$fm and $0.045\,$fm in steps of $0.005\,$fm and, for each of those, scan over $r_{\rm max}$ between $0.1\,$fm and $1\,$fm with the same spacing as for $r_{\rm min}$. This makes a total of 1086 different datasets, each one consisting on 500 fits due to our theoretical flat random scan over the parameters that specify the profile functions, for a total of over half a million individual fits for each perturbative order and setup. Before we show how results depend on the specific dataset, it is instructive to display the full set of best-fit values for the strong coupling as a whole for fits with single and multiple offsets. The outcome of such analyses is shown in the two panels of Fig.~\ref{fig:Histogram}, where histograms with 50 bins are used. We compute the mean and standard deviation for each of the ensembles and superimpose the corresponding Gaussian functions. For the single-offset ensemble, it follows the histogram rather well, while for multiple offsets one observes a non-Gaussian ``shoulder'' on the left side of the distribution that affects the agreement between histograms and the Gaussian. For single-offset fits the shoulder is absent, indicating that a thorough investigation is in order. In addition, we also show a Gaussian representing the result after having $\chi^2$-trimmed each dataset with the procedure described in Sec.~\ref{sec:errors}, what leaves also more than $5\times10^6$ results. It becomes clear that trimming barely affects the distribution, neither its average nor its dispersion as can in particular be seen by comparing the red and green Gaussians shown on both panels. In any case, as already stated, for the rest of our analyses we do not trim the results. For later reference, we quote the average and standard deviation for the full set of untrimmed results and a single (multiple) offset(s): $\langle\alpha_s\rangle = 0.3057$ and $\sigma_{\alpha_s} = 0.0028$ ($0.3052$ and $0.0035$), respectively.

\begin{figure}[t]
\centering
\begin{subfigure}{0.48\textwidth}
\centering
\includegraphics[width=\textwidth]{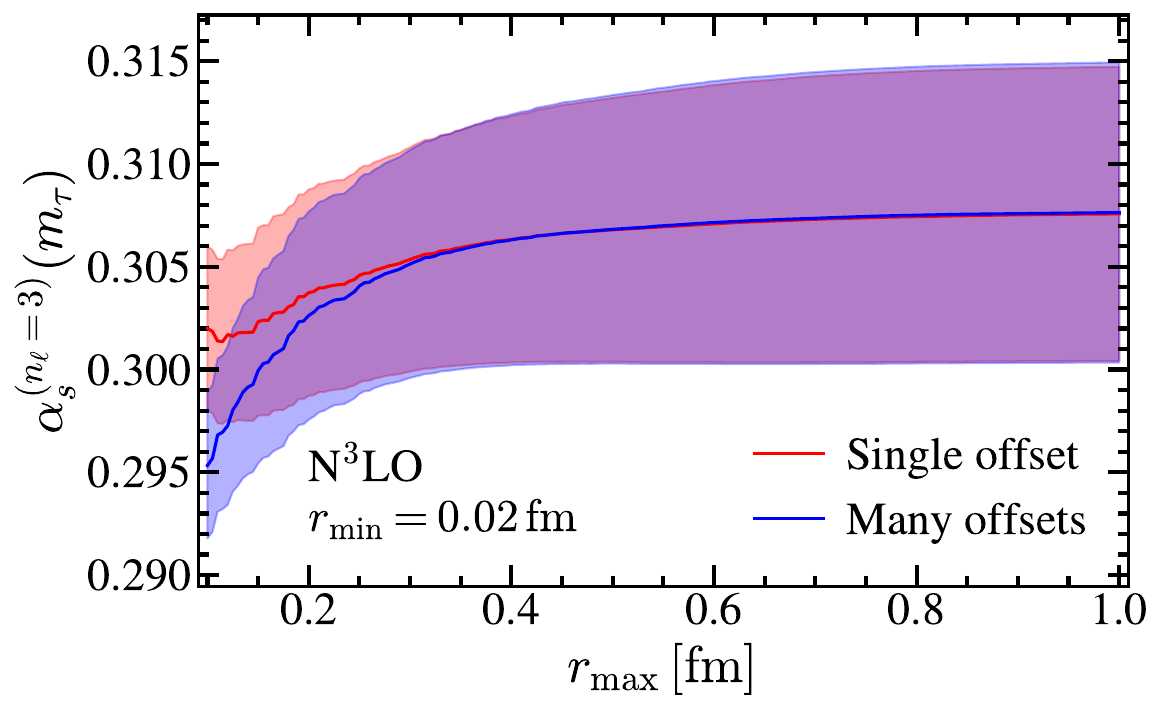}
\caption{}\label{fig:OffsetRange}
\end{subfigure}
\hfil
\begin{subfigure}{0.48\textwidth}
\centering
\includegraphics[width=\textwidth]{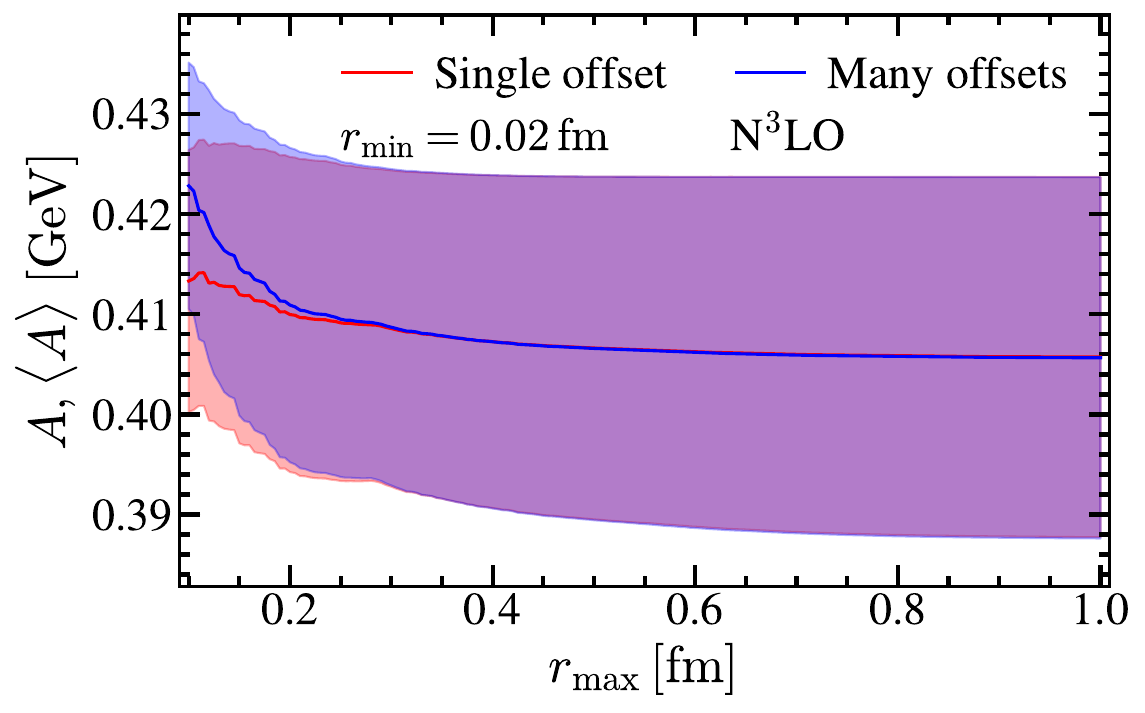}
\caption{}\label{fig:RangeOffset}
\end{subfigure}
\vskip 0.2cm
\begin{subfigure}{0.485\textwidth}
\centering
\includegraphics[width=\textwidth]{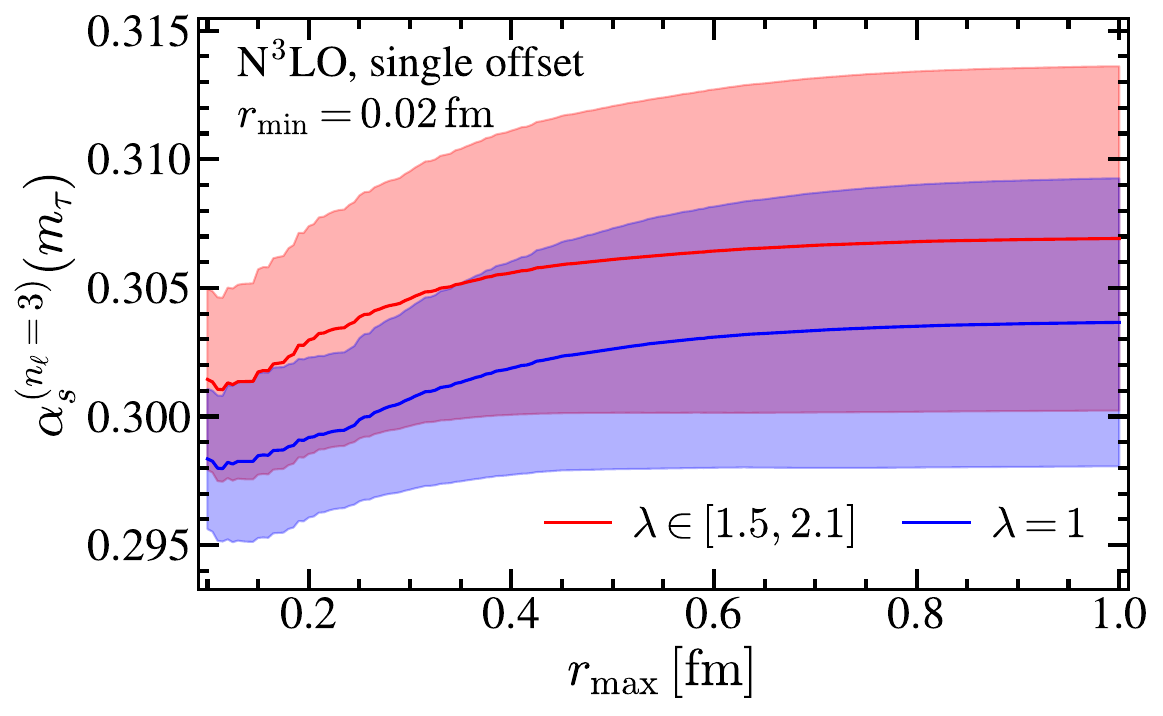}
\caption{}\label{fig:LambdaRange}
\end{subfigure}
\hfill
\begin{subfigure}{0.479\textwidth}
\centering
\includegraphics[width=\textwidth]{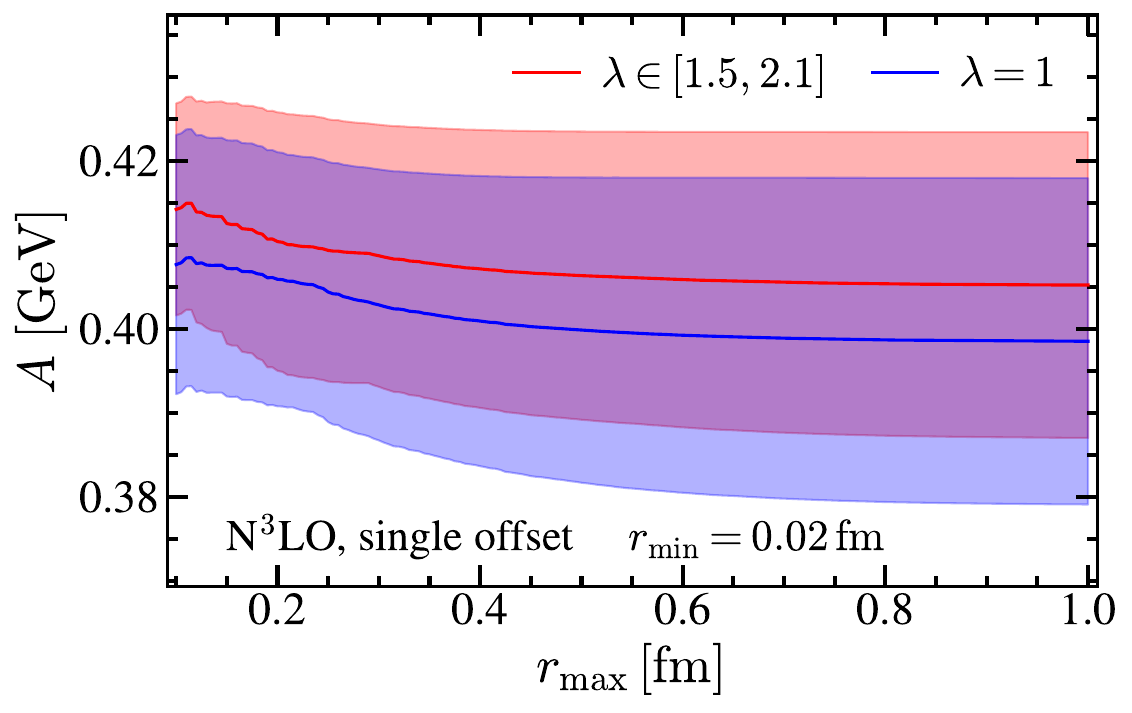}
\caption{}\label{fig:LambdaRangeA}
\end{subfigure}

\caption{\label{fig:Offset} Dependence of the strong coupling [\,(a) and (c)\,] and offset/average of offsets [\,(b) and (d)\,] with the maximal distance included in the dataset for N$^3$LO fits with \mbox{$r_{\rm min}=0.02\,$fm}. We show both central value and full uncertainty, which includes theoretical and lattice errors for the strong coupling, but only perturbative uncertainties for the offset. Upper panels: red and blue show results for single and multiple offsets, respectively. Lower panels: results obtained from single-offset fits, where blue corresponds to a fixed value of the R-revolution parameter $\lambda=1$, while red includes a scan as shown in the plot legend.}
\end{figure}

Given the findings in our previous analysis, we turn our attention to the difference between considering the various lattice ensembles' offsets as independent fit parameters versus identifying all of them into a single quantity to be determined. This is analyzed in the upper panels of Fig.~\ref{fig:Offset}, showing that the two approaches agree very well for \mbox{$r_{\rm max}>0.3\,$fm} but differ at smaller distances. In particular, the value of the strong coupling deeps down when multiple offsets are considered, but stays rather stable when only one of them is fit to data. These small values are responsible for the left-shoulder visible in the histogram shown in Fig.~\ref{fig:HistoMany}. One can also observe that the total $\alpha_s$ uncertainty drastically shrinks in the former case, but exhibits a moderate decrease in the latter. This behavior has its counterpart in the reduced $\chi^2$ skyrocketing in the single-offset setup as compared to a commensurate increase if multiple offsets are individually determined. This behavior is expected since lattice data at small distances has extraordinarily small (statistical) uncertainties as compared to those at larger $r$, which are the only ones that can be included in our $\chi^2$. We believe this bad behavior of multi-offset fits at small distances might be caused by an overparametrization of the $\chi^2$ that has the potential for creating a runaway direction causing the small value for the strong coupling. Unless otherwise stated, for the rest of the analyses carried out in this section we use single-offset fits only. As a precaution, for our final analysis we shall consider values of $r_{\max}\geq 0.35\,$fm, hence our results will not be affected by this choice.

\begin{figure}[t!]
\centering
\begin{subfigure}{0.47\textwidth}
\centering
\includegraphics[width=\textwidth]{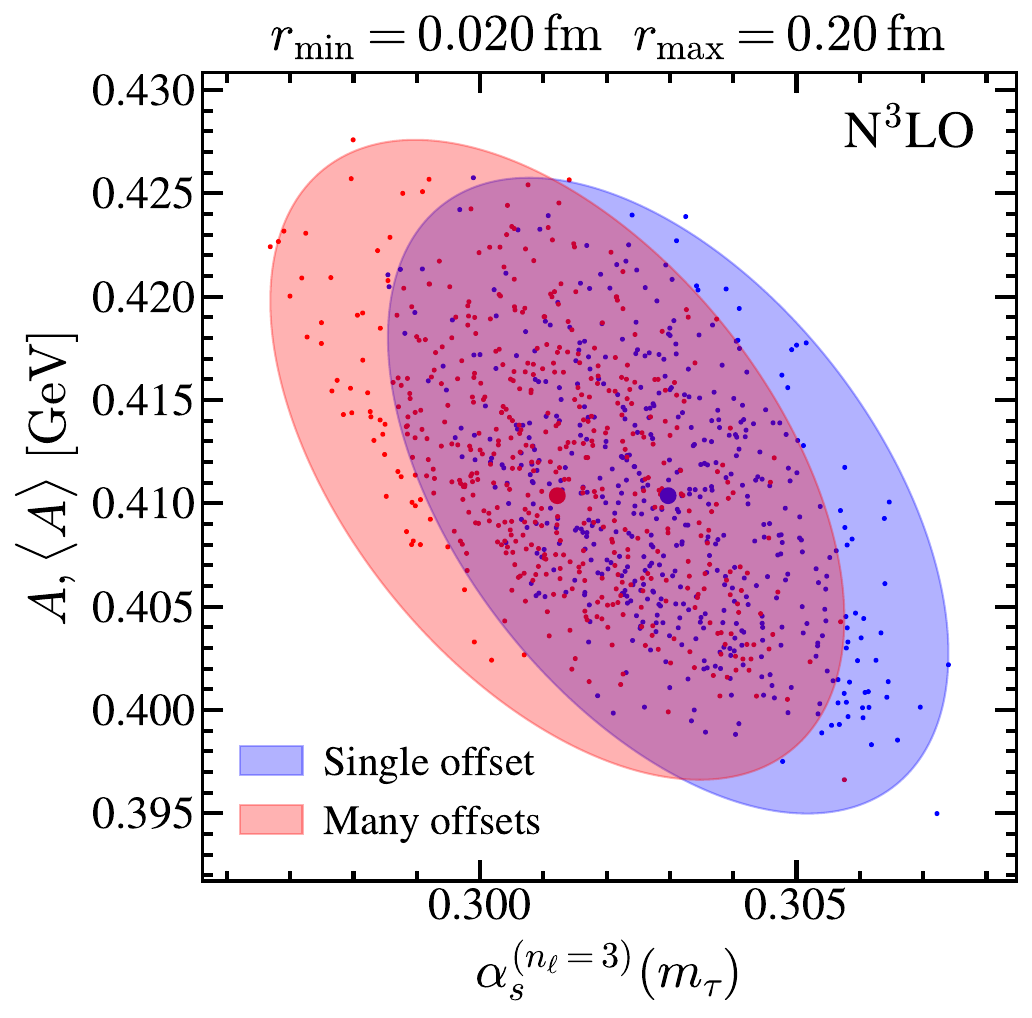}
\caption{}\label{fig:2DOffset}
\end{subfigure}
\hfill
\begin{subfigure}{0.48\textwidth}
\centering
\includegraphics[width=\textwidth]{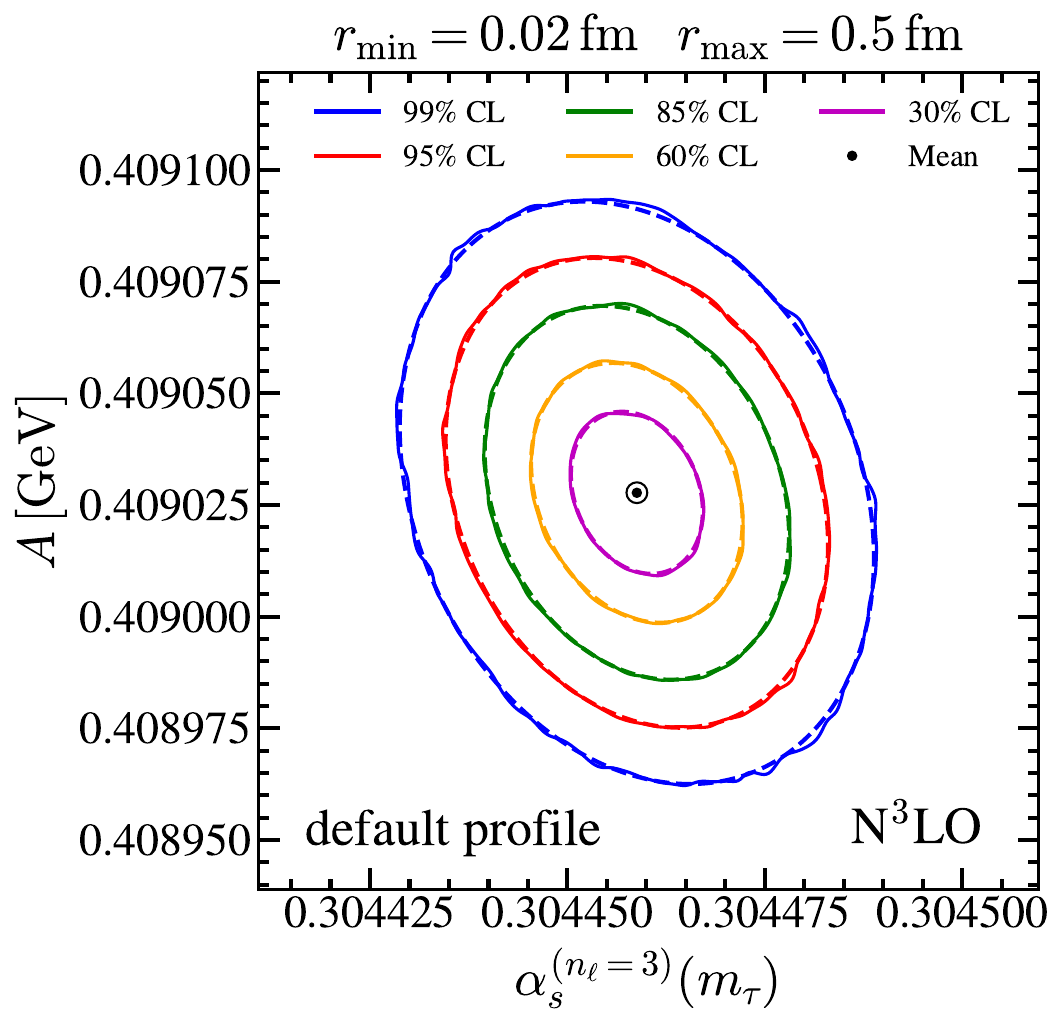}
\caption{}\label{fig:Contours}
\end{subfigure}

\caption{\label{fig:Alpha-Offset}
Left panel: distribution of $\alpha_s$-($A$/$\langle A\rangle$) best-fit pairs obtained from the 500 profiles in our random scan for $0.02\,{\rm fm}\leq r\leq 0.2\,{\rm fm}$, with red and blue corresponding with many- and single-offset fits, respectively. Large colored dots show the (39\% confidence-level) ellipses' center. Right panel: Numerical (solid lines) and expected analytical (dashed lines) contours for various confidence level corresponding to the half-million fits produced for the replica method. The mean value of $\alpha_s$ and $A$ is marked with a black dot, while the hollow circle corresponds to the best-fit values using the original dataset.}
\end{figure}

\begin{figure}[t]
\centering
\begin{subfigure}{0.495\textwidth}
\centering
\includegraphics[width=\textwidth]{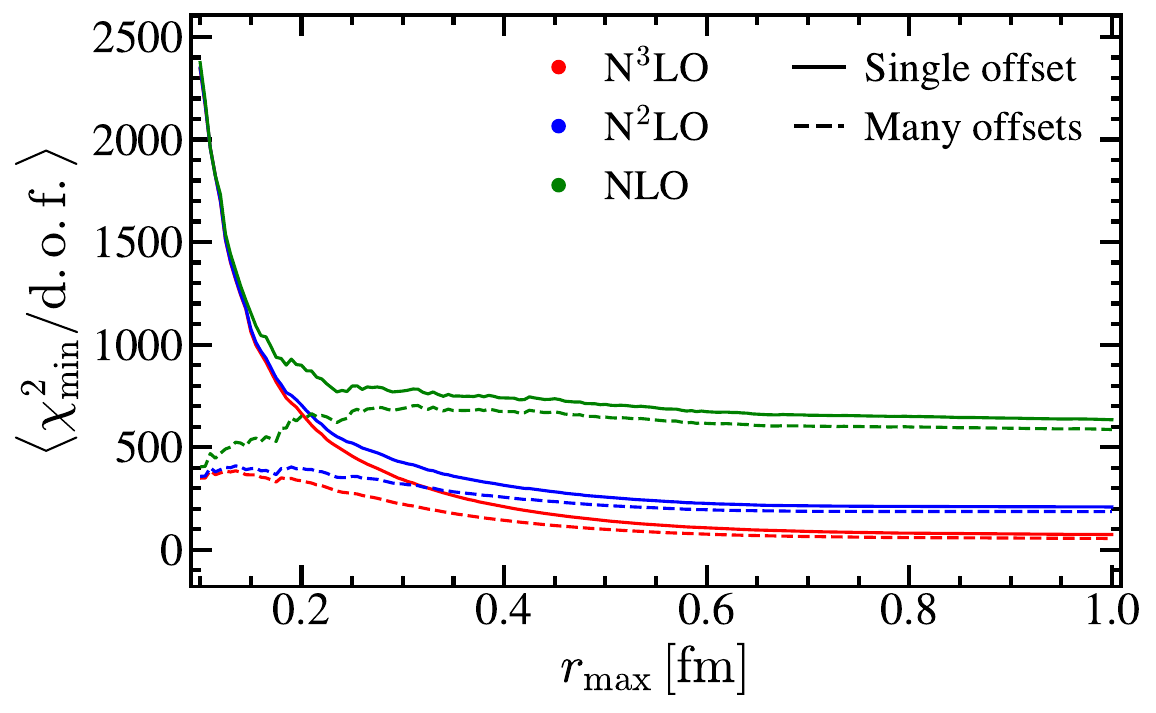}
\caption{}\label{fig:Chi2Range}
\end{subfigure}
\hfil
\begin{subfigure}{0.48\textwidth}
\centering
\includegraphics[width=\textwidth]{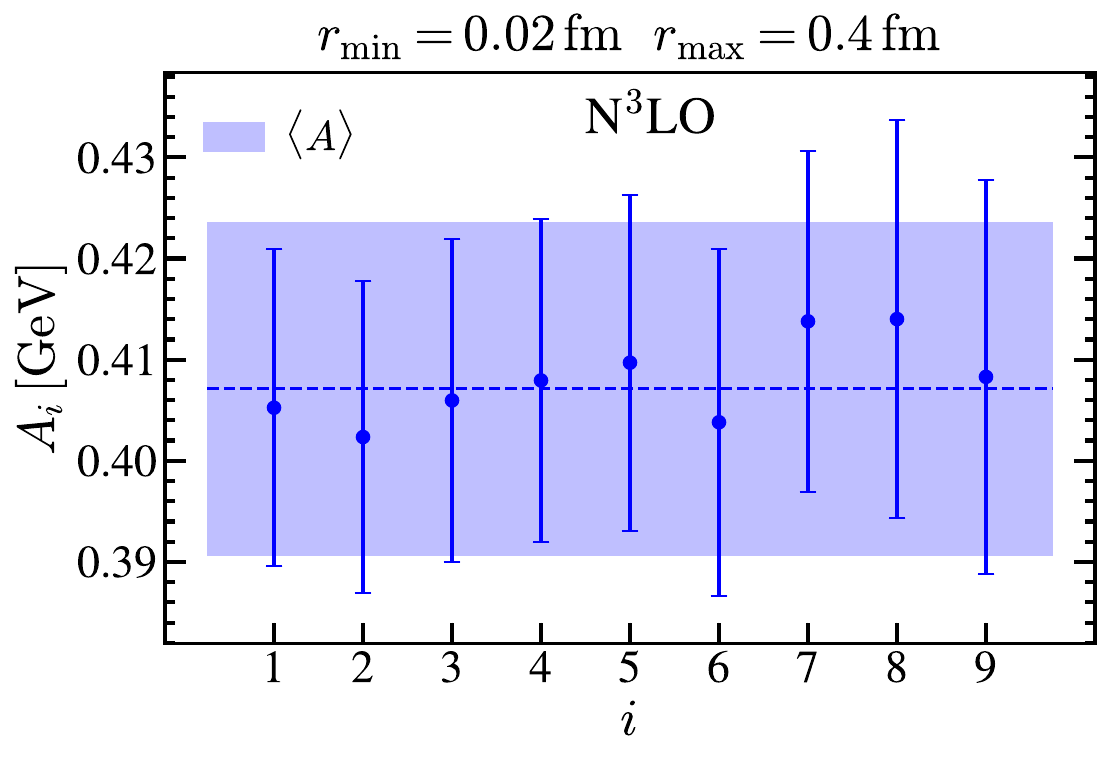}
\caption{}\label{fig:Offsets}
\end{subfigure}
\caption{\label{fig:Chi2Offset} Left panel: Dependence of reduced $\chi_{\rm min}^2$ with the maximal distance included in the dataset for fits at various orders. The mean of all reduced $\chi_{\rm min}^2$ values is shown, averaged over the $500$ profiles and the six $r_{\rm min}$ values considered. Panel (b): Results for the nine offsets $A_i$ corresponding to each lattice ensemble, where the error bars indicate the perturbative error. The translucent band represents the result for $\langle A\rangle$.}
\end{figure}

In the second part of this analysis we want to compare the single-offset with a weighted average of the multiple offsets which we denote by $\langle A\rangle$ and compute using the last line of Eq.~\eqref{eq:uncor} with $\alpha_s^{\rm BS}$ as obtained with the marginalized function with multiple offsets. The result of this exercise is shown in Fig.~\ref{fig:RangeOffset}, showcasing that $A$ and $\langle A\rangle$ are very similar for datasets with $r_{\rm max}>0.3\,$fm. Their difference below this threshold is not as pronounced as for the strong coupling. In Fig.~\ref{fig:2DOffset} we compare the random-scan spread of best-fit values in the $\alpha_s$-($A,\langle A\rangle$) plane for the dataset defined by $r\in[0.02, 0.2]\,{\rm fm}$. A comparison between the individual $A_i$ and the average $\langle A\rangle$ is shown in Fig.~\ref{fig:Offsets}, reinforcing the statistical compatibility among all offsets. Identical patterns are found at lower orders.

We also asses how much our non-standard variation for the R-evolution parameter $\lambda$ affects the outcome of the fits. This is displayed in the two lower panels of Fig.~\ref{fig:Offset}, where results for $\alpha_s$ and $A$ are shown in the left and right panels, respectively, for $\lambda=1$ in blue and $\lambda\in[1.5, 2.1]$ in red. Indeed, we observe that our choice of $\lambda$ variation yields a higher value for the strong coupling with a slightly larger perturbative uncertainty. The difference in central values can be interpreted as a bias due to using a non-representative $\lambda$.

As mentioned in Sec.~\ref{sec:data}, we only use uncorrelated statistical uncertainties to construct our $\chi^2$ function, since that is the only information
disclosed to us. Since this could inflict some bias to our central value, and to validate that our (tiny) lattice uncertainty is reliable despite the large values we find for the reduced $\chi^2_{\rm min}$, we apply the replica method. We carry out this test only to a representative dataset encompassing distances between $0.2\,$fm and $0.5\,$fm and for the default profile parameters,
but reach identical conclusions for other datasets and profiles.
In the absence of correlations, the replica method consists on generating pseudo-data (also known as synthetic datasets) by fluctuating the central values within the 1-$\sigma$ uncertainties following a Gaussian distribution. A fit to each replica is carried out and it is expected that the mean and variance of the best-fit values adhere to the central value and uncertainty obtained for the original dataset. To that end, we generate half a million replicas obtaining the same amount of best-fit $\alpha_s$ and $A$ values, as shown in the form of histograms in the lower panels of Fig.~\ref{fig:Histogram}. The first observation is that the 50 bins closely follow the superimposed Gaussian distributions obtained with the ensemble's means and variances. For $\alpha_s$, the mean of those and the central value of the original dataset differ by $5\times10^{-9}$, showcasing the robustness of our analysis. While the original fit uncertainty is $1.18\times 10^{-5}$ (identical to the mean of replica uncertainties), the replicas' variance is $1.017\times 10^{-5}$, almost identical to the estimate given by the hessian, again confirming that our fits are completely reliable. Identical conclusions hold for the distribution of offsets.
As a final check, we test if the two-dimensional distribution of results follows the expected pattern of a two-parameter Gaussian, which is uniquely determined by the central values and uncertainties of $\alpha_s$, $A$, and the correlation coefficient $\rho$ among both, which can be computed with the ensemble of results. The outcome of this analysis is displayed in Fig.~\ref{fig:Contours}, where we compare the numerically determined confidence-level contour plots for various values of $p$ with the theoretical expectation using Eq.~\eqref{eq:confidence}. The observed agreement is remarkable, even far out in the distribution's tail, again confirming the robustness of our fits.

\begin{figure}[t!]
\centering
\begin{subfigure}{0.47\textwidth}
\centering
\includegraphics[width=\textwidth]{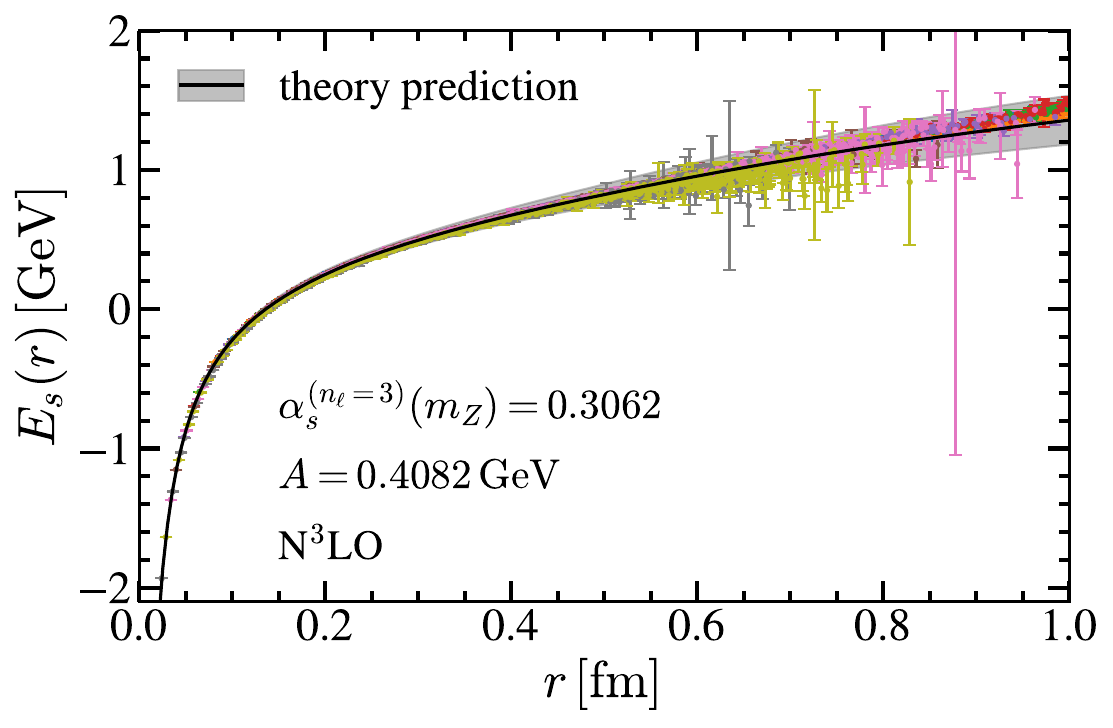}
\caption{}\label{fig:ThvsExp}
\end{subfigure}
\hfill
\begin{subfigure}{0.495\textwidth}
\centering
\includegraphics[width=\textwidth]{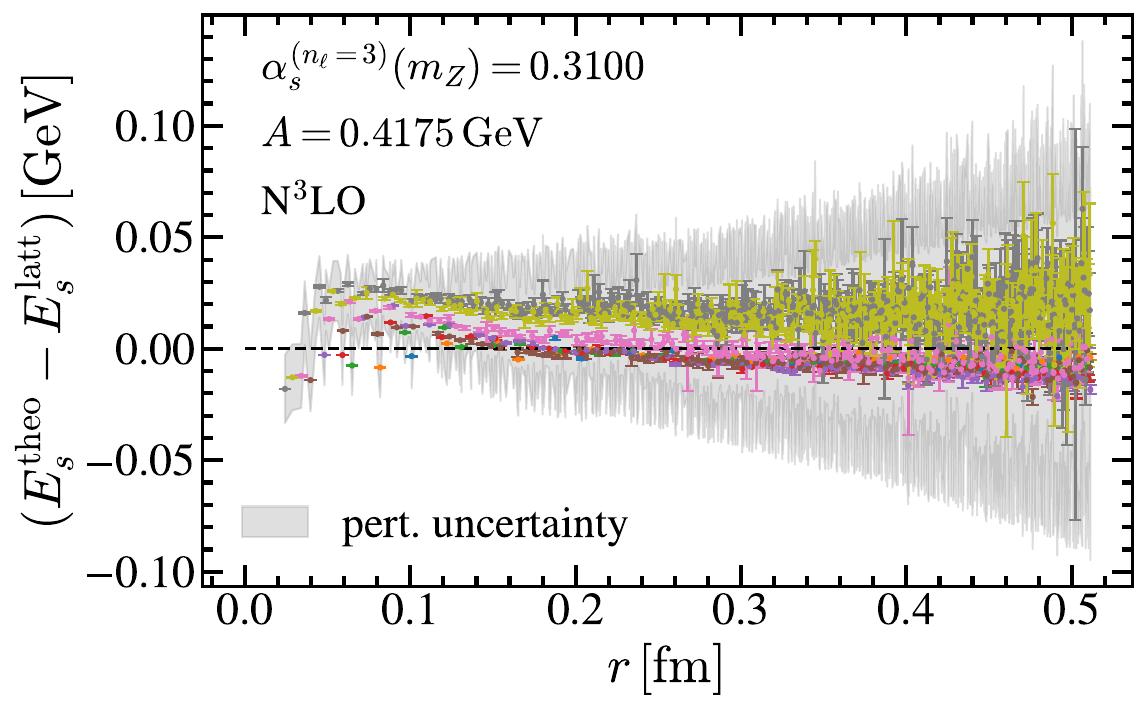}
\caption{}\label{fig:ThMinusExp}
\end{subfigure}

\caption{\label{fig:ThExp}
Comparison of lattice data (colored dots with error bars) with our best theoretical prediction using the central values for the strong coupling (shown in the figure) and offset as obtained in our final result in Eq.~\eqref{eq:final3}. Colors indicate different lattice ensembles, see legend of Fig.~\ref{fig:Cluster}. Error bars reflect only lattice statistical uncertainties, while gray bands correspond to the theoretical perturbative uncertainties. The left panel directly compares theory to data, while the right panel shows their difference.}
\end{figure}

Another potential point of concern, even if we already checked they do not affect our estimate of fit uncertainties, are the high values found for $\chi_{\rm min}^2/{\rm d.o.f.}$, much larger than unity. There are two reasons for this unwanted result: i)~we only know statistical uncertainties of lattice data, ignoring both correlations and systematic errors, ii)~our $\chi^2$ function does not include theoretical uncertainties. In Fig.~\ref{fig:ThExp} we compare lattice results for the various simulations shifted by a common offset with our best theoretical prediction including a theory uncertainty band generated with the random scan. For this comparison, $\alpha_s$ is obtained as the weighted average of the best-fit values corresponding to N$^3$LO single-offset fits for the datasets used in our final analysis, see Sec.~\ref{sec:final}. The common offset $A$ is obtained applying the last line of Eq.~\eqref{eq:uncor} to the dataset defined by $r\leq0.5\,$fm using the average value for the strong coupling.
Figure~\ref{fig:ThvsExp} shows that theory correctly reproduces lattice data within uncertainties, even beyond the region that has been used to obtain the strong coupling. In \ref{fig:ThMinusExp} we display the difference of theory minus experiment, which is closely tied to $\chi^2_{\rm min}$. While for data points at small distances the difference is many ``lattice'' sigma's away ---\,since statistical errors are extremely small\,--- once theoretical uncertainties are taken into account, every single difference is within errors. Hence, we conclude that our theory is precise enough to describe data, and that if perturbative uncertainties could be included, the reduced $\chi^2_{\rm min}$ should be around unity showcasing a good p-value.

\begin{figure}[t!]
\centering
\begin{subfigure}{0.489\textwidth}
\centering
\includegraphics[width=\textwidth]{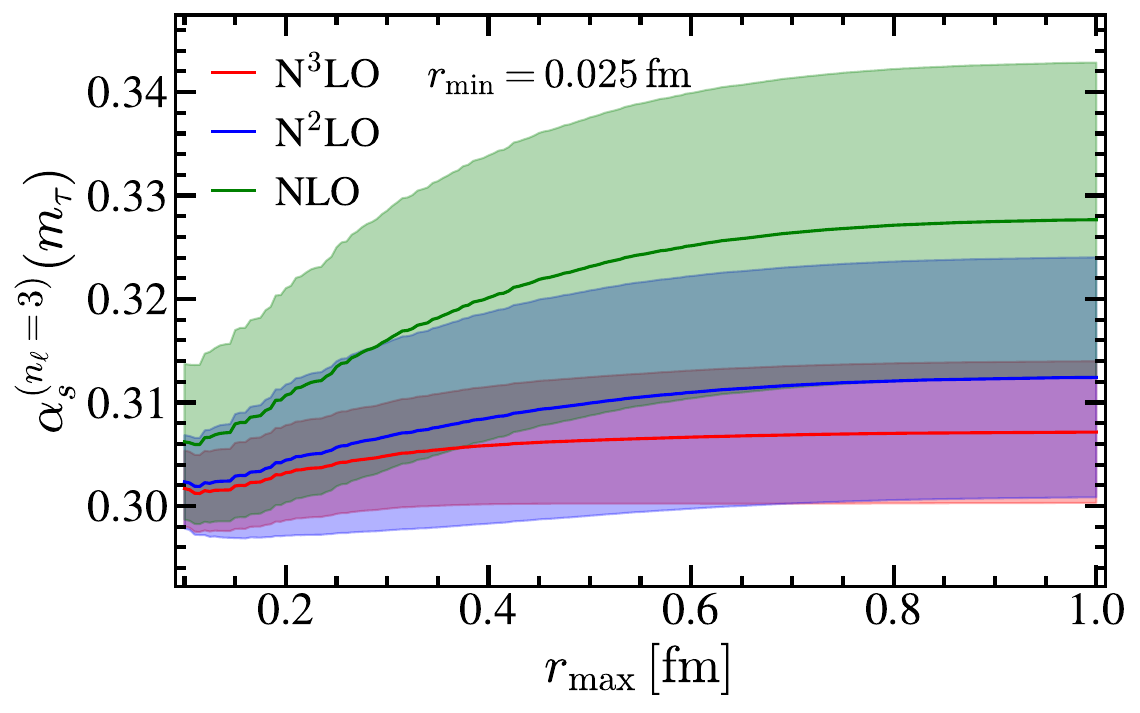}
\caption{}\label{fig:RangeOrders}
\end{subfigure}
\hfill
\begin{subfigure}{0.48\textwidth}
\centering
\includegraphics[width=\textwidth]{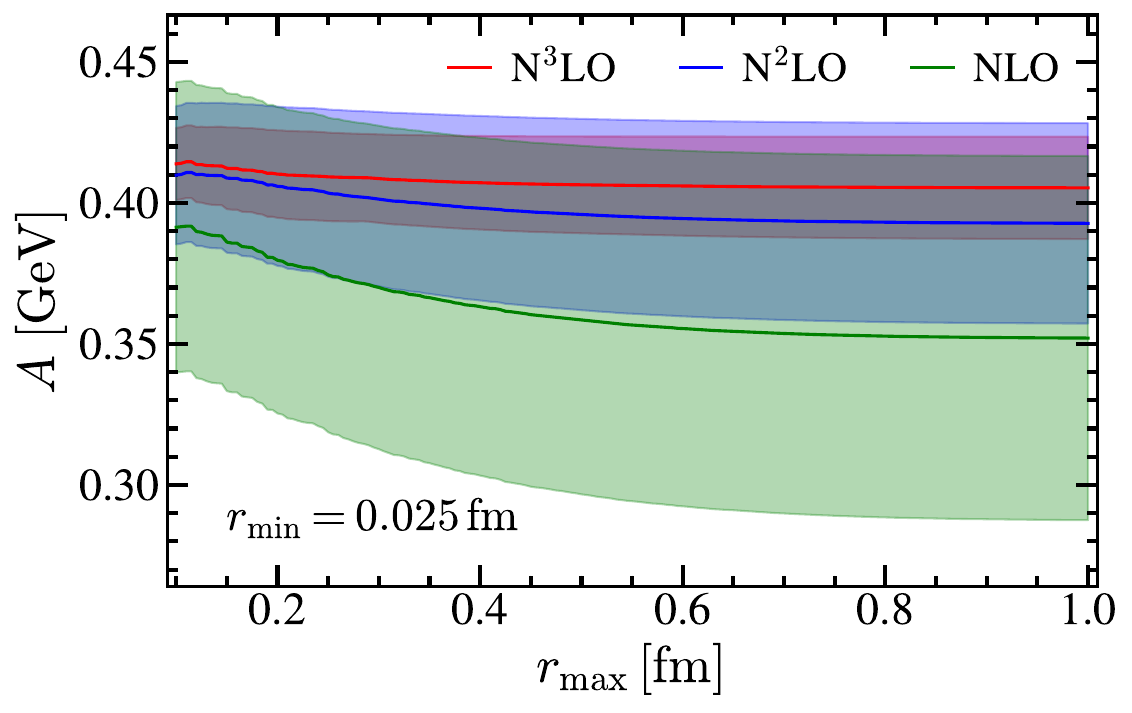}
\caption{}\label{fig:RangeOrdersA}
\end{subfigure}
\vskip 0.2cm
\begin{subfigure}{0.483\textwidth}
\centering
\includegraphics[width=\textwidth]{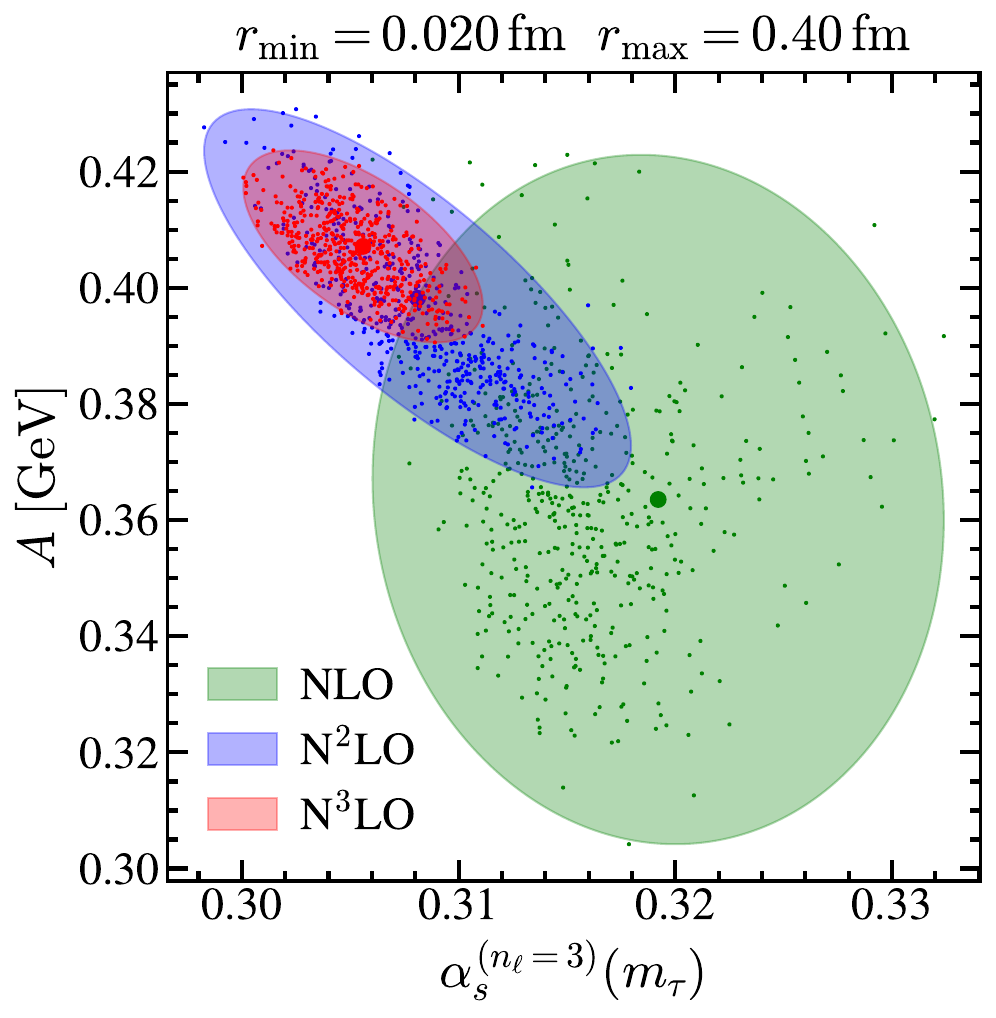}
\caption{}\label{fig:AvA}
\end{subfigure}
\hfill
\begin{subfigure}{0.489\textwidth}
\centering
\includegraphics[width=\textwidth]{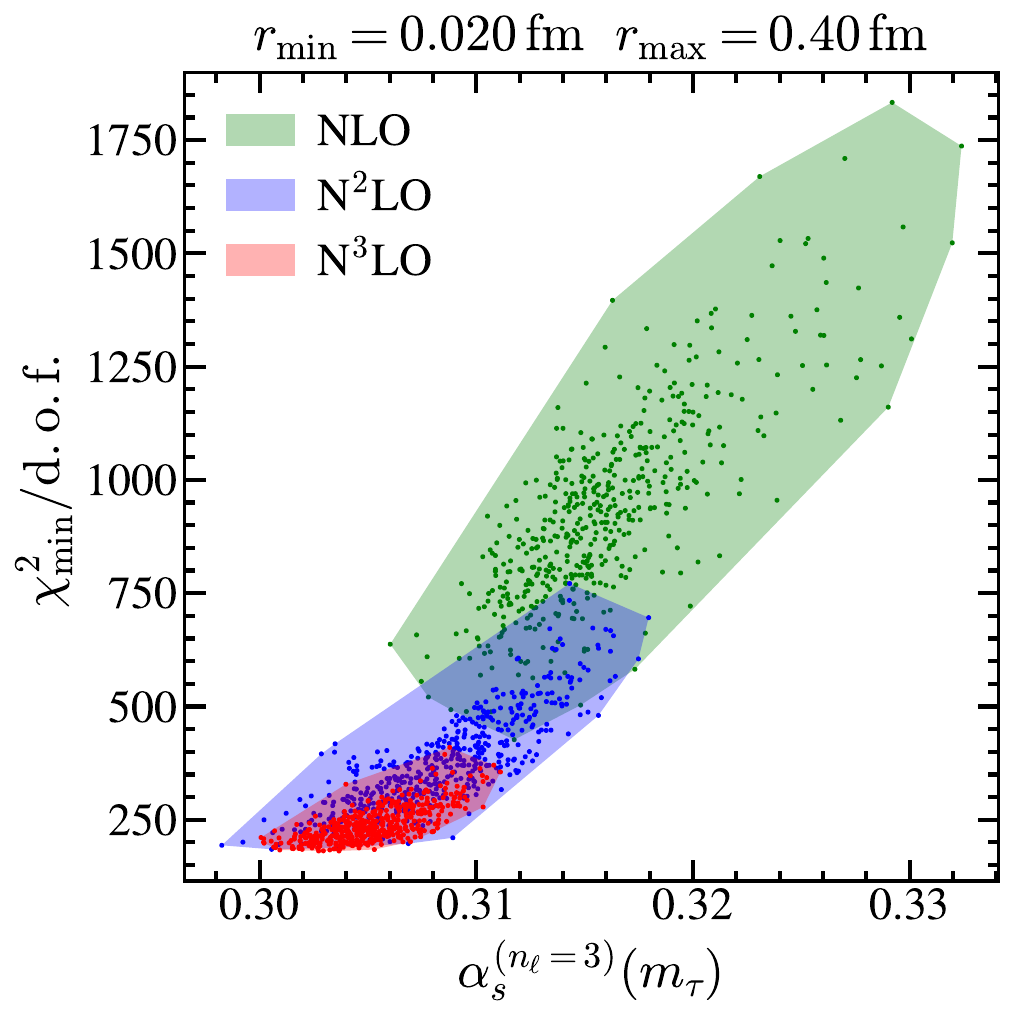}
\caption{}\label{fig:alphaChi2}
\end{subfigure}

\caption{\label{fig:OyO} All panels: results for single-offset fits where $\lambda$ has been varied at NLO (green), N$^2$LO (blue), and N$^3$LO (red). Upper panels: dependence of $\alpha_s$ (a) and $A$ (b) with $r_{\rm max}$ for a fixed value of $r_{\rm min}$. Lower panels: Distribution of best-fit points for the dataset defined by $r_{\rm min}=0.02\,$fm and $r_{\rm max}=0.4\,$fm. Panels (c) and (d) correspond to the $\alpha_s$-$A$ and $\alpha_s$-$(\chi_{\rm min}^2/{\rm d.o.f.})$ planes, respectively. On panel (c), ellipses represent $39\%$ confidence level and large thick colored dots indicate the ellipses' centers.}
\end{figure}
The last crucial feature our analysis must exhibit is order-by-order convergence and a reduction of the reduced $\chi^2_{\rm min}$ as more perturbative information gets added. For this exercise, as explained in Sec.~\ref{sec:pert}, we keep both the evolution for $\alpha_s$ and the MSR mass at the highest available order, five and four loops, respectively, and vary the $\lambda$ parameter as specified in Table~\ref{tab:profiles}. Moreover, we use single-offset fits. The results are shown in Fig.~\ref{fig:Chi2Range} and the two upper panels of Fig.~\ref{fig:OyO} as a function of $r_{\rm max}$ for fixed $r_{\rm min}=0.025\,$fm in the case of $\alpha_s$ and $A$, and for $\chi^2_{\rm min}$ as an average over the six values of $r_{\rm min}$ analyzed in this article. We observe that the reduced $\chi^2_{\rm min}$ indeed decreases when more perturbative information is included for any value of $r_{\rm max}$ in both kind of fits. The order-by-order convergence for $A$ is also excellent for all datasets, showing a reduction of the uncertainty while keeping the higher-order error bands contained within the lower-order ones. The same reduction of uncertainty is observed for $\alpha_s$, but bands are properly nested for $r_{\rm max}<0.5\,$fm only. Hence, to be on the safe side, in our final analysis we consider only $r_{\rm max}\leq 0.45\,$fm what, together with our previous requirement, fixes the range of values entering our final analysis to $0.35\,{\rm fm}\leq r_{\rm max} \leq0.45\,$fm. In Fig.~\ref{fig:OyO} we present more detailed results for the dataset defined by $r\in[0.02, 0.4]\,{\rm fm}$,
showing the scattered points in the \mbox{$\alpha_s$-$A$} and $\alpha_s$-$(\chi_{\rm min}^2/{\rm d.o.f.})$ planes, respectively. One observes a somewhat poor density of NLO points, due to an insufficient number of random profiles in the scan. Hence, the corresponding perturbative uncertainty could be slightly underestimated, being necessary more than the 500 currently used which, on the other hand, appear to suffice at the two highest orders. Hence, we are not concerned with this suboptimal behavior and still consider our perturbative uncertainties at N$^2$LO and N$^3$LO well estimated.

\begin{figure}[t!]
\centering
\begin{subfigure}{0.486\textwidth}
\centering
\includegraphics[width=\textwidth]{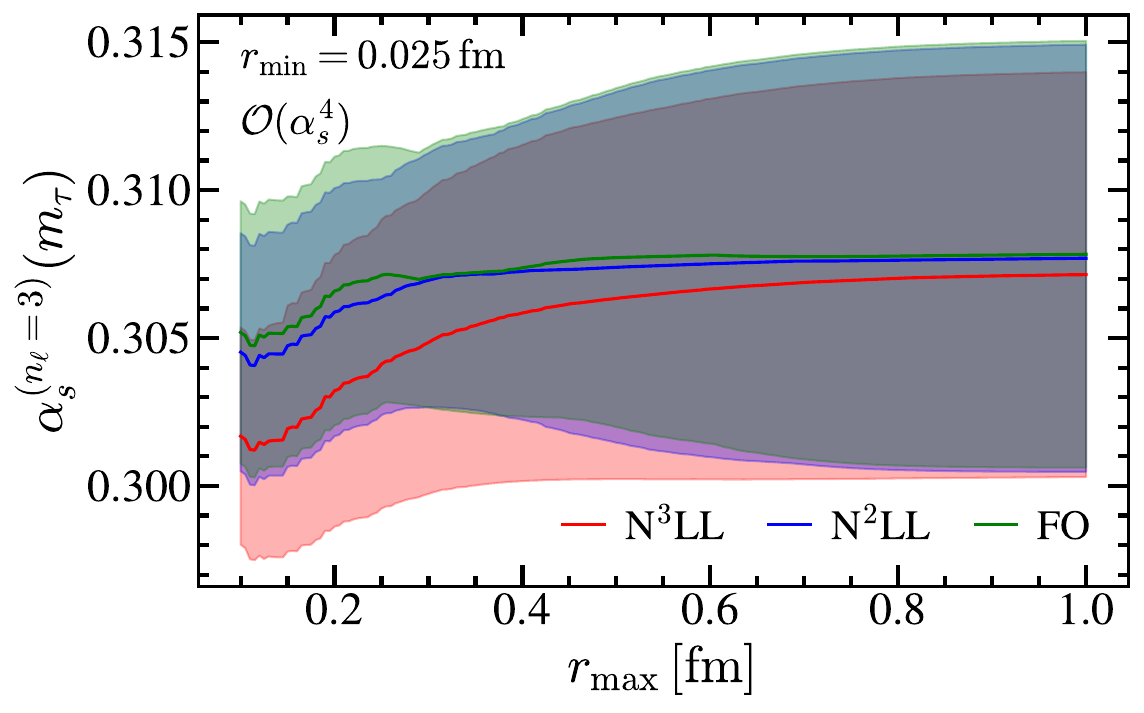}
\caption{}\label{fig:UsoftAlpha}
\end{subfigure}
\hfill
\begin{subfigure}{0.475\textwidth}
\centering
\includegraphics[width=\textwidth]{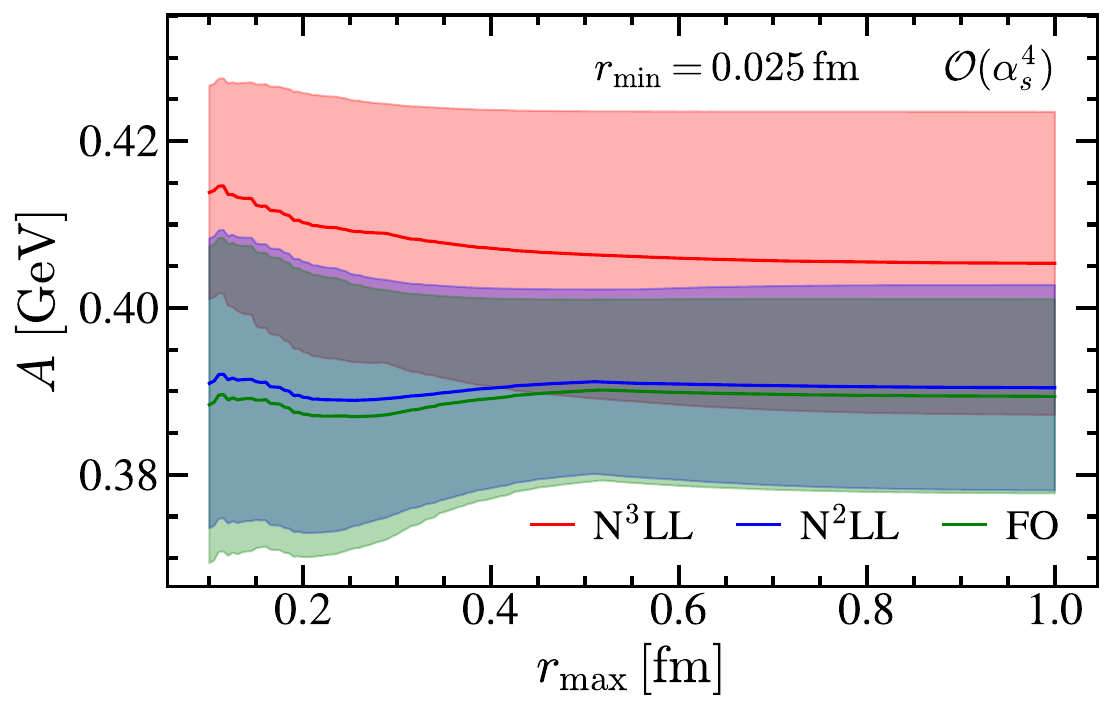}
\caption{}\label{fig:UsoftA}
\end{subfigure}

\caption{\label{fig:Usoft}
Dependence of $\mathcal{O}(\alpha_s^4)$ single-offset fits for $\alpha_s$ (left panel) and $A$ (right panel) with $r_{\rm max}$ fixing $r_{\rm min}=0.025\,$fm for various levels of ultrasoft resummation: fixed-order (green) N$^2$LL (blue) and N$^3$LL (red).}
\end{figure}

The effect of ultrasoft resummation is related to the previous study, that is, perturbative convergence. It is expected that resummation of potentially large logs makes perturbation theory more stable and theoretical errors smaller. To that end, we carry out N$^3$LO fits, which to avoid confusion correspond to our $\mathcal{O}(\alpha_s^4)$-precise prediction, turning completely off ultrasoft resummation (hence referred to as FO) and including it only at leading order or N$^2$LL, and compare the outcome to our default N$^3$LL analysis. This can be visualized in the two panels of Fig.~\ref{fig:Usoft}, showing a surprisingly big effect: while the outcome of FO and N$^2$LL fits are very similar for both $\alpha_s$ and $A$, the results at N$^3$LL for the strong coupling are roughly a sigma smaller for fits dominated by small distances, with a slightly smaller uncertainty band. This effect dilutes as $r_{\rm max}$ grows and has a marginal impact for the values used in our final analysis. In this moderate-$r_{\rm max}$ region, the uncertainties at N$^3$LL are slightly larger than for the rest of resummation setups. Even if our final prediction for the strong coupling will be barely affected, we believe it is worth further investigation in the future, where having the next resummation order available would be of great help.

\begin{figure}[t!]
\centering
\begin{subfigure}{0.488\textwidth}
\centering
\includegraphics[width=\textwidth]{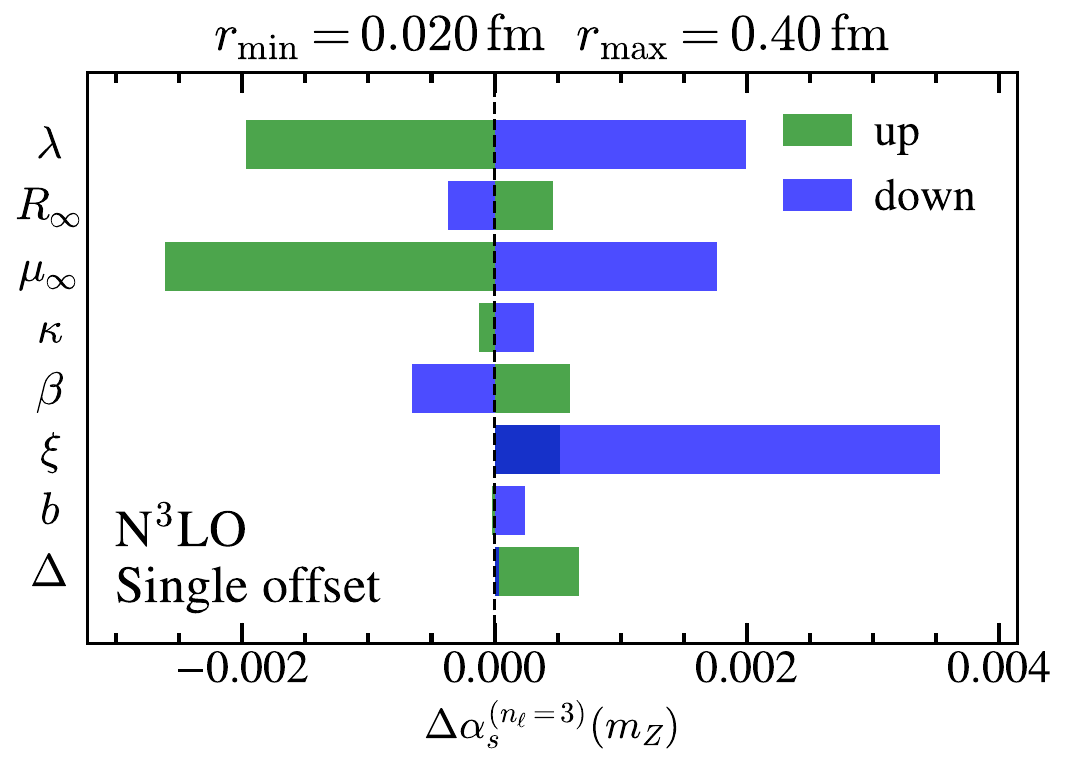}
\caption{}\label{fig:UpDown}
\end{subfigure}
\hfill
\begin{subfigure}{0.48\textwidth}
\centering
\includegraphics[width=\textwidth]{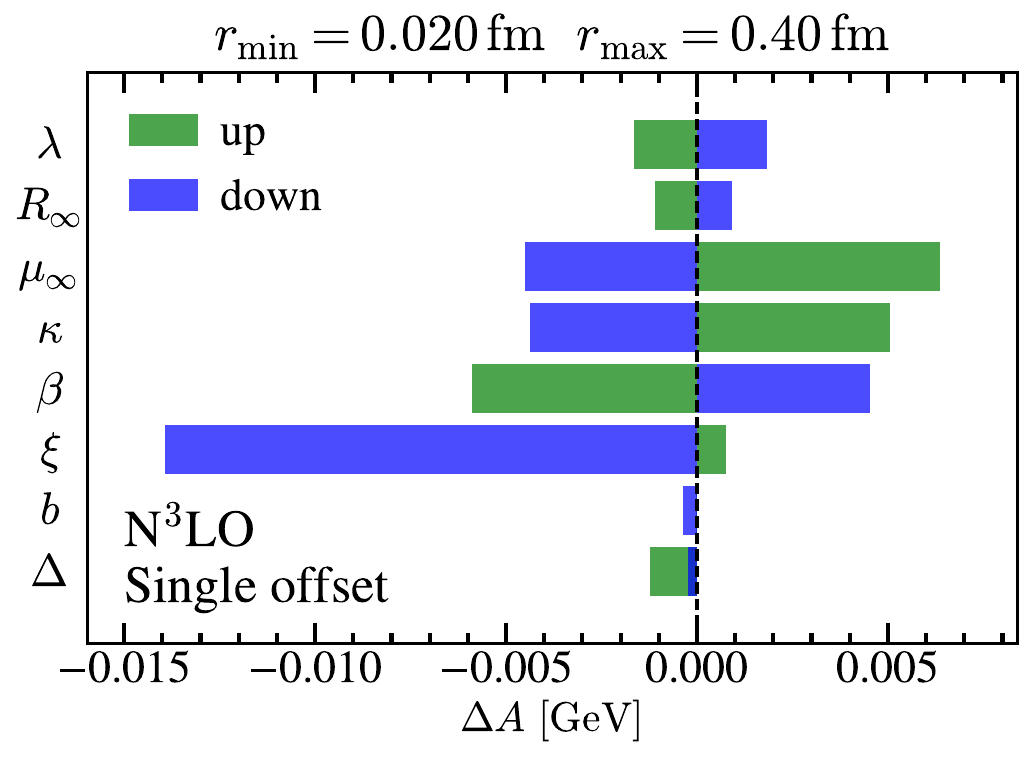}
\caption{}\label{fig:UpDownA}
\end{subfigure}

\caption{\label{fig:DownUp}
Shifts in central values caused by varying up or down one profile parameter at a time for N$^3$LO single-offset fits for $\alpha_s$ (left panel) and $A$ (right panel) using a dataset defined by $r_{\rm min}=0.02\,$fm and $r_{\rm max}=0.4\,$fm.}
\end{figure}

As explained in Sec.~\ref{sec:profiles}, perturbative uncertainties are estimated randomly varying a set of parameters that specify the dependence of renormalization scales with the distance~$r$. It is instructive to inspect the effect that each parameter has on the best-fit values. To that end, we do a set of fits in which all parameters but one are kept fixed to their default value, c.f.~Table~\ref{tab:profiles}, evaluating the floating parameter up and down to the limits of its variation range. We do this for all profile parameters, and compare the outcome to those best-fit values obtained from a fit in which all parameters stay at their respective defaults. We carry out this analysis for the dataset defined by $r\in[0.02, 0.4]\,\rm fm$, and the two panels of Fig.~\ref{fig:DownUp} show that, as expected, $\xi$ has the largest effect in the total uncertainty for both $\alpha_s$ and $A$. For the strong coupling, it is followed by $\lambda$ and $\mu_\infty$, that appear to be equally relevant, while the rest of parameters seem to have a much smaller effect. It turns out that $\kappa$, $\beta$ and $\mu_\infty$ give a sizable contribution to the uncertainty on $A$, while the impact of the remaining parameters is marginal. Whereas the effect of most parameters is roughly symmetric, for $\xi$ is notoriously asymmetric, being even one-sided on $\alpha_s$. We have investigated this issue and found that the dependence of the strong coupling's best-fit value with $\xi$ is not monotonic in its variation range. Using (in natural units) $\xi\in[1.5,2]$ makes the variation symmetric, but we do not find justified choosing such a non-standard variation, particularly when histograms like that of Fig.~\ref{fig:discard} showcase a distribution symmetric with respect to its mean, which is very similar to the best-fit value obtained with the default profiles. Finally, whereas for this dataset the theoretical error estimated by our random scan yields $\Delta_{\rm th}\alpha_s=0.0056$, adding in quadrature the single-parameter errors one gets $\Delta^{\rm up}_{\rm down}\alpha_s=0.0038\sim 2(\Delta_{\rm th}\alpha_s)/3$, showing that the scan is more conservative.\footnote{For this exercise we symmetrize the shifts of each parameter simply taking the average of absolute values.}

\begin{figure}[t!]
\centering
\begin{subfigure}{0.474\textwidth}
\centering
\includegraphics[width=\textwidth]{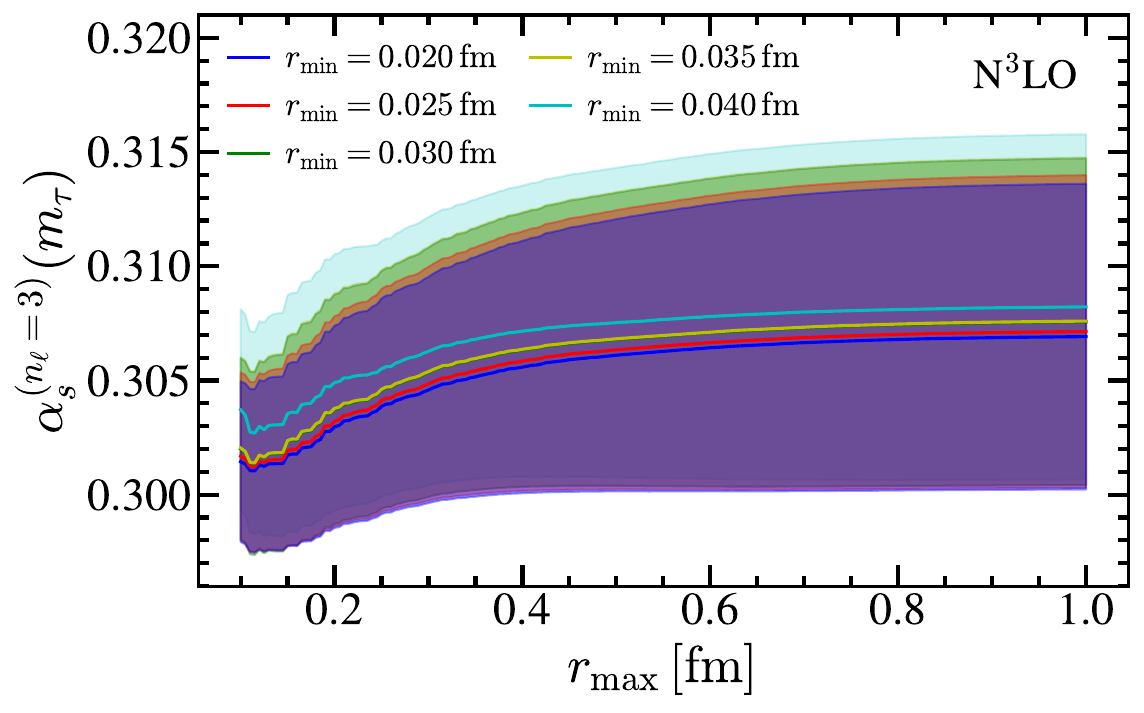}
\caption{}\label{fig:RangePlot}
\end{subfigure}
\hfill
\begin{subfigure}{0.495\textwidth}
\centering
\includegraphics[width=\textwidth]{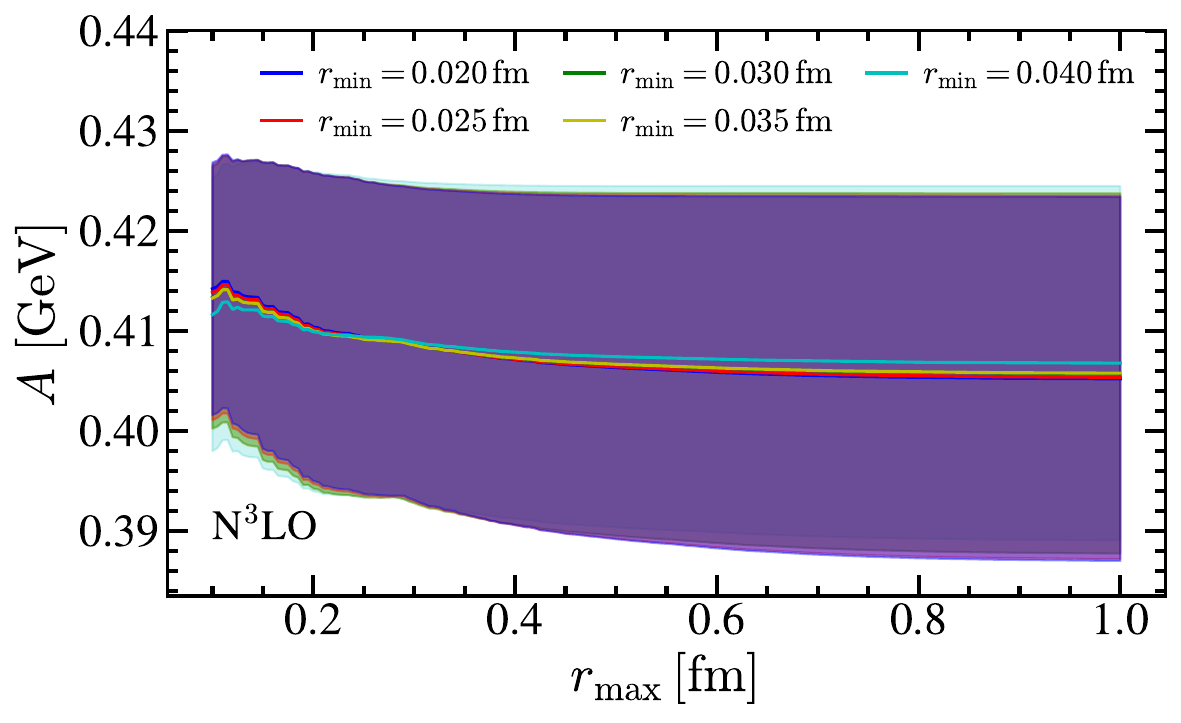}
\caption{}\label{fig:RangePlotA}
\end{subfigure}
\hfill
\begin{subfigure}{0.487\textwidth}
\centering
\includegraphics[width=\textwidth]{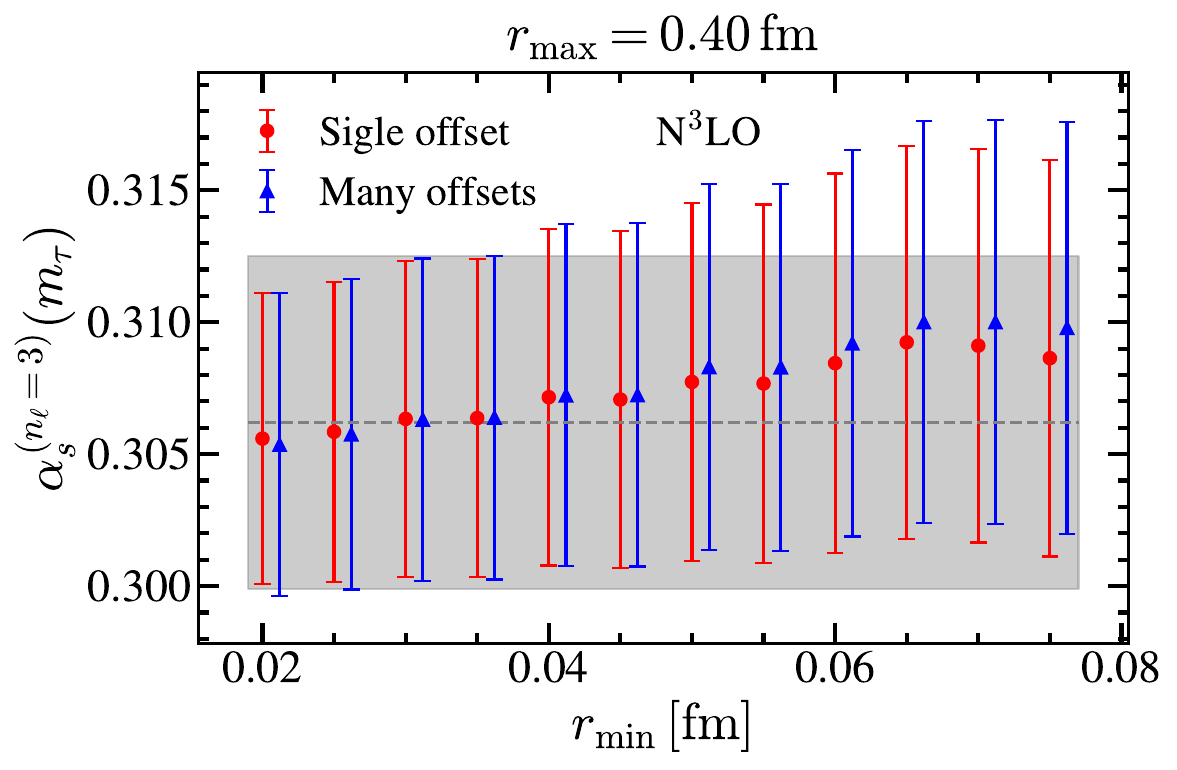}
\caption{}\label{fig:RminAlpha}
\end{subfigure}
\hfill
\begin{subfigure}{0.474\textwidth}
\centering
\includegraphics[width=\textwidth]{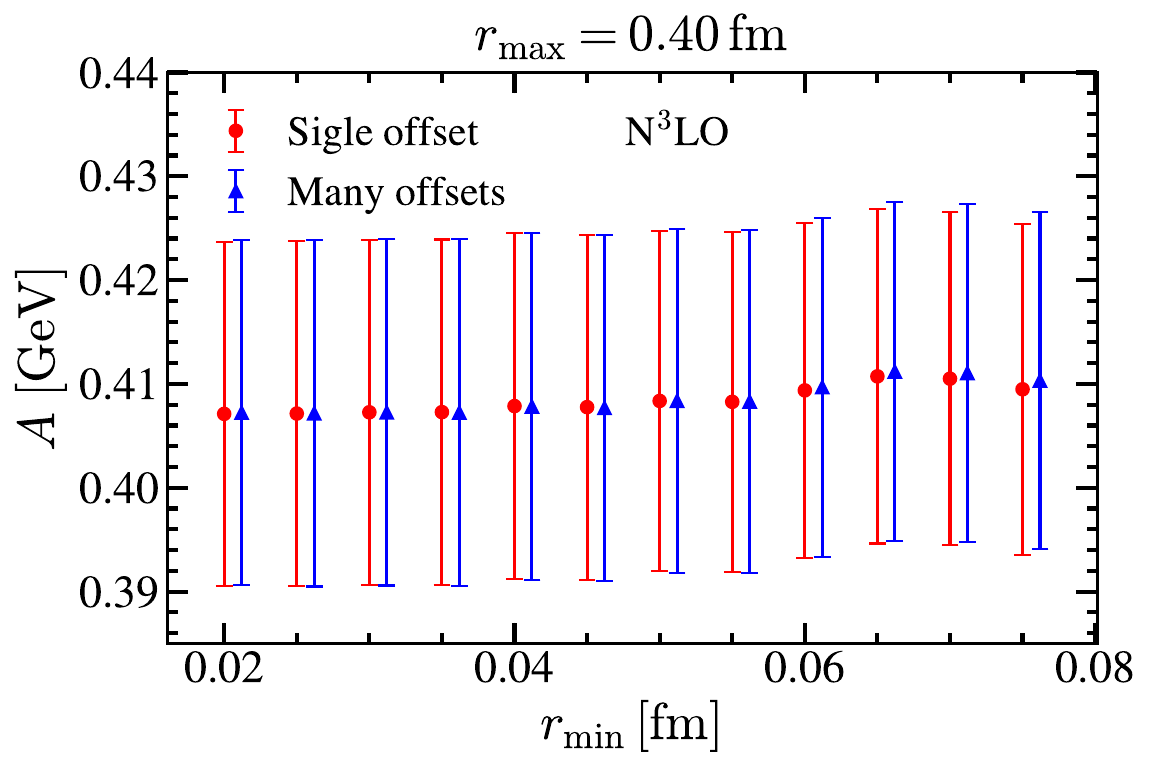}
\caption{}\label{fig:RminA}
\end{subfigure}
\caption{\label{fig:PlotRange}
Results for N$^3$LO fits for $\alpha_s$ (left panels) and $A$ (right panels). While strong-coupling uncertainties include perturbative and lattice errors added in quadrature, for the offset we only display the (overly dominant) theoretical error. Upper panels: Dependence of single-offset results with $r_{\rm max}$ for $r_{\rm min}=\{0.02, 0.025, 0.03, 0.035, 0.04\}\,$fm in blue, red, green, yellow and cyan, respectively. Lower panels: Dependence of single- (red) and many-offset (blue) results with $r_{\rm min}$ for $r_{\rm max}=0.4\,$fm. The gray translucent band in (c) indicates our final result for the strong coupling.}
\end{figure}
\section{Final Results}\label{sec:final}
Once we have verified that our fits are robust and well behaved, we present single-offset fit results at N$^3$LO for all datasets considered, that is, for various values of $r_{\rm min}$ and $r_{\rm max}$ as shown in the left (for $\alpha_s$) and right (for $A$) panels of Fig.~\ref{fig:PlotRange}. To avoid clutter, we do not show the results for $r_{\rm min}=0.045\,$fm, which are very similar to those with $r_{\rm min}=0.04\,$fm and do not add much to the discussion. We find that the offset (and its perturbative uncertainty) is rather insensitive to the value of $r_{\rm min}$ and exhibits a mild dependence with $r_{\rm max}$, decreasing for larger values of the upper limit defining the dataset. The strong coupling depends on $r_{\rm min}$ in a stronger way, yielding lower and more precise values of $\alpha_s$ for smaller minimal radii. The same observation holds for the upper limit of the dataset: larger central values and uncertainties for higher $r_{\rm max}$. One can also observe that the results for $r_{\rm min}=0.030$\,fm and $0.035\,$fm are identical, and that those of $r_{\rm min}=0.040$\,fm and $0.055\,$fm are quite similar for $r_{\rm max}>0.3\,$fm. For our final analysis, we will consider the range $0.02\leq r_{\rm min}\leq 0.04\,$fm.

In Ref.~\cite{Ayala:2020odx} lattice data with $r<0.0697\,$fm was discarded arguing that, according to Ref.~\cite{Bazavov:2019qoo}, at such short distances tree-level improvement on lattice data was not enough to make discretization errors smaller than statistical uncertainties. We believe that having a much larger dataset makes us less sensitive to such details, and to that end we have explored the dependence of fit results with a larger set of $r_{\rm min}$ values keeping $r_{\rm max}=0.4\,$fm, although identical conclusions are drawn for different $r_{\rm max}$. The outcome of this exercise is shown in Figs.~\ref{fig:RminAlpha} and \ref{fig:RminA} for the strong coupling and offset, respectively. We do not observe any abrupt change around $r_{\rm min}=0.07\,$fm that justifies discarding results below this value but, on the other hand, note an increase of the uncertainty caused by a reduction of the dataset's size. Therefore, for our final analysis we will use the range of $r_{\rm min}$ values stated in the previous paragraph.

\begin{figure}[t!]
\centering
\begin{subfigure}{0.48\textwidth}
\centering
\includegraphics[width=\textwidth]{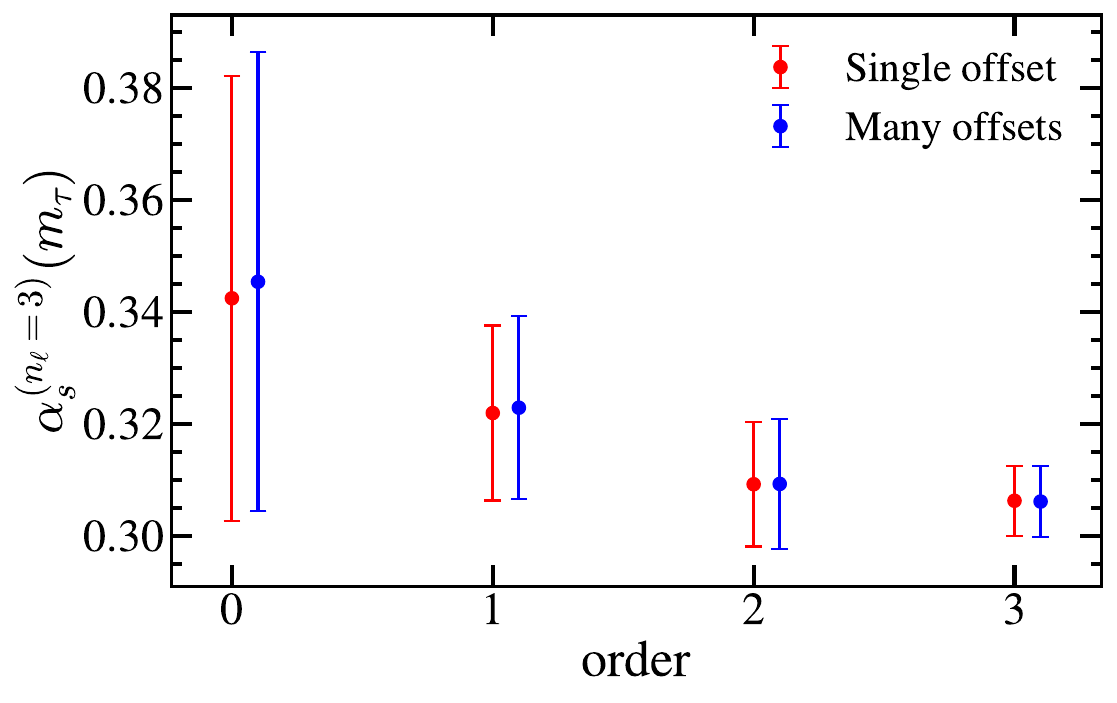}
\caption{}\label{fig:FinalAlphaOrders}
\end{subfigure}
\hfill
\begin{subfigure}{0.49\textwidth}
\centering
\includegraphics[width=\textwidth]{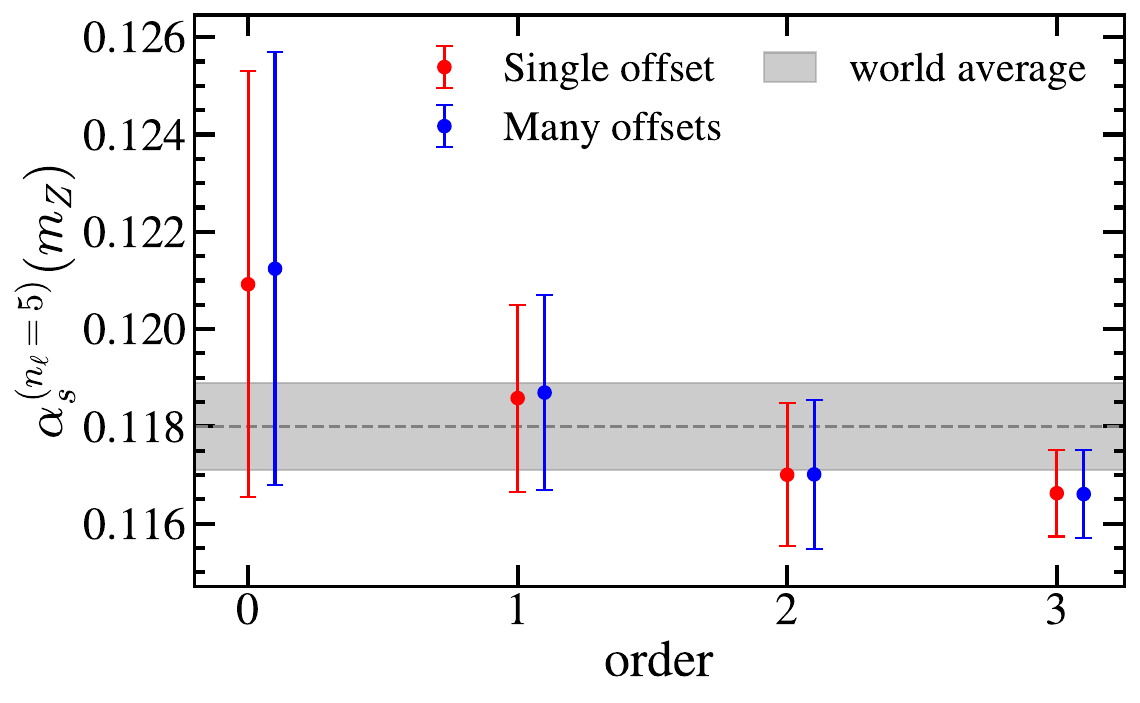}
\caption{}\label{fig:FinalAmZOrders}
\end{subfigure}

\caption{\label{fig:FinalOrders}
Final results for the strong coupling with three active flavors at the tau lepton mass (left panel) and with five active flavors at the $Z$-boson mass (right panel) at LO, NLO, N$^2$LO and N$^3$LO. In the right panel we show the current world average with a dashed gray line and a translucent band with the same color.}
\end{figure}

To obtain our final value we collect fit results from datasets fulfilling our requirements \mbox{$r_{\rm min}\in[0.02,0.04]\,$fm} and $r_{\rm max}\in[0.35,0.45]\,$fm, consisting on a set with 105 elements. The weighted average of best-fit values is adopted as our final central value, while the regular average of perturbative and lattice errors are taken as the corresponding final uncertainties. We add half of the difference between the largest and smaller central value in the set as an additional ``dataset'' incertitude. With this prescription we obtain at N$^3$LO order
\begin{align}\label{eq:final3}
\!\alpha_s^{(n_\ell=3)}(m_\tau) ={}& 0.3062\pm 0.0061_{\rm th}\pm0.00001_{\rm lat}\pm 0.0012_{\rm set}\pm 0.0014_{r_1} \\
={} &0.3062\pm 0.0063_{\rm tot}\,.\nonumber
\end{align}
where we have also accounted for the uncertainty on $r_1$. This number compares well with the regular average of all single-offset N$^3$LO fits $\langle\alpha_s^{(n_\ell=3)}(m_\tau)\rangle=0.3057$, and has an uncertainty twice as large as the standard deviation of the full sample, $\sigma_{\alpha_s}=0.0028$. Turning off ultrasoft resummation one gets \mbox{$\alpha_s^{(n_\ell=3)}(m_\tau)=0.3077(58)$}, where the error has shrank due to smaller perturbative and dataset incertitudes. The result using $r_1$ from Ref~\cite{Bazavov:2014pvz} is \mbox{$\alpha_s^{(n_\ell=3)}(m_\tau)=0.3100(62)$}, while with multiple offsets we get \mbox{$\alpha_s^{(n_\ell=3)}(m_\tau)=0.3061(65)$}. The result of this combined analysis at various orders is shown in Fig.~\ref{fig:FinalAlphaOrders}, where the outcome of using multi-offset fits is also displayed. Very nice convergence is showcased and, interestingly, the agreement between the two types of fits becomes better at higher orders. The result in Eq.~\eqref{eq:final3} can be translated to the QCD scale with three active flavors:
\begin{equation}\label{eq:LQCD}
\Lambda_{\rm{QCD}}^{(n_\ell=3)}= 318\pm12\,{\rm MeV}\,,
\end{equation}

Finally, we have to evolve our result to the $Z$-pole, and in order for that the charm and bottom thresholds must be crossed to match the strong coupling from three to four, and from four to five active flavors, respectively. We use the five-loop beta function and \mbox{four-loop} matching conditions implemented in \texttt{REvolver}. To quantify the errors caused by the matching procedure we consider two sources: a)~the uncertainty on the charm and bottom masses, b)~varying the matching scales. For the former (which turn out to be tiny) we take the PDG world averages~\cite{ParticleDataGroup:2024cfk}:
\begin{table}[t!]
\centering
\begin{tabular}{ c cccc } \toprule & LO & NLO & N$^2$LO & N$^3$LO \\\midrule
$\alpha_s^{(n_\ell=3)}(m_\tau)$ & $0.3417 \pm 0.0394$ & $0.3217 \pm 0.0155$ & $0.3091 \pm 0.0110$ & $0.3062 \pm 0.0062$\\[0.1cm]
$\alpha_s^{(n_\ell=5)}(m_Z)$ & $0.1208 \pm 0.0044$ & $0.1185 \pm 0.0019$ & $0.1170 \pm 0.0015$ & $0.1166 \pm 0.0009$ \\\bottomrule
\end{tabular}
\caption{Results for the strong coupling with three flavors at $m_\tau$ (middle line) and five flavors at $m_Z$ (bottom line) from fits to lattice data at various orders. The uncertainty is the sum in quadrature of the individual sources.\label{tab:loops}}
\end{table}
\begin{equation}
\overline m_c = 1.28\pm0.025\,{\rm GeV}\,,\qquad \overline m_b= 4.18\pm0.03\,{\rm GeV}\,.
\end{equation}
where the notation $\overline m_q\equiv\overline m_q(\overline m_q)$ is used. For the matching scales we consider \mbox{$\mu_c=2^{1\pm1}\overline m_c$} and $\mu_b=2^{\pm1}\overline m_b$. While for the bottom we use the standard factors of $2$ and $1/2$, this cannot be implemented for the charm given its small mass. Instead, we use the standard factors of two on twice the charm mass. Adding all errors in quadrature we obtain
\begin{align}\label{eq:amZ}
\alpha_s^{(n_\ell=5)}(m_Z) ={} &0.1166\pm 0.0008_{\rm th}\pm 0.0001_{\rm set}\pm0.0002_{r_1}\pm 0.0003_{\mu_c}\pm0.0002_{\mu_b} \\
={} &0.1166\pm 0.0009_{\rm total}\,,\nonumber
\end{align}
where we have shown only those errors larger or equal than $10^{-5}$. The lattice error amounts to $2\times10^{-6}$ while that of the \{charm, bottom\} mass is $\{1.4,4.3\}\times 10^{-5}$. If the matching at the charm threshold is performed at $\mu_c=\overline{m}_c$ the central value becomes \mbox{$\alpha_s^{(n_\ell=5)}(m_Z)=0.1170$}. Turning off ultrasoft resummation or keeping it at N$^2$LL one gets $\alpha_s^{(n_\ell=5)}(m_Z)=0.1168(8)$, essentially identical to Eq.~\eqref{eq:amZ}. If the R-evolution parameter is fixed to $\lambda=1$, the result reads $\alpha_s^{(n_\ell=5)}(m_Z)=0.1161(8)$, still compatible within less than a sigma with Eq.~\eqref{eq:amZ}. The result using $r_1$ from Ref~\cite{Bazavov:2014pvz} is $\alpha_s^{(n_\ell=5)}(m_Z)=0.1171(9)$ and the outcome of employing multi-offset fits is identical to Eq.~\eqref{eq:amZ} within the quoted decimal places. Finally, if we use the six ensembles with the shortest lattice spacing (that is, the sets used in Ref.~\cite{Bazavov:2019qoo}), we also obtain a result identical to that in \eqref{eq:amZ} within four significant figures. If only the three datasets with the largest $\beta$ are used for the fits, the central value stays unchanged but the total uncertainty grows to $0.0010$, and if only data with the smallest lattice spacing is used $\alpha_s^{(n_\ell=5)}(m_Z) =0.1165(10)$ is obtained. A summary of results for various datasets and types of fits is shown in Tables~\ref{tab:dataset} and \ref{tab:changes}.

The result of this final analysis at various orders is shown in Fig.~\ref{fig:FinalAmZOrders}, where a comparison to the current world average $0.1180(9)$ is also displayed. They can also be viewed in numerical form in Table~\ref{tab:loops}. Our final result has the same uncertainty as the PDG average and is $1.1$-$\sigma$ lower.

\section{Comparison with Previous Analyses}\label{sec:comparison}
It is instructive to compare our final results with recent determinations which also use the static energy, but to do so on equal footing it is crucial to translate the various results to a common value of $r_1$ and to ensure the same procedure is being used to RG evolve from three flavors to the $Z$-pole.

In Ref.~\cite{Ayala:2020odx}, using a dataset defined by $0.070\,{\rm fm}\leq r\leq 0.098\,{\rm fm}$ consisting on data taken from \cite{Bazavov:2019qoo} with $\beta=8.4$ and totaling 8 points, the value $\alpha_s^{(n_\ell=5)}(m_Z) =0.1181(9)$ was found, with the same uncertainty as in Eq.~\eqref{eq:final3}. Ref.~\cite{Ayala:2020odx} also provides \mbox{$\alpha_s^{(n_\ell=3)}(m_\tau)=0.3151(65)$} and \mbox{$\LQCD^{(n_\ell=3)}=338(12)\,$MeV}. If we employ the same dataset (and $r_1$) for carrying out fits within our theoretical setup we obtain $\alpha_s^{(n_\ell=3)}(m_\tau)=0.3123(80)$, hence it appears our estimate of uncertainties is more conservative. Using \texttt{REvolver} and their value for the strong coupling we obtain $\LQCD^{(n_\ell=3)}=335(12)\,$MeV, well within errors with their result. To obtain the quoted value for $\alpha_s^{(n_\ell=5)}(m_Z)$ from their $\alpha_s^{(n_\ell=3)}(m_\tau)$, the charm matching scale must be set to \mbox{$\mu_c=\overline{m}_c$}, which is half of our choice. Since Ref.~\cite{Ayala:2020odx} uses the `old' value $r_1=0.3106(17)\,$fm from Ref.~\cite{Bazavov:2014pvz}, to compare on equal footing to our findings we convert their results to $\LQCD^{(n_\ell=3)}=330(12)\,$MeV and $\alpha_s^{(n_\ell=3)}(m_\tau)=0.3112(65)$, where the latter has been obtained with \texttt{REvolver}. Evolving the strong coupling using our prescription to match three to four flavors we find $\alpha_s^{(n_\ell=5)}(m_Z)=0.1173(9)$ which is compatible within $0.55$-$\sigma$.

\begin{table}[t!]
\centering
\begin{tabular}{ c cccc } \toprule & full & $\beta\geq 7.737$ & $\beta\geq 8$ & $\beta=8.4$ \\\midrule
$\alpha_s^{(n_\ell=3)}(m_\tau)$ & $0.3062 \pm 0.0062$ & $0.3058 \pm 0.0060$ & $0.3059 \pm 0.0069$ & $0.3050 \pm 0.0071$\\[0.1cm]
$\alpha_s^{(n_\ell=5)}(m_Z)$ & $0.1166 \pm 0.0009$ & $0.1166 \pm 0.0009$ & $0.1166 \pm 0.0010$ & $0.1165 \pm 0.0010$ \\\bottomrule
\end{tabular}
\caption{Same as Table.~\ref{tab:loops} for fits to various lattice datasets. \label{tab:dataset}}
\end{table}

The latest HotQCD analysis in Ref.~\cite{Bazavov:2019qoo} from 2019 uses data constrained to the interval \mbox{$a\leq r \leq 0.073\,$fm} (according to Table~III therein, 17 d.o.f.\ which correspond to 19 data points) and quotes \mbox{$\alpha_s^{(n_\ell=5)}(m_Z) = 0.1166(8)$} using $r_1$ from \cite{Bazavov:2014pvz}, where we have symmetrized their uncertainties to ease comparison. The value $\LQCD^{(n_\ell=3)}=314(12)\,$MeV is also provided, which corresponds to \mbox{$\alpha_s^{(n_\ell=3)}(m_\tau)=0.3039(62)$}. Using our theoretical setup on the same dataset (and $r_1$) we obtain $\alpha_s^{(n_\ell=3)}(m_\tau)=0.3114(44)$.
Evolving their 3-flavor strong coupling to the $Z$-pole with the same prescription as that used in Ref.~\cite{Ayala:2020odx} reproduces the result for
$\alpha_s^{(n_\ell=5)}(m_Z)$ quoted in Ref.~\cite{Bazavov:2019qoo}. Converting to the updated $r_1$ we obtain \mbox{$\LQCD^{(n_\ell=3)}=307\,$MeV} which corresponds to $\alpha_s^{(n_\ell=3)}(m_\tau)=0.3002$. After RG evolving to the \mbox{$Z$-pole} with our prescription to cross the charm threshold we obtain $\alpha_s^{(n_\ell=5)}(m_Z) =0.1158(8)$, which is compatible with our Eq.~\eqref{eq:amZ} within $0.63$-$\sigma$.

The JQCD collaboration has also determined the strong coupling in Ref.~\cite{Takaura:2018vcy}, obtaining $\LQCD^{(n_\ell=3)}=334(22)\,$MeV from data in the interval $r\in[0.05, 0.4]\,{\rm fm}$ (totaling $\sim 14$ data points), which corresponds to $\alpha_s^{(n_\ell=3)}(m_\tau)=0.3144(118)$ that, after evolving to the $Z$-pole with \mbox{$\mu_c=\overline{m}_c$}, yields $\alpha_s^{(n_\ell=5)}(m_Z)=0.1180(15)$, in agreement with the result quoted in Eq.~(3.26) therein. The value \mbox{$r_1=0.311\,$fm} was employed, but converting to $r_1=0.3037¡\,$fm and evolving to the $Z$-pole using \mbox{$\mu_c=2\overline{m}_c$} we obtain $\alpha_s^{(n_\ell=5)}(m_Z)=0.1167(15)$, in complete agreement with Eq.~\eqref{eq:amZ} albeit less precise.

\section{Conclusions}\label{sec:conclusions}
The perturbative QCD prediction for the static energy has been improved through various theoretical innovations. The $u=1/2$ renormalon has been removed expressing the pole mass in terms of the low-scale short-distance MSR scheme at the fixed scale $R_0$. Since a proper renormalon removal requires expressing the subtraction series in terms of the same renormalization scale $\mu_s$ as that in the static potential, R-evolution has been used to sum up the associated large logarithms of $R_0/\mu_s$ at N$^3$LL. Ultrasoft logarithms, which are unrelated to the asymptotic behavior of the potential, are also summed up to N$^3$LL precision solving the corresponding pNRQCD RGE equation. Finally, radius-dependent renormalization scales ---\,dubbed profile functions\,--- have been used to extend the validity of perturbation theory up to distances around $r=0.5\,$fm. The final theoretical expression for the R-improved static energy at N$^3$LO includes up to $\mathcal{O}(\alpha_s^4)$ fixed-order corrections plus the two kind of resummation just described.

\begin{table}[t!]
\centering
\begin{tabular}{ c cccc } \toprule & $\mu_c=\overline m_c$ & $\lambda=1$ & FO & many offsets \\\midrule
$\alpha_s^{(n_\ell=3)}(m_\tau)$ & $0.3062 \pm 0.0062$ & $0.3025 \pm 0.0049$ & $0.3077 \pm 0.0056$ & $0.3061 \pm 0.0064$\\[0.1cm]
$\alpha_s^{(n_\ell=5)}(m_Z)$ & $0.1170 \pm 0.0009$ & $0.1161 \pm 0.0007$ & $0.1168 \pm 0.0008$ & $0.1166 \pm 0.0009$ \\\bottomrule
\end{tabular}
\caption{Same as Table.~\ref{tab:loops} for from various types of fits to lattice data. \label{tab:changes}}
\end{table}

We have made a thorough investigation of the perturbative properties of our theoretical result, confirming that the pole-scheme static energy lacks the accuracy necessary to carry precision studies. Concerning R-evolution, a numerical analysis has revealed that using the default parameter $\lambda=1$ can lead to biased results since, for $n_\ell=3$ and at N$^3$LL, it turns out to be extremely close to where the minimum with respect to this parameter lays. Hence, it is necessary to use a non-standard value for $\lambda$ and to vary that parameter to generate an additional uncertainty as a penalty for this uncommon behavior. We have also discarded subtraction schemes other than the MSR mass since they show a substantial sensitivity to ultrasoft effects and hence deteriorate the accuracy of our theoretical prediction. The parameters defining the profile functions are varied within certain fixed ranges to ensure one respects the hierarchy $\mu_s>\mu_{\rm us}$ at small distances, has perturbative stability ---\,absence of outliers and scales sufficiently larger than $\Lambda_{\rm{QCD}}$\,---, achieves order-by-order convergence, and renders a proper estimate of theoretical uncertainties. This requires varying the argument of perturbative fixed-order logs at small distances in the range $[0.7,2.2]$. We have also checked that
$\mathcal{O}(\alpha_s^k)$ corrections are smaller in absolute size than $\mathcal{O}(\alpha_s^i)$ ones if $k>i$ for the R-improved prediction. Finally, we observe that upon adding up all contributions, the characteristic Cornell-potential-like linear rising behavior at large $r$ is reproduced.

To determine the strong coupling, we compare our improved pQCD prediction with lattice simulations. Our analysis uses results from the HotQCD collaboration, which span a broad range of distances through a plethora of ensembles. We use their updated result for $r_1$, the parameter that sets the length and energy scales, and fit up to distances with $r\sim 0.45\,$fm. Given the large amount of lattice data, we have devised a method to cluster those into a new recombined dataset with a reduced number of points which is statistically compatible with the original set. For our final analysis, given that we have enough computer resources, the full HotQCD dataset is used. Along with $\alpha_s$ one must fit for the energy offsets, that is, the additive constants necessary to match pQCD and lattice predictions. While in principle the various HotQCD ensembles have been tuned to share, within uncertainties, a common origin of energies, given the high accuracy of our theoretical prediction we consider both fitting for a unique and multiple offsets. A comparison of both approaches reveals that their respective outcomes are very similar if data with $r\geq0.35\,$fm are included, and differ for more restricted datasets. Therefore, for our final analysis we consider only sets where the difference is marginal, which incidentally happen to diminish the (otherwise large) effect of N$^3$LL ultrasoft resummation. Good convergence among fits at one-, two- and three-loop accuracy is observed if data is restricted to $r\leq0.45\,$fm, and we use this condition to fully specify the datasets entering our final determination of $\alpha_s$ at N$^3$LO, which reads
\begin{equation}
\alpha_s^{(n_\ell=3)}(m_\tau) = 0.3062\pm 0.0063\,,\qquad
\alpha_s^{(n_\ell=5)}(m_Z) = 0.1166\pm 0.0009\,,
\end{equation}
where the quoted uncertainties include
all individual sources in the error budget. A caveat for our result is that we have used only uncorrelated statistical uncertainties as those are the only
publicly available. This has resulted in very large reduced $\chi^2_{\rm min}$ values. We cannot discard a shift in the central value once the full covariance matrix is used, along with an increase of the total uncertainty. We believe however these should be rather small and have checked that our fits are robust through the replica method. Our final result compares well with the current PDG world average and previous analyses using lattice data once we make sure the same value of $r_1$ and procedure to convert to the five-flavor scheme is employed.

Our work can be expanded in various ways. The most prominent is being able to use the full lattice covariance matrix including both correlations and systematic uncertainties. Once this information becomes public, we shall update our result. In another direction, we could subtract the $u=3/2$ renormalon to further improve our theoretical prediction, hopefully making our fits more precise and stable at larger distances. Finally, adapting the method of using theory nuisance parameters to construct a perturbative covariance matrix could lead to smaller reduced $\chi^2_{\rm min}$, along with a better assessment of the perturbative uncertainties.

\begin{acknowledgments}

The authors would like to thank P.~Petreczky and J.\,H.~Webber for providing us with HotQCD lattice data and for useful conversations.
This work was supported in part by the Spanish MICIU/AEI/10.13039/501100011033 grant No.\ PID2022-141910NB-I00, the JCyL grant SA091P24 under program EDU/841/2024, and the EU STRONG-2020 project under Program No.\ H2020-INFRAIA-2018-1, Grant Agreement No.\ 824093.
JMMV acknowledges funding from European Social Fund Plus, Programa Operativo de Castilla y León and Junta de Castilla y León through Consejería de educación.
\end{acknowledgments}
\\
\appendix

\section{The Static Potential in Momentum Space}\label{sec:AppPS}
In $d=3$ spacial dimensions, the angular integrals in Eq.~\eqref{eq:momV} can be performed trivially, such that the momentum-space soft static potential reads
\begin{equation}
\tilde V_{s}^{\rm soft}(q,\mu) = \frac{4 \pi}{q} \!\int_0^{\infty}\! \df r\,r \sin (q r) V^{\rm soft}_{\rm s}(r,\mu)\,.
\end{equation}
We
focus on the computation of $b_{i0}$ defined in Eq.~\eqref{eq:momV}, since the rest can be obtained by RG invariance. They are related to the position-space coefficients by
\begin{equation}
b_{i0} = \sum_{j = 0}^i a_{ij} I_j\,,\qquad {\rm with}\qquad I_j = j! \!\sum_{i = 0}^{ \left\lfloor \frac{j}{2} \right\rfloor} \kappa_{j - 2 i} \Bigl( \frac{\pi}{2} \Bigr)^{\!2i} \frac{(- 1)^i}{(2 i) !} \,,
\end{equation}
where the $\kappa_i$ coefficients have been defined in Eq.~\eqref{eq:kappa}. We find $ I_i=(1,0,-\pi^2/12, -2\zeta_3)$ for $i=(0,1,2,3)$.
We can also work out the inverse relation, that is, expressing the momentum-space static potential in terms of $\tilde V_{\rm QCD}$\,:
\begin{equation}
V_{s}^{\rm soft}(r,\mu) =\! \int \!\frac{\df^3 \vec{q}}{(2 \pi)^3} e^{i \vec{r} \cdot \vec{q}} \,\tilde{V}_{\rm s}^{\rm soft}(q,\mu)=
\frac{2}{r}\! \int_0^\infty\! \!\frac{\df q}{(2 \pi)^2} \,q\sin (q r)\tilde V_{s}^{\rm soft}(q,\mu)\,.
\end{equation}
Once again, we provide a relation to obtain only the $a_{i0}$ coefficients in terms of $b_{ij}$, since getting the rest is trivial:
\begin{equation}
a_{i0} = \sum_{j = 0}^i b_{ij} \hat I_j\,,\qquad {\rm with}\qquad
\hat I_j = j! \sum_{i = 0}^{ \left\lfloor \frac{j}{2} \right\rfloor} \kappa_{j - 2i} \Bigl( \frac{\pi}{2} \Bigr)^{\!2 i} \frac{(- 1)^{i+j}}{(2 i + 1) !} \,.
\end{equation}
We find $\hat I_i=(1,0,\pi^2/12, 2\zeta_3)$ for $i=(0,1,2,3)$. We finish this appendix providing the relation between the momentum-space static potential and the PS mass coefficients
\begin{equation}
c_i = \sum_{j = 0}^i j!\, b_{ij}\,.
\end{equation}

\section{\boldmath Fourier Transform of the Ultrasoft Potential in $d=3-2\varepsilon$}\label{sec:AppIR}
The $d=3-2\varepsilon$ Fourier transform of the ultrasoft potential can be easily performed provided one integrates first in $q$ and then in $\cos(\theta)$. The generic integral we need is
\begin{align}
\tilde{\mu}^{2 \varepsilon}\!\!\int\! \frac{\df^{3 - 2\varepsilon} \vec{q}}{(2 \pi)^{3 - 2 \varepsilon}} \biggl(\frac{q}{\mu}\biggr)^{\!\!\alpha} \frac{e^{i \vec{r} \cdot
\vec{q}}}{q^2} ={} &
\frac{4e^{\varepsilon\gamma_E}}{\Gamma (1 - \varepsilon)} \!\int \!\frac{\df q\, \df\! \cos (\theta)}{(4 \pi)^{2}}\biggl(\frac{q}{\mu}\biggr)^{\!\!\alpha-2\varepsilon}
[\sin(\theta)]^{-2\varepsilon} e^{i r q \cos (\theta)} \\
=\, & \frac{4 (\mu r)^{2 \varepsilon-\alpha}\Gamma (1+\alpha - 2 \varepsilon)e^{\varepsilon\gamma_E}}{(4 \pi)^{2}r
\Gamma (1 - \varepsilon)}\! \int_{-1}^1\df\!\cos (\theta) \frac{[- i \cos (\theta)]^{- 1+2
\varepsilon-\alpha}}{[1-\cos^2(\theta)]^{\varepsilon}} \nonumber\\
=\, & \frac{ \Gamma (1+\alpha - 2 \varepsilon) e^{\varepsilon\gamma_E}}{r
\Gamma\bigl(1-\frac{\alpha}{2}\bigr)\Gamma \bigl(1+\frac{\alpha}{2} - \varepsilon\bigr)}
\frac{(\mu r)^{2 \varepsilon-\alpha}}{4 \pi} \,. \nonumber
\end{align}
We can expand the above result around $\alpha=0$, where the first two expansion coefficients yield the Fourier transforms for the pieces that we need: $1$ and $\log(q/\mu)$. Higher-order terms in the expansion provide the transform of higher powers of the logarithm.

\section{Log-dependent R-evolution}\label{sec:Revo}
In this appendix we generalize the \texttt{REvolver} algorithm to carry out semi-analytic \mbox{R-evolution} to the case of R-anomalous dimensions dependent
on a single power of $\log(\alpha_s)$. Let us consider the generic case
\begin{equation}
\gamma_n (\alpha_s, b) = \alpha_s^n[ \,\log (\alpha_s) + b \,]\,,
\end{equation}
and proceed by changing variables from $R$ to $\alpha_s$ using the relation $R = \Lambda_{\rm{QCD}} e^{-G\!\bigl[- \frac{2 \pi}{\beta_0} \frac{1}{\alpha_s\!(R)}\bigr]}$,
followed by another change of variables from $\alpha_s$ to the $t$-variable introduced in~\cite{Hoang:2008yj,Hoang:2017suc} (explicit expressions for $t$ and $G$ can be
found in App.~A.2 of Ref.~\cite{Hoang:2021fhn})
\begin{align}\label{eq:usRV}
\Delta^{\rm us }_R (R_2, R_1, n, b) \, &= -\! \int_{R_1}^{R_2}\! \text{d} R \:
\gamma_n [\alpha_s (R), b] = - \Lambda_{\rm QCD}\! \int_{\alpha_1}^{\alpha_2}\!
\frac{\df \alpha}{\beta (\alpha)} e^{- G (\alpha)} \gamma_n(\alpha, b)\\
\, &= \Lambda_{\rm QCD}\! \int_{t_1}^{t_2}\! \df t\, \hat{b} (t) e^{- \tilde{G}(t)} \gamma_n\! \biggl(- \frac{2 \pi}{\beta_0} \frac{1}{t},b\biggr) e^{- t} (- t)^{- \hat{b}_1}\nonumber\\
\, & = \Lambda_{\rm QCD} \biggl( \frac{2 \pi}{\beta_0} \biggr)^{\!\!n} \sum_{j = 0} K_j\!
\int_{t_1}^{t_2} \df t \biggl[ \log\! \biggl( - \frac{2 \pi}{\beta_0}
\frac{1}{t} \biggr)\! + b \biggr] e^{- t} (- t)^{- n - \hat{b}_1 - j} \nonumber\\
\, & \equiv \Lambda_{\rm QCD} \biggl( \frac{2 \pi}{\beta_0} \biggr)^{\!\!n} \sum_{j = 0} K_j I_{\rm us} (\hat{b}_1, t_1, t_2, j + n - 1, b)\,.\nonumber
\end{align}
An expression for the functions $\hat b$ and $\tilde G $ can be found in Eqs.~(A.8) and (A.9) of Ref.~\cite{Hoang:2021fhn}, respectively. The $K_j$ coefficients
are defined from the expansion $\hat{b} (t) e^{- \tilde{G} (t)} \equiv \sum_{k = 0} K_k (- t)^{- k}$, and can be computed with the following relation
\begin{equation}
K_j = \sum_{i = 0}^j (- 1)^i \,\hat{b}_i \, \tilde{g}_{j - i}\,,
\end{equation}
where $\hat{b}_i$ and $\tilde{g}_i$ are defined in App.~A.2 of Ref.~\cite{Hoang:2021fhn} and can be computed from the recursive relations given therein.
Next, we discuss how to efficiently compute the integration kernel $I_{\rm us}$ defined in the last line of Eq.~\eqref{eq:usRV}. Our approach closely follows the
integration-by-parts strategy of Ref.~\cite{Hoang:2021fhn} and, in fact, we only need to apply a derivative with respect to $\hat b_1$ to Eqs.~(A.26) and (A.27)
therein to obtain the result, which now depends on two infinite sums. Defining $I_{\rm us} (\hat{b}_1, t_1, t_2, j, b) = I_{\rm us} (\hat{b}_1, t_2, j, b) - I_{\rm us} (\hat{b}_1, t_1, j, b)$ we find
\begin{align}
I_{\rm us} (\hat{b}_1, t, j, b) =&\, \frac{1}{(1 + \hat{b}_1)_j} \Biggl\{I_2 (\hat{b}_1, t) + e^{-
t} \sum_{i = 1}^j (1 + \hat{b}_1)_{i - 1} (- t)^{- i - \hat{b}_1} \sum_{k
= 1}^{i - 1} \frac{1}{\hat{b}_1 + k} \\
& \!+ \!\Biggl[ I_1 (\hat{b}_1, t) + e^{- t} \sum_{i = 1}^j (1 + \hat{b}_1)_{i - 1} (- t)^{- i - \hat{b}_1} \Biggr] \!\!\Biggl[ \log\! \biggl(\! -
\frac{2 \pi}{\beta_0} \frac{1}{t} \biggr) \!+ b -\! \sum_{k = 1}^j
\frac{1}{\hat{b}_1 + k} \Biggr] \!\Biggr\} ,\nonumber\\
I_k (\hat{b}_1, t) \equiv&\,- \sum_{i = 0}^{\infty} \frac{1}{i \, !} \frac{(-t)^{i - \hat{b}_1}}{(i - \hat{b}_1)^k} .\nonumber
\end{align}
In practice, the three infinite sums, namely $I_2 (\hat{b}_1, t_2)-I_2 (\hat{b}_1, t_1)$, $I_1 (\hat{b}_1, t1)$ and $I_1 (\hat{b}_1, t_2)$, are computed only once and then used for any value of $j$.
As an additional optimization, the sum over $k$ is also stored for consecutive values of the upper limit.

\bibliography{NRQCD}

\providecommand{\href}[2]{#2}\begingroup\raggedright\begin{thebibliography}{100}

\bibitem{Salam:2017qdl}
G.~P. Salam, \emph{{The strong coupling: a theoretical perspective}},  in
  \emph{From My Vast Repertoire ...: Guido Altarelli's Legacy} (A.~Levy,
  S.~Forte and G.~Ridolfi, eds.), pp.~101--121.
\newblock 2019.
\newblock \href{https://arxiv.org/abs/1712.05165}{{\ttfamily 1712.05165}}.
\newblock \href{http://dx.doi.org/10.1142/9789813238053_0007}{DOI}.

\bibitem{Pich:2018lmu}
A.~Pich, J.~Rojo, R.~Sommer and A.~Vairo, \emph{{Determining the strong
  coupling: status and challenges}},
  \href{http://dx.doi.org/10.22323/1.336.0035}{\emph{PoS} {\bfseries
  Confinement2018} (2018) 035},
  [\href{https://arxiv.org/abs/1811.11801}{{\ttfamily 1811.11801}}].

\bibitem{dEnterria:2022hzv}
D.~d'Enterria et~al., \emph{{The strong coupling constant: state of the art and
  the decade ahead}},
  \href{http://dx.doi.org/10.1088/1361-6471/ad1a78}{\emph{J. Phys. G}
  {\bfseries 51} (2024) 090501},
  [\href{https://arxiv.org/abs/2203.08271}{{\ttfamily 2203.08271}}].

\bibitem{Gehrmann:2010uax}
T.~Gehrmann, M.~Jaquier and G.~Luisoni, \emph{{Hadronization effects in event
  shape moments}},
  \href{http://dx.doi.org/10.1140/epjc/s10052-010-1288-4}{\emph{Eur. Phys. J.
  C} {\bfseries 67} (2010) 57--72},
  [\href{https://arxiv.org/abs/0911.2422}{{\ttfamily 0911.2422}}].

\bibitem{Abbate:2010xh}
R.~Abbate, M.~Fickinger, A.~H. Hoang, V.~Mateu and I.~W. Stewart, \emph{{Thrust
  at N$^3$LL with Power Corrections and a Precision Global Fit for
  $\alpha_s(m_Z)$}},
  \href{http://dx.doi.org/10.1103/PhysRevD.83.074021}{\emph{Phys. Rev.}
  {\bfseries D83} (2011) 074021},
  [\href{https://arxiv.org/abs/1006.3080}{{\ttfamily 1006.3080}}].

\bibitem{Abbate:2012jh}
R.~Abbate, M.~Fickinger, A.~H. Hoang, V.~Mateu and I.~W. Stewart,
  \emph{{Precision Thrust Cumulant Moments at N$^3$LL}},
  \href{http://dx.doi.org/10.1103/PhysRevD.86.094002}{\emph{Phys. Rev.}
  {\bfseries D86} (2012) 094002},
  [\href{https://arxiv.org/abs/1204.5746}{{\ttfamily 1204.5746}}].

\bibitem{Gehrmann:2012sc}
T.~Gehrmann, G.~Luisoni and P.~F. Monni, \emph{{Power corrections in the
  dispersive model for a determination of the strong coupling constant from the
  thrust distribution}},
  \href{http://dx.doi.org/10.1140/epjc/s10052-012-2265-x}{\emph{Eur. Phys. J.
  C} {\bfseries 73} (2013) 2265},
  [\href{https://arxiv.org/abs/1210.6945}{{\ttfamily 1210.6945}}].

\bibitem{Hoang:2015hka}
A.~H. Hoang, D.~W. Kolodrubetz, V.~Mateu and I.~W. Stewart, \emph{{Precise
  determination of $\alpha_s$ from the $C$-parameter distribution}},
  \href{http://dx.doi.org/10.1103/PhysRevD.91.094018}{\emph{Phys. Rev.}
  {\bfseries D91} (2015) 094018},
  [\href{https://arxiv.org/abs/1501.04111}{{\ttfamily 1501.04111}}].

\bibitem{Bell:2023dqs}
G.~Bell, C.~Lee, Y.~Makris, J.~Talbert and B.~Yan, \emph{{Effects of renormalon
  scheme and perturbative scale choices on determinations of the strong
  coupling from $e^+e^-$ event shapes}},
  \href{http://dx.doi.org/10.1103/PhysRevD.109.094008}{\emph{Phys. Rev. D}
  {\bfseries 109} (2024) 094008},
  [\href{https://arxiv.org/abs/2311.03990}{{\ttfamily 2311.03990}}].

\bibitem{Benitez:2024nav}
M.~A. Benitez, A.~H. Hoang, V.~Mateu, I.~W. Stewart and G.~Vita, \emph{{On
  Determining $\alpha_s(m_Z)$ from Dijets in $e^+e^-$ Thrust}},
  \href{http://dx.doi.org/10.1007/JHEP07(2025)249}{\emph{JHEP} {\bfseries 07}
  (2025) 249}, [\href{https://arxiv.org/abs/2412.15164}{{\ttfamily
  2412.15164}}].

\bibitem{Benitez:2025vsp}
M.~A. Benitez, A.~Bhattacharya, A.~H. Hoang, V.~Mateu, M.~D. Schwartz, I.~W.
  Stewart et~al., \emph{{A Precise Determination of $\alpha_s$ from the Heavy
  Jet Mass Distribution}},  \href{https://arxiv.org/abs/2502.12253}{{\ttfamily
  2502.12253}}.

\bibitem{ParticleDataGroup:2024cfk}
{\scshape Particle Data Group} collaboration, S.~Navas et~al., \emph{{Review of
  particle physics}},
  \href{http://dx.doi.org/10.1103/PhysRevD.110.030001}{\emph{Phys. Rev. D}
  {\bfseries 110} (2024) 030001}.

\bibitem{Davies:2008sw}
{\scshape HPQCD} collaboration, C.~T.~H. Davies, K.~Hornbostel, I.~D. Kendall,
  G.~P. Lepage, C.~McNeile, J.~Shigemitsu et~al., \emph{{Update: Accurate
  Determinations of $\alpha_s$ from Realistic Lattice QCD}},
  \href{http://dx.doi.org/10.1103/PhysRevD.78.114507}{\emph{Phys. Rev.}
  {\bfseries D78} (2008) 114507},
  [\href{https://arxiv.org/abs/0807.1687}{{\ttfamily 0807.1687}}].

\bibitem{McNeile:2010ji}
C.~McNeile, C.~T.~H. Davies, E.~Follana, K.~Hornbostel and G.~P. Lepage,
  \emph{{High-Precision c and b Masses, and QCD Coupling from Current-Current
  Correlators in Lattice and Continuum QCD}},
  \href{http://dx.doi.org/10.1103/PhysRevD.82.034512}{\emph{Phys. Rev.}
  {\bfseries D82} (2010) 034512},
  [\href{https://arxiv.org/abs/1004.4285}{{\ttfamily 1004.4285}}].

\bibitem{HPQCD:2014aca}
{\scshape HPQCD} collaboration, B.~Chakraborty, C.~T.~H. Davies, B.~Galloway,
  P.~Knecht, J.~Koponen, G.~C. Donald et~al., \emph{{High-precision quark
  masses and QCD coupling from $n_f=4$ lattice QCD}},
  \href{http://dx.doi.org/10.1103/PhysRevD.91.054508}{\emph{Phys. Rev.}
  {\bfseries D91} (2015) 054508},
  [\href{https://arxiv.org/abs/1408.4169}{{\ttfamily 1408.4169}}].

\bibitem{Nakayama:2016atf}
K.~Nakayama, B.~Fahy and S.~Hashimoto, \emph{{Short-distance charmonium
  correlator on the lattice with M{\"o}bius domain-wall fermion and a
  determination of charm quark mass}},
  \href{http://dx.doi.org/10.1103/PhysRevD.94.054507}{\emph{Phys. Rev.}
  {\bfseries D94} (2016) 054507},
  [\href{https://arxiv.org/abs/1606.01002}{{\ttfamily 1606.01002}}].

\bibitem{Maezawa:2016vgv}
Y.~Maezawa and P.~Petreczky, \emph{{Quark masses and strong coupling constant
  in 2+1 flavor QCD}},
  \href{http://dx.doi.org/10.1103/PhysRevD.94.034507}{\emph{Phys. Rev.}
  {\bfseries D94} (2016) 034507},
  [\href{https://arxiv.org/abs/1606.08798}{{\ttfamily 1606.08798}}].

\bibitem{Petreczky:2019ozv}
P.~Petreczky and J.~H. Weber, \emph{{Strong coupling constant and heavy quark
  masses in \mbox{(2+1)-flavor} QCD}},
  \href{http://dx.doi.org/10.1103/PhysRevD.100.034519}{\emph{Phys. Rev.}
  {\bfseries D100} (2019) 034519},
  [\href{https://arxiv.org/abs/1901.06424}{{\ttfamily 1901.06424}}].

\bibitem{Petreczky:2020tky}
P.~Petreczky and J.~H. Weber, \emph{{Strong coupling constant from moments of
  quarkonium correlators revisited}},
  \href{http://dx.doi.org/10.1140/epjc/s10052-022-09998-0}{\emph{Eur. Phys. J.
  C} {\bfseries 82} (2022) 64},
  [\href{https://arxiv.org/abs/2012.06193}{{\ttfamily 2012.06193}}].

\bibitem{Boito:2019pqp}
D.~Boito and V.~Mateu, \emph{{Precise $\alpha_s$ determination from charmonium
  sum rules}},
  \href{http://dx.doi.org/10.1016/j.physletb.2020.135482}{\emph{Phys. Lett. B}
  {\bfseries 806} (2020) 135482},
  [\href{https://arxiv.org/abs/1912.06237}{{\ttfamily 1912.06237}}].

\bibitem{Boito:2020lyp}
D.~Boito and V.~Mateu, \emph{{Precise determination of $\alpha_s$ from
  relativistic quarkonium sum rules}},
  \href{http://dx.doi.org/10.1007/JHEP03(2020)094}{\emph{JHEP} {\bfseries 03}
  (2020) 094}, [\href{https://arxiv.org/abs/2001.11041}{{\ttfamily
  2001.11041}}].

\bibitem{Blossier:2013ioa}
{\scshape ETM} collaboration, B.~Blossier, P.~Boucaud, M.~Brinet, F.~De~Soto,
  V.~Morenas, O.~Pene et~al., \emph{{High statistics determination of the
  strong coupling constant in Taylor scheme and its OPE Wilson coefficient from
  lattice QCD with a dynamical charm}},
  \href{http://dx.doi.org/10.1103/PhysRevD.89.014507}{\emph{Phys. Rev.}
  {\bfseries D89} (2014) 014507},
  [\href{https://arxiv.org/abs/1310.3763}{{\ttfamily 1310.3763}}].

\bibitem{Bruno:2017gxd}
{\scshape ALPHA} collaboration, M.~Bruno, M.~Dalla~Brida, P.~Fritzsch,
  T.~Korzec, A.~Ramos, S.~Schaefer et~al., \emph{{QCD Coupling from a
  Nonperturbative Determination of the Three-Flavor $\Lambda$ Parameter}},
  \href{http://dx.doi.org/10.1103/PhysRevLett.119.102001}{\emph{Phys. Rev.
  Lett.} {\bfseries 119} (2017) 102001},
  [\href{https://arxiv.org/abs/1706.03821}{{\ttfamily 1706.03821}}].

\bibitem{Zafeiropoulos:2019flq}
S.~Zafeiropoulos, P.~Boucaud, F.~De~Soto, J.~Rodr{\'\i}guez-Quintero and
  J.~Segovia, \emph{{Strong Running Coupling from the Gauge Sector of Domain
  Wall Lattice QCD with Physical Quark Masses}},
  \href{http://dx.doi.org/10.1103/PhysRevLett.122.162002}{\emph{Phys. Rev.
  Lett.} {\bfseries 122} (2019) 162002},
  [\href{https://arxiv.org/abs/1902.08148}{{\ttfamily 1902.08148}}].

\bibitem{DallaBrida:2022eua}
{\scshape ALPHA} collaboration, M.~Dalla~Brida, R.~H\"ollwieser, F.~Knechtli,
  T.~Korzec, A.~Nada, A.~Ramos et~al., \emph{{Determination of $\alpha _s(m_Z)$
  by the non-perturbative decoupling method}},
  \href{http://dx.doi.org/10.1140/epjc/s10052-022-10998-3}{\emph{Eur. Phys. J.
  C} {\bfseries 82} (2022) 1092},
  [\href{https://arxiv.org/abs/2209.14204}{{\ttfamily 2209.14204}}].

\bibitem{FlavourLatticeAveragingGroupFLAG:2024oxs}
{\scshape Flavour Lattice Averaging Group (FLAG)} collaboration, Y.~Aoki
  et~al., \emph{{FLAG Review 2024}},
  \href{https://arxiv.org/abs/2411.04268}{{\ttfamily 2411.04268}}.

\bibitem{Bazavov:2012ka}
A.~Bazavov, N.~Brambilla, X.~Garcia~i Tormo, P.~Petreczky, J.~Soto and
  A.~Vairo, \emph{{Determination of $\alpha_s$ from the QCD static energy}},
  \href{http://dx.doi.org/10.1103/PhysRevD.86.114031}{\emph{Phys. Rev.}
  {\bfseries D86} (2012) 114031},
  [\href{https://arxiv.org/abs/1205.6155}{{\ttfamily 1205.6155}}].

\bibitem{Bazavov:2011nk}
A.~Bazavov et~al., \emph{{The chiral and deconfinement aspects of the QCD
  transition}}, \href{http://dx.doi.org/10.1103/PhysRevD.85.054503}{\emph{Phys.
  Rev. D} {\bfseries 85} (2012) 054503},
  [\href{https://arxiv.org/abs/1111.1710}{{\ttfamily 1111.1710}}].

\bibitem{Bazavov:2014soa}
A.~Bazavov, N.~Brambilla, X.~Garcia~i Tormo, P.~Petreczky, J.~Soto et~al.,
  \emph{{Determination of $\alpha_s$ from the QCD static energy: An update}},
  \href{http://dx.doi.org/10.1103/PhysRevD.90.074038}{\emph{Phys. Rev.}
  {\bfseries D90} (2014) 074038},
  [\href{https://arxiv.org/abs/1407.8437}{{\ttfamily 1407.8437}}].

\bibitem{Bazavov:2019qoo}
{\scshape TUMQCD} collaboration, A.~Bazavov, N.~Brambilla, X.~Garcia~i Tormo,
  P.~Petreczky, J.~Soto, A.~Vairo et~al., \emph{{Determination of the QCD
  coupling from the static energy and the free energy}},
  \href{http://dx.doi.org/10.1103/PhysRevD.100.114511}{\emph{Phys. Rev.}
  {\bfseries D100} (2019) 114511},
  [\href{https://arxiv.org/abs/1907.11747}{{\ttfamily 1907.11747}}].

\bibitem{Bazavov:2014pvz}
{\scshape HotQCD} collaboration, A.~Bazavov et~al., \emph{{Equation of state in
  (2+1)-flavor QCD}},
  \href{http://dx.doi.org/10.1103/PhysRevD.90.094503}{\emph{Phys. Rev.}
  {\bfseries D90} (2014) 094503},
  [\href{https://arxiv.org/abs/1407.6387}{{\ttfamily 1407.6387}}].

\bibitem{Bazavov:2017dsy}
A.~Bazavov, P.~Petreczky and J.~H. Weber, \emph{{Equation of State in 2+1
  Flavor QCD at High Temperatures}},
  \href{http://dx.doi.org/10.1103/PhysRevD.97.014510}{\emph{Phys. Rev.}
  {\bfseries D97} (2018) 014510},
  [\href{https://arxiv.org/abs/1710.05024}{{\ttfamily 1710.05024}}].

\bibitem{Ayala:2020odx}
C.~Ayala, X.~Lobregat and A.~Pineda, \emph{{Determination of $\alpha(m_Z)$ from
  an hyperasymptotic approximation to the energy of a static quark-antiquark
  pair}}, \href{http://dx.doi.org/10.1007/JHEP09(2020)016}{\emph{JHEP}
  {\bfseries 09} (2020) 016},
  [\href{https://arxiv.org/abs/2005.12301}{{\ttfamily 2005.12301}}].

\bibitem{Ananthanarayan:2020umo}
B.~Ananthanarayan, D.~Das and M.~S.~A. Alam~Khan, \emph{{QCD static energy
  using optimal renormalization and asymptotic Pad{\'e}-approximant methods}},
  \href{http://dx.doi.org/10.1103/PhysRevD.102.076008}{\emph{Phys. Rev. D}
  {\bfseries 102} (2020) 076008},
  [\href{https://arxiv.org/abs/2007.10775}{{\ttfamily 2007.10775}}].

\bibitem{Bazavov:2018wmo}
{\scshape TUMQCD} collaboration, A.~Bazavov, N.~Brambilla, P.~Petreczky,
  A.~Vairo and J.~H. Weber, \emph{{Color screening in (2+1)-flavor QCD}},
  \href{http://dx.doi.org/10.1103/PhysRevD.98.054511}{\emph{Phys. Rev. D}
  {\bfseries 98} (2018) 054511},
  [\href{https://arxiv.org/abs/1804.10600}{{\ttfamily 1804.10600}}].

\bibitem{Takaura:2018lpw}
H.~Takaura, T.~Kaneko, Y.~Kiyo and Y.~Sumino, \emph{{Determination of
  $\alpha_s$ from static QCD potential with renormalon subtraction}},
  \href{http://dx.doi.org/10.1016/j.physletb.2018.12.060}{\emph{Phys. Lett.}
  {\bfseries B789} (2019) 598--602},
  [\href{https://arxiv.org/abs/1808.01632}{{\ttfamily 1808.01632}}].

\bibitem{Takaura:2018vcy}
H.~Takaura, T.~Kaneko, Y.~Kiyo and Y.~Sumino, \emph{{Determination of
  $\alpha_s$ from static QCD potential: OPE with renormalon subtraction and
  lattice QCD}}, \href{http://dx.doi.org/10.1007/JHEP04(2019)155}{\emph{JHEP}
  {\bfseries 04} (2019) 155},
  [\href{https://arxiv.org/abs/1808.01643}{{\ttfamily 1808.01643}}].

\bibitem{Sumino:2020mxk}
Y.~Sumino and H.~Takaura, \emph{{On renormalons of static QCD potential at
  $u=1/2$ and $3/2$}},
  \href{http://dx.doi.org/10.1007/JHEP05(2020)116}{\emph{JHEP} {\bfseries 05}
  (2020) 116}, [\href{https://arxiv.org/abs/2001.00770}{{\ttfamily
  2001.00770}}].

\bibitem{Mateu:2018zym}
V.~Mateu, P.~G. Ortega, D.~R. Entem and F.~Fern{\'a}ndez, \emph{{Calibrating
  the Na{\"\i}ve Cornell Model with NRQCD}},
  \href{http://dx.doi.org/10.1140/epjc/s10052-019-6808-2}{\emph{Eur. Phys. J.
  C} {\bfseries 79} (2019) 323},
  [\href{https://arxiv.org/abs/1811.01982}{{\ttfamily 1811.01982}}].

\bibitem{Hoang:2008yj}
A.~H. Hoang, A.~Jain, I.~Scimemi and I.~W. Stewart, \emph{{Infrared
  Renormalization Group Flow for Heavy Quark Masses}},
  \href{http://dx.doi.org/10.1103/PhysRevLett.101.151602}{\emph{Phys. Rev.
  Lett.} {\bfseries 101} (2008) 151602},
  [\href{https://arxiv.org/abs/0803.4214}{{\ttfamily 0803.4214}}].

\bibitem{Hoang:2017suc}
A.~H. Hoang, A.~Jain, C.~Lepenik, V.~Mateu, M.~Preisser, I.~Scimemi et~al.,
  \emph{{The MSR mass and the $
  \mathcal{O}\left({\Lambda}_{\mathrm{QCD}}\right) $ renormalon sum rule}},
  \href{http://dx.doi.org/10.1007/JHEP04(2018)003}{\emph{JHEP} {\bfseries 04}
  (2018) 003}, [\href{https://arxiv.org/abs/1704.01580}{{\ttfamily
  1704.01580}}].

\bibitem{Beneke:1998rk}
M.~Beneke, \emph{{A Quark mass definition adequate for threshold problems}},
  \href{http://dx.doi.org/10.1016/S0370-2693(98)00741-2}{\emph{Phys. Lett.}
  {\bfseries B434} (1998) 115--125},
  [\href{https://arxiv.org/abs/hep-ph/9804241}{{\ttfamily hep-ph/9804241}}].

\bibitem{Hagiwara:2003da}
K.~Hagiwara, A.~D. Martin, D.~Nomura and T.~Teubner, \emph{{Predictions for g-2
  of the muon and $\alpha_{QED}(M_Z^2)$}},
  \href{http://dx.doi.org/10.1103/PhysRevD.69.093003}{\emph{Phys. Rev.}
  {\bfseries D69} (2004) 093003},
  [\href{https://arxiv.org/abs/hep-ph/0312250}{{\ttfamily hep-ph/0312250}}].

\bibitem{Dehnadi:2011gc}
B.~Dehnadi, A.~H. Hoang, V.~Mateu and S.~M. Zebarjad, \emph{{Charm Mass
  Determination from QCD Charmonium Sum Rules at Order $\alpha_{s}^{3}$}},
  \href{http://dx.doi.org/10.1007/JHEP09(2013)103}{\emph{JHEP} {\bfseries 09}
  (2013) 103}, [\href{https://arxiv.org/abs/1102.2264}{{\ttfamily 1102.2264}}].

\bibitem{python3}
G.~Van~Rossum and F.~L. Drake, \emph{Python 3 Reference Manual}.
\newblock CreateSpace, Scotts Valley, CA, 2009.

\bibitem{gccmakeNTU}
NTU, ``Gcc and make: Compiling, linking and building c/c++ applications.''
  \url{https://www3.ntu.edu.sg/home/ehchua/programming/cpp/gcc_make.html},
  2023.

\bibitem{gfortran}
\emph{{GFortran, Gnu compiler collection (gcc), Version 14.2.0}}.
\newblock Copyright (C) 2024 Free Software Foundation, Inc., 2024.

\bibitem{Hoang:2021fhn}
A.~H. Hoang, C.~Lepenik and V.~Mateu, \emph{{REvolver: Automated running and
  matching of couplings and masses in QCD}},
  \href{http://dx.doi.org/10.1016/j.cpc.2021.108145}{\emph{Comput. Phys.
  Commun.} {\bfseries 270} (2022) 108145},
  [\href{https://arxiv.org/abs/2102.01085}{{\ttfamily 2102.01085}}].

\bibitem{Larin:1993tp}
S.~A. Larin and J.~A.~M. Vermaseren, \emph{{The three-loop QCD $\beta$ function
  and anomalous dimensions}}, {\emph{Phys. Lett.} {\bfseries B303} (1993)
  334--336}, [\href{https://arxiv.org/abs/hep-ph/9302208}{{\ttfamily
  hep-ph/9302208}}].

\bibitem{vanRitbergen:1997va}
T.~van Ritbergen, J.~A.~M. Vermaseren and S.~A. Larin, \emph{{The four-loop
  beta function in quantum chromodynamics}},
  \href{http://dx.doi.org/10.1016/S0370-2693(97)00370-5}{\emph{Phys. Lett.}
  {\bfseries B400} (1997) 379--384},
  [\href{https://arxiv.org/abs/hep-ph/9701390}{{\ttfamily hep-ph/9701390}}].

\bibitem{Czakon:2004bu}
M.~Czakon, \emph{{The four-loop QCD beta-function and anomalous dimensions}},
  \href{http://dx.doi.org/10.1016/j.nuclphysb.2005.01.012}{\emph{Nucl. Phys.}
  {\bfseries B710} (2005) 485--498},
  [\href{https://arxiv.org/abs/hep-ph/0411261}{{\ttfamily hep-ph/0411261}}].

\bibitem{Luthe:2017ttg}
T.~Luthe, A.~Maier, P.~Marquard and Y.~Schroder, \emph{{The five-loop Beta
  function for a general gauge group and anomalous dimensions beyond Feynman
  gauge}}, \href{http://dx.doi.org/10.1007/JHEP10(2017)166}{\emph{JHEP}
  {\bfseries 10} (2017) 166},
  [\href{https://arxiv.org/abs/1709.07718}{{\ttfamily 1709.07718}}].

\bibitem{Chetyrkin:1997un}
K.~G. Chetyrkin, B.~A. Kniehl and M.~Steinhauser, \emph{{Decoupling relations
  to $\mathcal{O}(\alpha_s^3)$ and their connection to low-energy theorems}},
  \href{http://dx.doi.org/10.1016/S0550-3213(98)81004-3,
  10.1016/S0550-3213(97)00649-4}{\emph{Nucl. Phys.} {\bfseries B510} (1998)
  61--87}, [\href{https://arxiv.org/abs/hep-ph/9708255}{{\ttfamily
  hep-ph/9708255}}].

\bibitem{Chetyrkin:2005ia}
K.~G. Chetyrkin, J.~H. Kuhn and C.~Sturm, \emph{{QCD decoupling at four
  loops}}, \href{http://dx.doi.org/10.1016/j.nuclphysb.2006.03.020}{\emph{Nucl.
  Phys.} {\bfseries B744} (2006) 121--135},
  [\href{https://arxiv.org/abs/hep-ph/0512060}{{\ttfamily hep-ph/0512060}}].

\bibitem{Schroder:2005hy}
Y.~Schroder and M.~Steinhauser, \emph{{Four-loop decoupling relations for the
  strong coupling}},
  \href{http://dx.doi.org/10.1088/1126-6708/2006/01/051}{\emph{JHEP} {\bfseries
  01} (2006) 051}, [\href{https://arxiv.org/abs/hep-ph/0512058}{{\ttfamily
  hep-ph/0512058}}].

\bibitem{swig}
D.~M. Beazley, \emph{SWIG: Simplified Wrapper and Interface Generator}.
\newblock The Regents of the University of California, 2024.

\bibitem{harris2020array}
C.~R. Harris, K.~J. Millman, S.~J. van~der Walt, R.~Gommers, P.~Virtanen,
  D.~Cournapeau et~al., \emph{Array programming with {NumPy}},
  \href{http://dx.doi.org/10.1038/s41586-020-2649-2}{\emph{Nature} {\bfseries
  585} (Sept., 2020) 357--362}.

\bibitem{2020SciPy-NMeth}
P.~Virtanen, R.~Gommers, T.~E. Oliphant, M.~Haberland, T.~Reddy, D.~Cournapeau
  et~al., \emph{{{SciPy} 1.0: Fundamental Algorithms for Scientific Computing
  in Python}}, \href{http://dx.doi.org/10.1038/s41592-019-0686-2}{\emph{Nature
  Methods} {\bfseries 17} (2020) 261--272}.

\bibitem{Hunter:2007}
J.~D. Hunter, \emph{Matplotlib: A 2d graphics environment},
  \href{http://dx.doi.org/10.1109/MCSE.2007.55}{\emph{Computing in Science \&
  Engineering} {\bfseries 9} (2007) 90--95}.

\bibitem{Pineda:1997bj}
A.~Pineda and J.~Soto, \emph{{Effective field theory for ultrasoft momenta in
  NRQCD and NRQED}},
  \href{http://dx.doi.org/10.1016/S0920-5632(97)01102-X}{\emph{Nucl. Phys.
  Proc. Suppl.} {\bfseries 64} (1998) 428--432},
  [\href{https://arxiv.org/abs/hep-ph/9707481}{{\ttfamily hep-ph/9707481}}].

\bibitem{Brambilla:1999xf}
N.~Brambilla, A.~Pineda, J.~Soto and A.~Vairo, \emph{{Potential NRQCD: An
  Effective theory for heavy quarkonium}},
  \href{http://dx.doi.org/10.1016/S0550-3213(99)00693-8}{\emph{Nucl. Phys.}
  {\bfseries B566} (2000) 275},
  [\href{https://arxiv.org/abs/hep-ph/9907240}{{\ttfamily hep-ph/9907240}}].

\bibitem{Lepage:1987gg}
G.~P. Lepage and B.~A. Thacker, \emph{{Effective Lagrangians for Simulating
  Heavy Quark Systems}},
  \href{http://dx.doi.org/10.1016/0920-5632(88)90102-8}{\emph{Nucl. Phys. Proc.
  Suppl.} {\bfseries 4} (1988) 199}.

\bibitem{Luke:1999kz}
M.~E. Luke, A.~V. Manohar and I.~Z. Rothstein, \emph{{Renormalization group
  scaling in nonrelativistic QCD}},
  \href{http://dx.doi.org/10.1103/PhysRevD.61.074025}{\emph{Phys. Rev.}
  {\bfseries D61} (2000) 074025},
  [\href{https://arxiv.org/abs/hep-ph/9910209}{{\ttfamily hep-ph/9910209}}].

\bibitem{Appelquist:1977tw}
T.~Appelquist, M.~Dine and I.~J. Muzinich, \emph{{The Static Potential in
  Quantum Chromodynamics}},
  \href{http://dx.doi.org/10.1016/0370-2693(77)90651-7}{\emph{Phys. Lett.}
  {\bfseries B69} (1977) 231--236}.

\bibitem{Appelquist:1977es}
T.~Appelquist, M.~Dine and I.~J. Muzinich, \emph{{The Static Limit of Quantum
  Chromodynamics}},
  \href{http://dx.doi.org/10.1103/PhysRevD.17.2074}{\emph{Phys. Rev.}
  {\bfseries D17} (1978) 2074}.

\bibitem{Fischler:1977yf}
W.~Fischler, \emph{{$Q\overline Q$ Potential in QCD}},
  \href{http://dx.doi.org/10.1016/0550-3213(77)90026-8}{\emph{Nucl. Phys.}
  {\bfseries B129} (1977) 157--174}.

\bibitem{Billoire:1979ih}
A.~Billoire, \emph{{How Heavy Must Be Quarks in Order to Build Coulombic $q
  \bar q$ Bound States}},
  \href{http://dx.doi.org/10.1016/0370-2693(80)90279-8}{\emph{Phys. Lett.}
  {\bfseries B92} (1980) 343--347}.

\bibitem{Schroder:1998vy}
Y.~Schroder, \emph{{The Static potential in QCD to two loops}},
  \href{http://dx.doi.org/10.1016/S0370-2693(99)00010-6}{\emph{Phys. Lett.}
  {\bfseries B447} (1999) 321--326},
  [\href{https://arxiv.org/abs/hep-ph/9812205}{{\ttfamily hep-ph/9812205}}].

\bibitem{Pineda:1997hz}
A.~Pineda and F.~J. Yndurain, \emph{{Calculation of quarkonium spectrum and
  $m_b$, $m_c$ to order $\alpha_s^4$}},
  \href{http://dx.doi.org/10.1103/PhysRevD.58.094022}{\emph{Phys. Rev.}
  {\bfseries D58} (1998) 094022},
  [\href{https://arxiv.org/abs/hep-ph/9711287}{{\ttfamily hep-ph/9711287}}].

\bibitem{Brambilla:1999qa}
N.~Brambilla, A.~Pineda, J.~Soto and A.~Vairo, \emph{{The Infrared behavior of
  the static potential in perturbative QCD}},
  \href{http://dx.doi.org/10.1103/PhysRevD.60.091502}{\emph{Phys. Rev.}
  {\bfseries D60} (1999) 091502},
  [\href{https://arxiv.org/abs/hep-ph/9903355}{{\ttfamily hep-ph/9903355}}].

\bibitem{Kniehl:2002br}
B.~A. Kniehl, A.~A. Penin, V.~A. Smirnov and M.~Steinhauser, \emph{{Potential
  NRQCD and heavy quarkonium spectrum at next-to-next-to-next-to-leading
  order}}, \href{http://dx.doi.org/10.1016/S0550-3213(02)00403-0}{\emph{Nucl.
  Phys.} {\bfseries B635} (2002) 357--383},
  [\href{https://arxiv.org/abs/hep-ph/0203166}{{\ttfamily hep-ph/0203166}}].

\bibitem{Penin:2002zv}
A.~A. Penin and M.~Steinhauser, \emph{{Heavy quarkonium spectrum at
  $\mathcal{O}(\alpha_s^5 m_q)$ and bottom / top quark mass determination}},
  \href{http://dx.doi.org/10.1016/S0370-2693(02)02040-3}{\emph{Phys. Lett.}
  {\bfseries B538} (2002) 335--345},
  [\href{https://arxiv.org/abs/hep-ph/0204290}{{\ttfamily hep-ph/0204290}}].

\bibitem{Smirnov:2008pn}
A.~V. Smirnov, V.~A. Smirnov and M.~Steinhauser, \emph{{Fermionic contributions
  to the three-loop static potential}},
  \href{http://dx.doi.org/10.1016/j.physletb.2008.08.070}{\emph{Phys. Lett.}
  {\bfseries B668} (2008) 293--298},
  [\href{https://arxiv.org/abs/0809.1927}{{\ttfamily 0809.1927}}].

\bibitem{Smirnov:2009fh}
A.~V. Smirnov, V.~A. Smirnov and M.~Steinhauser, \emph{{Three-loop static
  potential}},
  \href{http://dx.doi.org/10.1103/PhysRevLett.104.112002}{\emph{Phys. Rev.
  Lett.} {\bfseries 104} (2010) 112002},
  [\href{https://arxiv.org/abs/0911.4742}{{\ttfamily 0911.4742}}].

\bibitem{Anzai:2009tm}
C.~Anzai, Y.~Kiyo and Y.~Sumino, \emph{{Static QCD potential at three-loop
  order}}, \href{http://dx.doi.org/10.1103/PhysRevLett.104.112003}{\emph{Phys.
  Rev. Lett.} {\bfseries 104} (2010) 112003},
  [\href{https://arxiv.org/abs/0911.4335}{{\ttfamily 0911.4335}}].

\bibitem{Lee:2016cgz}
R.~N. Lee, A.~V. Smirnov, V.~A. Smirnov and M.~Steinhauser, \emph{{Analytic
  three-loop static potential}},
  \href{http://dx.doi.org/10.1103/PhysRevD.94.054029}{\emph{Phys. Rev.}
  {\bfseries D94} (2016) 054029},
  [\href{https://arxiv.org/abs/1608.02603}{{\ttfamily 1608.02603}}].

\bibitem{Mateu:2017hlz}
V.~Mateu and P.~G. Ortega, \emph{{Bottom and Charm Mass determinations from
  global fits to $Q\overline{Q}$ bound states at N$^3$LO}},
  \href{http://dx.doi.org/10.1007/JHEP01(2018)122}{\emph{JHEP} {\bfseries 01}
  (2018) 122}, [\href{https://arxiv.org/abs/1711.05755}{{\ttfamily
  1711.05755}}].

\bibitem{Kniehl:1999ud}
B.~A. Kniehl and A.~A. Penin, \emph{{Ultrasoft effects in heavy quarkonium
  physics}}, \href{http://dx.doi.org/10.1016/S0550-3213(99)00564-7}{\emph{Nucl.
  Phys.} {\bfseries B563} (1999) 200--210},
  [\href{https://arxiv.org/abs/hep-ph/9907489}{{\ttfamily hep-ph/9907489}}].

\bibitem{Brambilla:2006wp}
N.~Brambilla, X.~Garcia~i Tormo, J.~Soto and A.~Vairo, \emph{{The Logarithmic
  contribution to the QCD static energy at N$^4$LO}},
  \href{http://dx.doi.org/10.1016/j.physletb.2007.02.015}{\emph{Phys. Lett.}
  {\bfseries B647} (2007) 185--193},
  [\href{https://arxiv.org/abs/hep-ph/0610143}{{\ttfamily hep-ph/0610143}}].

\bibitem{Pineda:2000gza}
A.~Pineda and J.~Soto, \emph{{The Renormalization group improvement of the QCD
  static potentials}},
  \href{http://dx.doi.org/10.1016/S0370-2693(00)01261-2}{\emph{Phys. Lett.}
  {\bfseries B495} (2000) 323--328},
  [\href{https://arxiv.org/abs/hep-ph/0007197}{{\ttfamily hep-ph/0007197}}].

\bibitem{Brambilla:2009bi}
N.~Brambilla, A.~Vairo, X.~Garcia~i Tormo and J.~Soto, \emph{{The QCD static
  energy at NNNLL}},
  \href{http://dx.doi.org/10.1103/PhysRevD.80.034016}{\emph{Phys. Rev.}
  {\bfseries D80} (2009) 034016},
  [\href{https://arxiv.org/abs/0906.1390}{{\ttfamily 0906.1390}}].

\bibitem{Tormo:2013tha}
X.~Garcia~i Tormo, \emph{{Review on the determination of $\alpha_s$ from the
  QCD static energy}},
  \href{http://dx.doi.org/10.1142/S0217732313300280}{\emph{Mod. Phys. Lett.}
  {\bfseries A28} (2013) 1330028},
  [\href{https://arxiv.org/abs/1307.2238}{{\ttfamily 1307.2238}}].

\bibitem{Pineda:1998id}
A.~Pineda, \emph{{Heavy quarkonium and nonrelativistic effective field
  theories}}.
\newblock PhD thesis, Barcelona U., 1998.

\bibitem{Hoang:1998nz}
A.~H. Hoang, M.~C. Smith, T.~Stelzer and S.~Willenbrock, \emph{{Quarkonia and
  the pole mass}},
  \href{http://dx.doi.org/10.1103/PhysRevD.59.114014}{\emph{Phys. Rev.}
  {\bfseries D59} (1999) 114014},
  [\href{https://arxiv.org/abs/hep-ph/9804227}{{\ttfamily hep-ph/9804227}}].

\bibitem{Hoang:2009yr}
A.~H. Hoang, A.~Jain, I.~Scimemi and I.~W. Stewart, \emph{{R-evolution:
  Improving perturbative QCD}},
  \href{http://dx.doi.org/10.1103/PhysRevD.82.011501}{\emph{Phys. Rev.}
  {\bfseries D82} (2010) 011501},
  [\href{https://arxiv.org/abs/0908.3189}{{\ttfamily 0908.3189}}].

\bibitem{Gracia:2021nut}
N.~G. Gracia and V.~Mateu, \emph{{Toward massless and massive event shapes in
  the large-\ensuremath{\beta}$_{0}$ limit}},
  \href{http://dx.doi.org/10.1007/JHEP07(2021)229}{\emph{JHEP} {\bfseries 07}
  (2021) 229}, [\href{https://arxiv.org/abs/2104.13942}{{\ttfamily
  2104.13942}}].

\bibitem{Brambilla:2010pp}
N.~Brambilla, X.~Garcia~i Tormo, J.~Soto and A.~Vairo, \emph{{Precision
  determination of $r_0\Lambda_{\rm \overline{MS}}$ from the QCD static
  energy}}, \href{http://dx.doi.org/10.1103/PhysRevLett.105.212001,
  10.1103/PhysRevLett.108.269903}{\emph{Phys. Rev. Lett.} {\bfseries 105}
  (2010) 212001}, [\href{https://arxiv.org/abs/1006.2066}{{\ttfamily
  1006.2066}}].

\bibitem{Beneke:1994rs}
M.~Beneke, \emph{{More on ambiguities in the pole mass}},
  \href{http://dx.doi.org/10.1016/0370-2693(94)01505-7}{\emph{Phys. Lett.}
  {\bfseries B344} (1995) 341--347},
  [\href{https://arxiv.org/abs/hep-ph/9408380}{{\ttfamily hep-ph/9408380}}].

\bibitem{Pineda:2001zq}
A.~Pineda, \emph{{Determination of the bottom quark mass from the
  $\Upsilon(1S)$ system}},
  \href{http://dx.doi.org/10.1088/1126-6708/2001/06/022}{\emph{JHEP} {\bfseries
  06} (2001) 022}, [\href{https://arxiv.org/abs/hep-ph/0105008}{{\ttfamily
  hep-ph/0105008}}].

\bibitem{Bali:2003jq}
G.~S. Bali and A.~Pineda, \emph{{QCD phenomenology of static sources and
  gluonic excitations at short distances}},
  \href{http://dx.doi.org/10.1103/PhysRevD.69.094001}{\emph{Phys. Rev.}
  {\bfseries D69} (2004) 094001},
  [\href{https://arxiv.org/abs/hep-ph/0310130}{{\ttfamily hep-ph/0310130}}].

\bibitem{Hoang:2008fs}
A.~H. Hoang and S.~Kluth, \emph{{Hemisphere Soft Function at
  $\mathcal{O}(\alpha_s^2)$ for Dijet Production in $e^+e^-$ Annihilation}},
  \href{https://arxiv.org/abs/0806.3852}{{\ttfamily 0806.3852}}.

\bibitem{Jain:2008gb}
A.~Jain, I.~Scimemi and I.~W. Stewart, \emph{{Two-loop Jet-Function and
  Jet-Mass for Top Quarks}},
  \href{http://dx.doi.org/10.1103/PhysRevD.77.094008}{\emph{Phys. Rev.}
  {\bfseries D77} (2008) 094008},
  [\href{https://arxiv.org/abs/0801.0743}{{\ttfamily 0801.0743}}].

\bibitem{Clavero:2024yav}
A.~M. Clavero, R.~Br{\"u}ser, V.~Mateu and M.~Stahlhofen, \emph{{Three-loop jet
  function for boosted heavy quarks}},
  \href{http://dx.doi.org/10.1007/JHEP04(2025)040}{\emph{JHEP} {\bfseries 04}
  (2025) 040}, [\href{https://arxiv.org/abs/2412.06881}{{\ttfamily
  2412.06881}}].

\bibitem{Butenschoen:2016lpz}
M.~Butenschoen, B.~Dehnadi, A.~H. Hoang, V.~Mateu, M.~Preisser and I.~W.
  Stewart, \emph{{Top Quark Mass Calibration for Monte Carlo Event
  Generators}},
  \href{http://dx.doi.org/10.1103/PhysRevLett.117.232001}{\emph{Phys. Rev.
  Lett.} {\bfseries 117} (2016) 232001},
  [\href{https://arxiv.org/abs/1608.01318}{{\ttfamily 1608.01318}}].

\bibitem{Hoang:2018zrp}
A.~H. Hoang, S.~Pl{\"a}tzer and D.~Samitz, \emph{{On the Cutoff Dependence of
  the Quark Mass Parameter in Angular Ordered Parton Showers}},
  \href{http://dx.doi.org/10.1007/JHEP10(2018)200}{\emph{JHEP} {\bfseries 10}
  (2018) 200}, [\href{https://arxiv.org/abs/1807.06617}{{\ttfamily
  1807.06617}}].

\bibitem{Dehnadi:2023msm}
B.~Dehnadi, A.~H. Hoang, O.~L. Jin and V.~Mateu, \emph{{Top quark mass
  calibration for Monte Carlo event generators {\textemdash} an update}},
  \href{http://dx.doi.org/10.1007/JHEP12(2023)065}{\emph{JHEP} {\bfseries 12}
  (2023) 065}, [\href{https://arxiv.org/abs/2309.00547}{{\ttfamily
  2309.00547}}].

\bibitem{Bachu:2020nqn}
B.~Bachu, A.~H. Hoang, V.~Mateu, A.~Pathak and I.~W. Stewart, \emph{{Boosted
  top quarks in the peak region with N$^3$LL resummation}},
  \href{http://dx.doi.org/10.1103/PhysRevD.104.014026}{\emph{Phys. Rev. D}
  {\bfseries 104} (2021) 014026},
  [\href{https://arxiv.org/abs/2012.12304}{{\ttfamily 2012.12304}}].

\bibitem{Hoang:2019ceu}
A.~H. Hoang, S.~Mantry, A.~Pathak and I.~W. Stewart, \emph{{Nonperturbative
  Corrections to Soft Drop Jet Mass}},
  \href{http://dx.doi.org/10.1007/JHEP12(2019)002}{\emph{JHEP} {\bfseries 12}
  (2019) 002}, [\href{https://arxiv.org/abs/1906.11843}{{\ttfamily
  1906.11843}}].

\bibitem{Boronat:2019cgt}
M.~Boronat, E.~Fullana, J.~Fuster, P.~Gomis, A.~Hoang, V.~Mateu et~al.,
  \emph{{Top quark mass measurement in radiative events at electron-positron
  colliders}},
  \href{http://dx.doi.org/10.1016/j.physletb.2020.135353}{\emph{Phys. Lett. B}
  {\bfseries 804} (2020) 135353},
  [\href{https://arxiv.org/abs/1912.01275}{{\ttfamily 1912.01275}}].

\bibitem{Beneke:2014pta}
M.~Beneke, A.~Maier, J.~Piclum and T.~Rauh, \emph{{The bottom-quark mass from
  non-relativistic sum rules at NNNLO}},
  \href{http://dx.doi.org/10.1016/j.nuclphysb.2014.12.001}{\emph{Nucl. Phys.}
  {\bfseries B891} (2015) 42--72},
  [\href{https://arxiv.org/abs/1411.3132}{{\ttfamily 1411.3132}}].

\bibitem{Beneke:2005hg}
M.~Beneke, Y.~Kiyo and K.~Schuller, \emph{{Third-order Coulomb corrections to
  the S-wave Green function, energy levels and wave functions at the origin}},
  \href{http://dx.doi.org/10.1016/j.nuclphysb.2005.02.028}{\emph{Nucl. Phys.}
  {\bfseries B714} (2005) 67--90},
  [\href{https://arxiv.org/abs/hep-ph/0501289}{{\ttfamily hep-ph/0501289}}].

\bibitem{Beneke:2015kwa}
M.~Beneke, Y.~Kiyo, P.~Marquard, A.~Penin, J.~Piclum and M.~Steinhauser,
  \emph{{Next-to-Next-to-Next-to-Leading Order QCD Prediction for the Top
  Antitop $S$-Wave Pair Production Cross Section Near Threshold in $e^+e^-$
  Annihilation}},
  \href{http://dx.doi.org/10.1103/PhysRevLett.115.192001}{\emph{Phys. Rev.
  Lett.} {\bfseries 115} (2015) 192001},
  [\href{https://arxiv.org/abs/1506.06864}{{\ttfamily 1506.06864}}].

\bibitem{Hoang:2014wka}
A.~H. Hoang, D.~W. Kolodrubetz, V.~Mateu and I.~W. Stewart,
  \emph{{$C$-parameter distribution at N$^3$LL$^\prime$ including power
  corrections}},
  \href{http://dx.doi.org/10.1103/PhysRevD.91.094017}{\emph{Phys. Rev.}
  {\bfseries D91} (2015) 094017},
  [\href{https://arxiv.org/abs/1411.6633}{{\ttfamily 1411.6633}}].

\bibitem{Boito:2025pwg}
D.~Boito, A.~Eiben, M.~Golterman, K.~Maltman, L.~M. Mansur and S.~Peris,
  \emph{{Strong coupling from hadronic $\tau$-decay data including
  $\tau\to\pi^-\pi^0\nu_\tau$ from Belle}},
  \href{http://dx.doi.org/10.1103/PhysRevD.111.074010}{\emph{Phys. Rev. D}
  {\bfseries 111} (2025) 074010},
  [\href{https://arxiv.org/abs/2502.08147}{{\ttfamily 2502.08147}}].

\bibitem{Bazavov:2016uvm}
A.~Bazavov, N.~Brambilla, H.~T. Ding, P.~Petreczky, H.~P. Schadler, A.~Vairo
  et~al., \emph{{Polyakov loop in 2+1 flavor QCD from low to high
  temperatures}},
  \href{http://dx.doi.org/10.1103/PhysRevD.93.114502}{\emph{Phys. Rev.}
  {\bfseries D93} (2016) 114502},
  [\href{https://arxiv.org/abs/1603.06637}{{\ttfamily 1603.06637}}].

\bibitem{Leino:2025pvl}
{\scshape TUMQCD} collaboration, V.~Leino, A.~Bazavov, N.~Brambilla, A.~S.
  Kronfeld, J.~Mayer-Steudte, P.~Petreczky et~al., \emph{{Strong coupling in
  (2+1+1)-flavor QCD}}, \href{http://dx.doi.org/10.22323/1.466.0298}{\emph{PoS}
  {\bfseries LATTICE2024} (2025) 298},
  [\href{https://arxiv.org/abs/2502.01453}{{\ttfamily 2502.01453}}].

\bibitem{Dehnadi:2015fra}
B.~Dehnadi, A.~H. Hoang and V.~Mateu, \emph{{Bottom and Charm Mass
  Determinations with a Convergence Test}},
  \href{http://dx.doi.org/10.1007/JHEP08(2015)155}{\emph{JHEP} {\bfseries 08}
  (2015) 155}, [\href{https://arxiv.org/abs/1504.07638}{{\ttfamily
  1504.07638}}].

\bibitem{DAgostini:1993arp}
G.~D'Agostini, \emph{{On the use of the covariance matrix to fit correlated
  data}}, \href{http://dx.doi.org/10.1016/0168-9002(94)90719-6}{\emph{Nucl.
  Instrum. Meth.} {\bfseries A346} (1994) 306--311}.

\bibitem{Tackmann:2024kci}
F.~J. Tackmann, \emph{{Beyond scale variations: perturbative theory
  uncertainties from nuisance parameters}},
  \href{http://dx.doi.org/10.1007/JHEP08(2025)098}{\emph{JHEP} {\bfseries 08}
  (2025) 098}, [\href{https://arxiv.org/abs/2411.18606}{{\ttfamily
  2411.18606}}].

\end{thebibliography}\endgroup
\bibliographystyle{JHEP}

\end{document}